\documentstyle[pre,aps,epsf,multicol,amsmath,amssymb]{revtex}

\begin{document}

%
\newcommand{\fig}[2]{\epsfxsize=#1\epsfbox{#2}}
%
%
%


 \newcommand{\passage}{
         \end{multicols}\widetext%
                \vspace{-.5cm}\noindent\rule{8.8cm}{.1mm}\rule{.1mm}{.4cm}} 
 \newcommand{\retour}{
         \vspace{-.5cm}\noindent\rule{9.1cm}{0mm}\rule{.1mm}{.4cm}\rule[.4cm]{8.8cm}{.1mm}%
         \begin{multicols}{2} }
 \newcommand{\unecol}{\end{multicols}}
 \newcommand{\deuxcol}{\begin{multicols}{2}}
%

\tolerance 2000

\title{Glass transition of a particle in a random potential,
front selection in non linear RG
and entropic phenomena in Liouville and SinhGordon models}
\author{David Carpentier {$^{1,2}$} and Pierre Le Doussal {$^2$}}
\address{{$^1$} Institute for Theoretical Physics, University of
California, Santa Barbara, CA 93106--4030}
\address{{$^2$} CNRS-Laboratoire de Physique Th{\'e}orique de l'Ecole Normale 
Sup{\'e}rieure, 24 Rue Lhomond, 75231 Paris\cite{lptens}
}

\maketitle

\begin{abstract}
We study via RG, numerics, exact bounds and qualitative arguments
the equilibrium Gibbs measure of a particle in a 
$d$-dimensional gaussian random potential with {\it translationally
invariant logarithmic} spatial correlations. We show that for any $d \ge 1$
it exhibits a transition at $T=T_c>0$. The 
low temperature glass phase has a non trivial structure,
being dominated by {\it a few} distant states (with
replica symmetry breaking phenomenology). In finite dimension
this transition exists 
only in this ``marginal glass'' case
(energy fluctuation exponent $\theta=0$) and disappears 
if correlations grow faster (single ground state dominance
$\theta>0$) or slower (high temperature phase).
The associated extremal statistics problem for correlated energy 
landscapes exhibits universal features which we describe using
a non linear (KPP) RG equation. These include the tails of the distribution of the
minimal energy (or free energy) and the finite size corrections
which are universal. The glass transition is closely related 
to Derrida's random energy models. In $d=2$, the connexion between this problem
and Liouville and sinh-Gordon models is discussed. The glass transition
of the particle exhibits interesting similarities with the
weak to strong coupling transition in Liouville ($c=1$
barrier) and with a transition that we conjecture for the sinh-Gordon model,
with correspondence in some exact results and RG analysis. 
Glassy freezing of the particle is associated with the generation under
RG of new local operators and of non-smooth configurations  
in Liouville. Applications to Dirac fermions in random 
magnetic fields at criticality reveals a peculiar ``quasi-localized'' regime 
(corresponding to the glass phase for the particle) where eigenfunctions 
are concentrated over {\it a finite number} of distant regions,
and allows to recover the multifractal spectrum in the
delocalized regime.
\end{abstract}

\deuxcol


\section{introduction}

Despite significant progress in the last two decades,
disordered systems continue to pose considerable theoretical
challenges. Two important questions still largely open,
are, respectively, to which extent the (better understood) mean field models
are relevant to describe low dimensional physical systems,
and, in the special case of two dimension, to which extent
the powerful field theoretic treatments developed for
pure models can be adapted to treat disordered models.

A celebrated controversy is whether the 
structure found in the solution of mean field 
models for spin glasses and other complex disordered 
systems, both in the statics \cite{parisi}
and in the dynamics
\cite{cuku}, has any counterpart in the world of experimentally
relevant low dimensional models. Specifically
it has been vigorously questionned \cite{fisher_huse} whether the breaking of the
phase space in ``many pure states'', predicted to occur in mean field,
may also occur in short range models, and
how it can be properly defined \cite{newman_stein,replica_ref}.
The unusual nature of the technique
used to solve the statics, i.e the replica method with 
a hierarchical breaking of the permutation symmetry
between replicas in the limit $n \to 0$ (RSB), did not contribute
to make the physics transparent. A distinct structure, 
which remarkably parallels
the one in the statics, has been found \cite{cuku}
to occur in the nonequilibrium dynamics. The dynamical
problem can be studied by a priori better defined methods 
and leads to predictions which are in principle
directly testable in experiments, such as a non trivial
generalization of the fluctuation dissipation relations. 
Even so, it has been emphasized that
mean field models, which usually involve infinite range
or infinite number of component limits, may not capture
physical processes important in low dimensions. The alternative
``droplet picture'' in its simplest form \cite{fisher_huse}
postulates the existence of a single
ground state with excitations (droplets) of (free) energy $\Delta E$ scaling 
with their size $x$ as $\Delta E \sim x^\theta$, $\theta>0$.
It provides a more conventional scaling description
of the glass physics, as being controlled by zero temperature RG
fixed points where temperature is formally irrelevant
(with eigenvalue $-\theta$).

Another important advance was the exact solution of 
simpler prototype models, such as the random energy model (REM)
\cite{derrida81}, where one consider simply a collection of
independently distributed energy levels, as well as
its generalization, the GREM \cite{derrida85},
or the Directed Polymer on the Cayley Tree (DPCT)
with disorder \cite{derrida88}. These solutions being direct
with no use of replica, their results can be fully relied upon.
They exhibit a similar physics though, with a glass transition
and in the glass phase, an exponential tail for the distribution
of the free energy $P(f) \sim e^{ \beta_c f}$ for negative
$f$. This feature is crucial to recover the same physics,
and indeed many observables were found to be similar \cite{derrida_rw}.
In fact the alternative solution of the REM using replicas,
given in \cite{derrida81} or the one of the DPCT 
\cite{derrida_dpct_replica} {\it do} involve RSB.
In the REM model the structure of the glass phase is particularly
transparent as being dominated by {\it a few} states
\cite{derrida81,derrida_rw}.

It is important to go beyond models defined in mean field
or on hierarchical (or ultrametric) structures and to
study simple yet non trivial (and non artificial) finite $d$ models
with full statistical translational invariance. In this paper
we study the model of a particle in a gaussian random
potential $V(\bf{r})$ with spatial correlations which
are {\it invariant by translation} and
which grow as the {\it logarithm} of the distance.
We consider this model in any dimension $d$, but in $d=2$
it has also been studied recently since it is of 
direct relevance for several physical systems
\cite{chamon96,tsvelik_prl,castillo97,nattermann_xy_lowtemp,scheidl_xy_lowtemp,%
tang_xy_lowtemp,korshunov_nattermann_energy,carpentier_xy_prl,carpentier_pld_long}.
One example is a spin model with $XY$ symmetry 
and random gauge quenched disorder, which arises naturally
in describing Josephson junction arrays \cite{arrays} or
2D crystalline structures with smooth disorder, e.g.
flux lattices in superconductors \cite{giamarchi_book_young},
or electrons at the surface of Helium \cite{electrons}.
In this model, a single topological defect (a $XY$ vortex), 
or a single neutral pair, sees precisely a random potential 
with logarithmic correlations \cite{nattermann_xy_lowtemp,scheidl_xy_lowtemp,%
tang_xy_lowtemp,korshunov_nattermann_energy,carpentier_xy_prl}.
Another example arises in a model of localization of Dirac fermions
in a random magnetic field, motivated by quantum Hall
physics. There, the zero energy $E=0$ normalized wavefunction 
is identical to the Boltzmann weight of the particle studied here
\cite{chamon96,tsvelik_prl,castillo97}. This wave function is
``critical'' in a sense discussed below. 

Here we study this model using a renormalization group
(RG) approach, bounds, numerical
methods and qualitative arguments. We show that it 
exhibits a transition at $T=T_c>0$ in any $d \ge 1$.
We find that in the high temperature phase
the particle is essentially delocalized over the whole system,
while in the low temperature glass phase the Gibbs measure is concentrated
in a few minima. The fact that such a simple (finite $d$) model exhibits 
a genuine glass transition is already surprising. Indeed, as we
argue, this transition exists {\it only} for such a ``marginal'' type
of correlations (which correspond to $\theta=0$ in the glass
scaling mentionned above \cite{footnote_theta}).
It disappears (for gaussian $V({\bf{r}})$)
if correlations grow faster (with only a low temperature phase and 
single ground state dominance) or slower (with only a high temperature phase).
Logarithmic growth of correlations thus produces exactly the right balance
between the depth of the energy wells and their number
(entropy). Note that for slower growing correlations
one can recover a transition but only by artificially rescaling
the disorder variance with the size of the system: in the extreme
case of uncorrelated variables, it is the REM model. Here by contrast, there is
a genuine phase transition in the thermodynamic limit, with no
need for rescaling. Most interestingly, the glass 
phase is non trivial. The Gibbs measure is concentrated 
in {\it a few distant minima}
which remain in a finite number in the thermodynamic limit. This is
because the extrema of random variables with such correlations
exhibit an interesting property of ``return near the minimum'':
there is, with a finite probability in a sample of size $L \to + \infty$,
at least one second minimum far away (at distances
of order $L$), and with a finite energy difference with the
absolute minimum. And there are not too many (a thermodynamic number)
of these secondary minima, leading to a zero entropy.
As in the REM, this property leads here naturally to a non trivial
ground state structure, reminiscent as we discuss
of a genuine property of replica symmetry breaking
in the replicated theory. The low temperature limit 
corresponds to a non trivial problem of extremal statistics of
{\it correlated} variables, studied here.

Another interesting property of this model is its relation to the
Liouville model (LM) and the sinhGordon model (SGM) 
in $d=2$ (and their boundary restriction in $d=1$):
$V({\bf r})$ turns out to be the Liouville field while
the LM and SGM partition functions arise simply as generating
functions of the probability distribution of the partition
sum $Z[V]=\int_{\bf r} e^{- \beta V({\bf r})}$ of a single particle.
The high temperature phase for the particle
corresponds to the weak coupling regime for the LM and SGM, where
we find that known exact results compare well with results for the particle.
In the SGM we predict here the existence of a
transition (more appropriately, a ``change of behaviour'').
It corresponds to the glass transition for the particule,
which also exhibits
interesting similarities with the weak to strong coupling transition
in the Liouville theory (and the so called $c=1$ barrier).
The glassy freezing of the particule is associated, in the
LM and SGM, to new local operators and non smooth configurations
being generated under RG. 

To study the model we will introduce a RG approach based on a
Coulomb gas renormalization a la Kosterlitz.
It leads to a non linear RG equation (of the so-called
Kolmogorov KPP type) for the full probability
distribution of the ``local disorder''. Indeed, a separation
between the long range part of the disorder and the local,
short range part, arises naturally in our approach. The 
RG equation has traveling wave types of solutions.
The corresponding well known problem, in such non linear equations,
of the selection of the velocity of the traveling front, and its freezing
for $T \leq T_c$ is related to glassy freezing of the particle free energy and,
in the LM or SGM, to the ``selection'' of the anomalous dimensions
(and at the transition dimension degeneracy leads to logarithmic
operators). When temperature is lowered the local disorder
becomes broadly distributed and the freezing occurs when its tails 
become relevant. Our RG method indicates that the physics depends only
weakly of $d$. We will take advantage of this fact and check our
results using simulations in $d=1$.

It is important to compare the present work to previous studies of the
model. The existence of a freezing transition in $d=2$ has been conjectured
previously \cite{korshunov_nattermann_energy,scheidl_xy_lowtemp,%
tang_xy_lowtemp}
based on an approximation which completely neglect spatial
correlations (REM approximation, see below). Stronger arguments 
were given in \cite{castillo97}, but did not fully establish
the existence of a transition, which is done here (see Appendix
(\ref{proof})). 
The present work is thus motivated by the need to go 
beyond the REM approximation to
describe this problem. In particular one wants to know what is the
precise universality class of the model, which we hope can be determined
from the RG method introduced here. This RG method yields some remarkable
universal features of the probability distribution of the free energy
and of its finite size corrections, different from the REM. It 
shows that the problem is more closely related to the 
directed polymer on a Cayley tree. A {\it qualitative}
analogy between the present model and the DPCT was in fact
cleverly guessed recently in Ref. \cite{tang_xy_lowtemp,chamon96}. It
is based on 
the observation that the energy of polymer configuration 
on a tree also scale logarithmically with the overlap distance defined 
on the tree (see Fig \ref{figdpct}). It is remarkable that this
connection naturally emerges here from the Kosterlitz type RG
performed on this problem, via the KPP equation.
It is all the more surprising, since the model studied here has
statistical translational invariance, while a tree has
a hierarchical structure. The solution of Derrida and Spohn \cite{derrida88}
(and the mapping onto the DPCT proposed in \cite{tang_xy_lowtemp,chamon96})
would be exact
for random variables $V({\bf r})$ correlated with a {\it hierarchical} 
(i.e ultrametric) matrix 
of correlation. Here instead the correlations are translationally invariant
and it is thus important to understand the origin of the analogy with the DPCT
and to which extent it holds. The RG procedure developed in this paper is
a first attempt to address these questions. The result is that we can make
the mapping precise: at least for the universal
observables studied here (e.g. the tails of the free energy distribution),
the mapping is onto a {\it continuum
branching process}, i.e a continuum limit of a Cayley tree
(whereas \cite{tang_xy_lowtemp,chamon96} could not be so specific).

\begin{figure}[thb]  
\centerline{\fig{3cm}{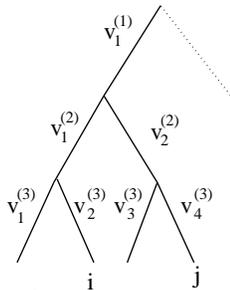}}
\caption{\narrowtext Directed polymer approximation: sites are the tree
endpoints. The $v^{p}_q$ are uncorrelated of variance $2 \sigma$. 
The random potential at $i$ is $V_i=v^{(1)}_1 + v^{(2)}_1 + v^{(3)}_2$ and
at $j$ it is $V_j=v^{(1)}_1 + v^{(2)}_2 + v^{(3)}_4$. Thus
$\overline{(V_i - V_j)^2}= 4 \sigma d(i,j)$ where $d(i,j) \sim \ln|i-j|$ is the
distance (in generations) on the tree \label{figdpct}}
\end{figure}

The present model has also been studied in the context of random Dirac
problems  
and localization. An early study \cite{grinstein} of the $E=0$ wavefunction 
established that it was critical (in the sense of corresponding
to a ``delocalized'' wave function, while $E \neq 0$ has finite
localization length). However this study missed the glass
transition. Later studies \cite{castillo97} computed the multifractal spectrum
based on the REM approximation and noticed the existence of a 
strong disorder regime. These and other studies \cite{chamon96,tsvelik_prl}
however focused on properties of the high temperature phase: it was conjectured
\cite{tsvelik_prl} that
the (conformal) Liouville field theory (LFT) (i.e a continuum limit of the LM)
was able to describe all spatial correlations of the model in the
high temperature phase. These works call for more investigations.
First the {\it glass transition} and
the peculiar physical properties of the low T (i.e strong disorder) phase have not
been addressed, even at the most qualitative level. We thus find
it useful to present the problem from a different perspective by
comparing with other types of correlated disorder, or by recasting it 
as a problem of extremal statistics. Although well known
properties \cite{galambos} of extremal statistics
of {\it uncorrelated} variables were often used to
study model disordered systems (see e.g. \cite{rammal,vinokur,bouchaud_mezard})
a lot remains to be understood about the (more realistic) case 
of correlated variables. Second, the question
of the universality class is in our opinion far from established. Evidence
for the LFT description mostly 
comes from reproducing the multifractal spectrum as given
by the REM approximation and one would like to check it against more
detailed predictions. The present RG procedure is a step towards
clarifying the connexion between this model and
solvable models such as Derrida's REM and Derrida-Spohn DPCT.
In this respect finite size corrections are important to understand,
as they are found to exhibit universal prefactors allowing to distinguish
between various universality classes. In addition, they determine the
anomalous dimensions, and thus control the critical behaviour, in the full disordered XY model
as shown in \cite{carpentier_xy_prl}. Since they are found to be very large, they
are also crucial in order to analyze the results of 
numerical simulations. In particular, although we confirm the result of
\cite{chamon96}, we also conclude that the sizes used in the numerical study of \cite{chamon96}
were in fact vastly insufficient for drawing firm conclusions:
we do perform here a more detailed finite size analysis
on much larger samples to confirm analytical predictions.

The model studied here is thus related to a surprising number of
interesting problems. Let us mention for completeness that it 
also has connexions to problems such as two dimensional interfaces, or films, confined 
between two walls (for $\beta=+\infty$ it is the confinement
entropy of a film), wetting transitions 
\cite{fisher_wetting}, extremal statistics of correlated variables useful
e.g. for problems of ``persistence'' in nonequilibrium dynamics
\cite{majumdar_sire}, and finally, to the clumping transition of a 
self gravitating planar gas \cite{tabar}. We will not explore these connections in
details here.

This paper is organized as follows:
in section \ref{part-Model}, the single particle model is defined and
in section \ref{part-REMapprox} the random energies approximation (REM)
is applied, which amounts to neglect the spatial correlations of the
random potential. The full problem, with correlations taken into
account, is related to the description of extremal statistics in 
\ref{part-extremalstats}, and three different classes of correlations
are identified in \ref{qualitative} from qualitative arguments.  
 A new renormalization (RG) technique is applied to this problem in section 
\ref{renormgroup}. The resulting non-linear scaling equation for the
distribution of the local disorder is studied in
section \ref{part-RGanalysis}, and found to be related
to the Kolmogorov KPP equation, which admits front solutions.
This connection between front solutions of non-linear equations and
the renormalization of disordered models is pursued in section
\ref{fronts}, where a solution to the REM is found via a similar non-linear RG
(details in Appendix C).
The non trivial nature of the glass phase is discussed in \ref{part-RSB}
together with its relations to replica symetry breaking.
 In part \ref{numerics} we present a numerical analysis of the problem
of the particle in a random potential in $d=1$. Section \ref{part-Liouville}
is devoted to the connexion between the particle model 
-and its transition- and entropic phenomena in the Liouville and Sinh-Gordon models.
A direct  RG analysis in  Section \ref{part-RGLiouville}
allows to recover the corresponding change of behaviours in
these models. Section \ref{part:dirac} contains the applications to 
the properties of the critical wave function of a Dirac fermion in a random magnetic
field, in particular the multifractal properties and the property of quasi-localization.
Appendix A contains an outline of the proof of the existence of
a transition, Appendix B is a review of well known (and not so well known) results about
extremal statistics, Appendix D contains an extended model which exhibits three
phases.

\section{Model and qualitative analysis}
\label{part-Model}

In this Section we define the model of a single particle in 
a correlated random potential. Then we describe the random energy model
(REM) approximation used in previous studies which consist in neglecting
correlations. We then pose the new questions which we want to address here
for the true model and present a qualitative analysis showing
physically why we expect that logarithmic correlations (as opposed
to faster growing or slower growing correlations) is the only case 
which leads to (i) a glass transition (ii) a low temperature phase
with a non trivial structure of quasi degenerate distant minima 

\subsection{the model}

The equilibrium problem of a single particle in a $d$-dimensional 
random potential is defined by the canonical partition function

\begin{equation} \label{def-1part}
Z[V]=\sum_{{\bf r}}  e^{-\beta V ({\bf r})}  
\end{equation}
where $\beta=1/T$ is the inverse temperature, 
in a sample of finite size $L$ (and total number of sites $L^d$)
and for a given configuration of the random variables $V({\bf r})$.
The equilibrium Gibbs measure, or probability distribution for the
position of the particle is:
\begin{eqnarray} \label{def-gibbs}
p({\bf r}) = e^{-\beta V ({\bf r})}/Z[V] 
\end{eqnarray}

We are interested here in cases where the random variables $V({\bf r})$
can be correlated. As discussed below, the statics (and dynamics) 
of this problem in the limit of large sizes depends on the type of correlations,
the distribution of the disorder and the dimensionality of space $d$.
Some of these cases and their dynamical aspects (such as the Sinai model)
have been extensively studied, e.g. in the context of diffusion in random 
media \cite{comtet98}. Even logarithmic correlations in $d=2$ were
studied then \cite{diff}, but it was not realized at that time
that a static glass transition could exist in that case.

Correlated random potentials $V({\bf r})$ are most 
conveniently studied for Gaussian distributions, on 
which we focus, parametrized by the correlator 
$\Gamma({\bf r},{\bf r}')= \overline{V({\bf r}) V({\bf r}')}$
(and we choose $\overline{V({\bf r})}=0$). Non gaussian
extensions will be mentionned. Unless specified
otherwise the correlations will be chosen translationally invariant 
$\Gamma({\bf r},{\bf r}')= \Gamma_{L}({\bf r}-{\bf r}')$
with cyclic boundary conditions, or in (discrete) Fourier space 
$\overline{V({\bf q}) V({\bf p}) }  = \Gamma({\bf q}, {\bf p} )= 
\Gamma({\bf q}) \delta_{{\bf p},-{\bf q}}$. We will
often denote
$\overline{(V ({\bf r})-V ({\bf r}'))^2} = \tilde{\Gamma}({\bf r}-{\bf r}')
= 2 \int_{\bf q} \Gamma({\bf q}) (1 - \cos({\bf q}.({\bf r}-{\bf r}')))$
(with $\int_{\bf q}= \frac{1}{L} \sum_{\bf q} 
\to \int \frac{d^d {\bf q}}{(2 \pi)^d}$).

One important quantity is the free energy:
\begin{eqnarray}
F[V]= - T \ln Z[V]
\end{eqnarray}
and, since it fluctuates from configuration to configuration,
as $F[V] = \overline{F[V]} + \delta F[V]$
we will be interested in its average $F=\overline{F[V]}$
and in its distribution. From the convexity of the 
logarithm follows the well known exact bound for $F$ in terms
of the annealed free energy $F_A$:
\begin{eqnarray}
&& - T \overline{ \ln Z} = F  \ge F_A = - T \ln \overline{Z} \\
&& F  \ge - ( T d \ln L + \frac{1}{2 T } \overline{V(\bf{r})^2})
\end{eqnarray}
for the Gaussian case.

In this paper we will mainly focus on the case 
of correlations growing {\it logarithmically} with
distance:
\begin{eqnarray}  \label{correlog}
\overline{(V ({\bf r})-V ({\bf r}'))^{2}} \sim 4 \sigma
\ln \frac{|{\bf r}-{\bf r}'|}{a}   \qquad  a \ll |{\bf r}-{\bf r}'| \ll L
\end{eqnarray}
which also requires a small distance ultraviolet (UV) cutoff $a$ (we
can set here 
$a=1$ in accordance with the definition \ref{def-1part} of a discrete model,
but in the following Sections we will consider a continuum version 
and vary $a$). This behaviour is
achieved in $d$ dimension by choosing a propagator in Fourier
space $\Gamma({\bf q}) \sim \frac{2 \sigma (2 \pi)^d}{S_d q^d}$. The 
$d=2$ case is also of special interest as the propagator
is the usual Coulomb one:
\begin{eqnarray} \label{coulomb}
\Gamma({\bf q}) \sim \frac{4 \pi \sigma}{q^2} 
\end{eqnarray}
and boundary conditions must be specified later on.
It is important to note that for LR correlations
the single site variance $\overline{V ({\bf r})^2} = \Gamma_L(0)$
diverges with the system size, e.g. for (\ref{correlog})
one has $\Gamma_L(0) \sim 2 \sigma  \ln (L/a)$.

For such logarithmic correlations (as well as for weaker correlations
\cite{footnote3}) one will find that $F$ scales as $d \ln L$
(consistent with the number of degree of freedom being $L^d$ in this problem).
Thus it is natural to define the intensive free energy:
\begin{eqnarray}
f(\beta)= \lim_{L \to + \infty} \frac{F[V]}{d \ln L}
\end{eqnarray}
which will be found to be self averaging. 
The above bound gives:
\begin{eqnarray}
f(\beta) \ge  - \left(\frac{1}{\beta} + \frac{\sigma}{d} \beta \right) 
\end{eqnarray}
Thus we will find that $F[V] \sim f(\beta) d \ln L$ with
subdominant corrections. These corrections have a non fluctuating
universal $O(\ln (\ln L))$ piece, as well as an $O(1)$ fluctuating
part $\delta F[V]$ which we will both study.

\subsection{the REM approximation}
\label{part-REMapprox}

A useful {\it approximation} to the problem studied 
here, which can be called the REM approximation,
consists in neglecting all correlations but keeping the
on site variance exact \cite{korshunov_nattermann_energy,castillo97}:
\begin{eqnarray}
\Gamma_L({\bf r}) \to \Gamma^{{\rm REM}}_L({\bf r}) 
= \Gamma_L(0) \delta_{{\bf r},{\bf r}'}
= 2 \sigma \ln\left(\frac{L}{a}\right) \delta_{{\bf r},{\bf r}'}
\label{remapp}
\end{eqnarray}
The corresponding Gaussian REM model can then be solved,
being identical to \cite{derrida81}, and one finds that
it exhibits a transition at $\beta_c = \sqrt{d/\sigma}$
with:

\begin{subequations}
\begin{eqnarray}
&& f(\beta) = - \left(\frac{1}{\beta} + \frac{\sigma}{d} \beta \right) 
\qquad \beta < \beta_c \\
&& f(\beta) = - \frac{2}{\beta_c} \qquad \beta > \beta_c
\label{remfree}
\end{eqnarray}
\end{subequations}
Most previous studies of the original model (all in $d=2$) amount to 
study the REM approximation and argue that it is a good
approximation. Indeed, as we will also find here,
this REM approximation appears to give the exact result for 
some observables (e.g. for $f(\beta)$). In particular, it does seem
to give correctly the transition temperature $\beta_c$.

\subsection{beyond the REM approximation: extremal statistics of 
correlated variables}
\label{part-extremalstats}

Since it is not obvious a priori why logarithmic correlations 
can be considered so weak as to be neglected,
one would like to go beyond the REM approximation and describe
the effect of the neglected correlations
\cite{footnote4}. One would like to understand
why this approximation works for some observables (and for which ones)
and whether it gives exactly the universality class of the model
(i.e all universal behaviour of observables). The answer to the latter
is negative: as our analysis will reveal, the correlations do matter for the
more detailed behaviour and the original model
(\ref{def-1part},\ref{correlog}) is {\it not} in the same universality
class as the REM model.

In fact, the problem at hand is related to describing universal 
features of the extremal value statistics for a set of 
{\it correlated} random variables. Indeed, the zero temperature limit
($T=0$ for fixed $L$) of the problem defined
by (\ref{def-1part}) amounts to finding the distribution
of the {\it minimum} $- \lim_{T \to 0} T \ln Z_L = V_{min}
= min_{r}(\{V_{r}\})$ of a set of
{\it correlated} random variables. In the case of 
uncorrelated (or short range correlated) variables 
a lot is known in probability theory on this problem (see
e.g. \cite{gnedenko}), 
some of which is summarized in Appendix \ref{galambos-app}.
For the type of distributions considered here 
(gaussian and some extensions) the distribution of
the minimum $V_{min}$ has a strong universality property, 
being given, up to - non trivial - rescaling and shift
(see Appendix \ref{galambos-app} and below),
by the Gumbell distribution: 
\begin{eqnarray}
\text{Proba}( y < x ) = {\cal G}(x) = \exp(- e^{x})
\label{gumbell0}
\end{eqnarray}
The Gumbell distribution thus appears as the distribution of the
zero temperature free energy in the REM.
For the case of a Gaussian distribution
the standard probability theory results are usually given 
in terms of a variable $X_{r}$ such that $\overline{X_{r}^2}=1$. 
One can simply rescale $V_{r} = \sqrt{2 \sigma \ln L} X_{r}$ 
from Appendix \ref{galambos-app} and get:
\begin{eqnarray}
V_{min} = 2 \sqrt{\sigma d} \ln L 
- \frac{1}{2} \sqrt{\frac{\sigma}{d}} \ln(4 \pi d \ln L) 
+ \sqrt{\frac{\sigma}{d}} y
\label{remgumbell}
\end{eqnarray}
where $y$ is distributed as in (\ref{gumbell0}).

Much less is known in the case of variables with stronger 
correlations studied here, though it is more important
in practice. The statistics of $V_{min}$ in the logarithmically correlated
case is thus one of the open issues discussed here.
One key question is to determine what is universal
in the distribution of the minimum of correlated variables.
Here, we can formulate the question as follows: given Gaussian
random variables satisfying (\ref{correlog}), what in the distribution
of the minimum (i.e of the ground state energy for fixed large $L$)
is universal, i.e depends only on $\sigma$ and not on
the details of the correlator $\Gamma({\bf r})$ 
at short scale. Writing
\begin{eqnarray}
V_{min} \sim \overline{V_{min}} + \delta V_{\min}
\end{eqnarray}
one finds, for the logarithmic correlator, that
the averaged ground state energy must satisfy:
\begin{eqnarray}
\overline{V_{\min}} \ge  - 2 \sqrt{\sigma d} \ln L
\end{eqnarray}
which follows from the 
above annealed bound, together with the fact that
$\partial f/\partial T = - S \leq 0$. Furthermore
one will find here that $V_{\min} \sim e_{min} d \ln L$
up to a positive subdominant - universal - piece
and that $e_{min}= - 2 \sqrt{\sigma/d}$ saturates the bound.
In the distribution of $\delta V_{\min} \sim O(1)$
we can clearly expect {\it less} universality
that in the problem of random variables with short range 
correlations \cite{footnote5}. 

\subsection{Qualitative study of a particle in a random potential}
\label{qualitative}

Before describing the RG method which allows to go beyond the
REM approximation, let us give some simple qualitative arguments
and numerical results which illustrate the main physics 
of the thermodynamics of a particle in a correlated random potential.
To put things in context we 
discuss several types of correlations (short range, long range, and
marginal). We focus on $d=1$ for simplicity but the arguments extend to 
any {\it finite} $d$.

Whether there is a single phase or not here comes simply from
whether the entropy of typical sites wins or not over the
energy of the low energy sites. When there is a low temperature phase,
to decide its structure one must pay special attention to
distant secondary local minima.

Indeed, when there is a low temperature phase, it is
controlled by the regions with most negative potential.
To investigate its structure one can start, for 
a given system of size $L$, with the $T=0$ state
which is determined by the absolute minimum over the system,
denoted $V_{min}$ and located at ${\bf r}_{min}$. At $T$ very small but 
strictly positive, each (low lying) secondary local minima $V$ 
will also be occupied with a probability $\sim e^{-(V - V_{\min})/T}$ 
which is very small except when $V-V_{min} \sim O(T)$.
Thus to characterize the low temperature phase we need to know 
how many of these secondary minima exist and where they are located.
For a smooth enough disorder (see e.g. Fig \ref{figlr}) there will
always be ``trivial'' secondary local minima in the vicinity of $r_{min}$.
To eliminate these, we define $V_{min2}(R)$ as the next lowest minimum
constrained to be at a distance at least $R$ of the absolute minimum.
An interesting quantity to study is then the distribution
$P_{R,L}(\Delta E)$ of $\Delta E(R,L) = V_{min2}(R) - V_{min}$ over 
environments (which a priori depends on $R$ and $L$)

\begin{figure}[thb]  
\centerline{\fig{8cm}{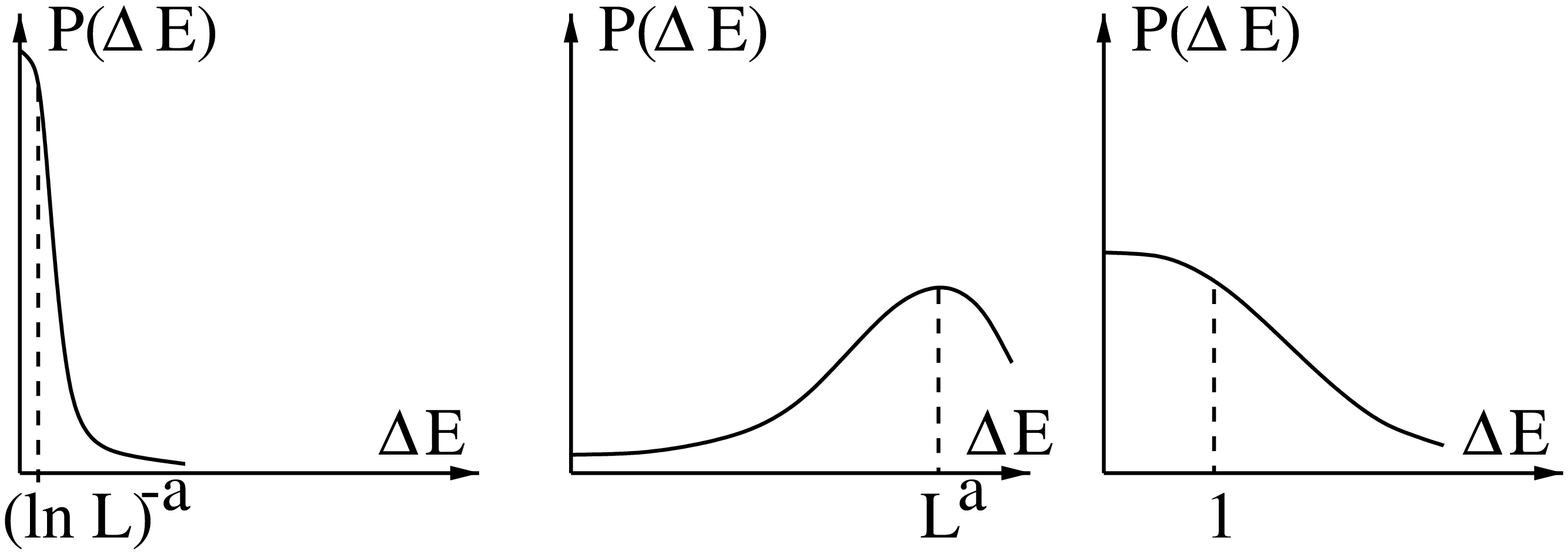}}
\caption{\narrowtext Three cases for the distribution of energy difference
$\Delta E$ between absolute and secondary minimum (separated at least
by $R\sim L^c$) in a system of size $L$:
(a) short range correlated potentials, $\Delta E_{typ} \to 0$
logarithmically with size
(b) algebraically growing correlations $\Delta E_{typ} \to + \infty$
(c) logarithmic correlations, $\Delta E_{typ}$ remains constant
as the system size increases. \label{fig1}}
\end{figure}

We now distinguish three main cases, according to the behaviour
of the  correlator $\overline{(V ({\bf r})-V ({\bf r}'))^{2}} 
= \tilde{\Gamma}({\bf r}-{\bf r}')$ at large scale
(we restrict to gaussian potentials \cite{footnote6}).
In these three cases the distribution $P_{R,L}(\Delta E)$ has markedly 
different behaviours as illustrated in Fig. \ref{fig1}:

\medskip

(i) {\it short range correlations}. i.e 
$\tilde{\Gamma}({\bf r}) \to Cst$ at large ${\bf r}$, equivalently
$\Gamma_{L}({\bf r}) \to 0$ at large ${\bf r}$ 
(or e.g. $\Gamma_{{\bf q}} \sim q^{- d + \delta}$ with $\delta > 0$).
In this case it is clear that {\it there is only a high temperature phase}
in any finite $d$ and no phase transition. 
The entropy $T d \ln L$ of typical sites (of energy typically $\sim O(1)$)
always wins over the energy of optimal sites ($V_{\min} \sim
\sqrt{2 \sigma d \ln L}$ for gaussian distributions with on-site variance $\sigma$
). The optimal
energy $V_{min}$ can be estimated using $1/L^d =
\int_{-\infty}^{V_{min}} P_1(V) dV$  
in terms of the single site distribution $P_1(V)$, which yields the 
exact leading behavior for uncorrelated disorder \cite{footnote15}
(and also for  
weak enough correlations - see Appendix \ref{galambos-app}).
Thus, the particle is delocalized over 
the system for all $T>0$.
One estimates the number of states within $\Delta E$
of the minimum as $N(\Delta E) \sim L^d \int_{V_{\min}}^{V_{\min} +
\Delta E} P_1(V) dV
\sim \exp(\Delta E \sqrt{2 d \ln L}/\sqrt{\sigma})$ for a Gaussian
distribution. Thus there is a large number
of sites almost degenerate with the absolute minimum 
$V_{min}$, separated by finite barriers, and $\Delta E_{typ}$ 
decays to 0 as a power of $1/\ln L$ ($1/\sqrt{\ln L}$ for a Gaussian)
\cite{footnote7}. These minima however are irrelevant for the
thermodynamics of the system at a fixed finite temperature.

For these minima to play a role and to obtain a transition
even for SR disorder one needs to perform some {\it artificial rescaling},
as in the REM model \cite{derrida81}: 
either, at fixed size, to concentrate on the very low $T$ region, 
(e.g. take $\beta \sim \ln L$ in the Gaussian case), or 
equivalently, to rescale disorder with the system size.
By making disorder larger as the system increases,
for instance using $P_1(V) \sim e^{- |V/V_{typ}|^{\alpha}}$
with $V_{typ} \sim (\ln L)^{1- 1/\alpha}$ one recovers
artificially a transition \cite{bouchaud_mezard}. For $\alpha=2$ and 
uncorrelated $V(\bf{r})$ this is exactly the REM 
studied in \cite{derrida81}. There, the simple argument for
the transition is that the averaged density of sites 
at energy $E=V$ is $\overline{\Omega(E)} = L^d 
e^{-E^2/(2 \sigma_L)}/\sqrt{2 \pi\sigma_L}$
(related to the annealed partition sum via
$\overline{Z} =\int_E e^{-\beta E} \overline{\Omega(E)}$).
If $\sigma_L=\sigma$ is not rescaled the average energy is $O(1)$
and the huge entropy of these states always wins. 
If $\sigma$ scales with $L$ as $\sigma_L \sim 2 \sigma \ln L$
then there is a transition at $\beta_c = \sqrt{d/\sigma}$.
Indeed, $\overline{\Omega(E)} \sim \exp(d \ln L (1 - (e/e_{min})^2))$
where $e=E/(d \ln L)$ and $e_{min}=E_{min}/(d \ln L) = - 2 \sqrt{\sigma/d}$
and there is a saddle point in 
$\overline{Z}$ at $\langle E \rangle /(d \ln L)= e_{sp}
=- \beta {e_{min}}^2/2$:
since $e_{sp}$ must be larger than $e_{min}$ (as 
$\overline{\Omega(\langle E \rangle)}$ cannot become smaller than $1$)
the saddle point cannot be valid below $T_c=1/\beta_c=e_{min}/2=\sqrt{\sigma /d}$
and the system freezes in low lying states.
Although this argument implicitly relies on
using $\ln \overline{\Omega(E)}$ instead of $\overline{\ln \Omega(E)}$
it does give the correct picture for the REM, as shown in \cite{derrida81}.

This picture generalizes to correlated potentials provided 
$\Gamma_L(r)$ decreases fast enough at large $r$. The decay
must be faster than $1/\ln r$ (which is a rather slow decay)
as indicated by the theorems recalled in Appendix \ref{galambos-app}
or also by a simple argument given in Appendix \ref{limrange}.
Finally, let us point out also that another way to obtain a 
transition for SR disorder is to take the $d =\infty$ limit before
taking the large $L$ limit: there the model (even without rescaling)
always exhibit a transition (in the statics and in the dynamics).

\medskip

(ii) {\it long range correlations}: when the typical
$V({\bf r})  - V({\bf r}')$ grows with distance as a power law 
$\tilde{\Gamma}({\bf r}) \sim |{\bf r}|^\delta$, 
{\it there is only a low temperature phase} and no transition.
The particle is now always localized near the absolute minimum 
of the potential in the system at ${\bf r}_{min}$.
The typical minimum energy $V_{\min}$ grows as $\sim - L^{\delta/2}$ 
and thus overcomes the entropy $\sim T d \ln L$ which is never sufficient
to delocalize the particle. The structure of this single low temperature phase 
is simple: there are no quasi degenerate minima separated by infinite distance
(and thus also by infinite barriers) in the thermodynamic limit. As can be seen
on Fig. \ref{figlr} there is typically a single minimum, with many 
secondary ones near it, but none far away. 
More precisely, as $L \to \infty$, the probability 
that the lowest energy excitation
$\Delta E(R,L)$ above the ground state (a distance at least $R \sim L^c$ 
from ${\bf r}_{min}$) be smaller than a fixed finite (arbitrary) value decays 
algebraically to $0$ with $L$ (and $\Delta E_{typ}$ and $\overline{\Delta E}$
increase algebraically with $L$). This is the scenario familiar from the
droplet picture \cite{fisher_huse},
with $Proba(\Delta E < T) \sim T L^{-\delta/2}$
(i.e in some configurations which become more and more rare as
$L \to +\infty$, there are two far away quasidegenerate ground states).
In some cases, e.g. in $d=1$ Sinai's model ($\delta=1$)
the distribution of rare events with quasi degenerate minima has been
studied extensively
\cite{golosov,laloux_pld_sinai,fisher_pld_monthus}. For instance it has been shown 
\cite{golosov,laloux_pld_sinai} that
there is a well defined limit 
distribution $Q(R) dR$ (when $L \to + \infty$) to find quasi degenerate
minima \cite{footnote_x2} at fixed distance between $R$ and $R+dR$,
with $Q(R) \sim R^{-3/2}$ at large $R$.

\begin{figure}[thb]  
\centerline{\fig{8cm}{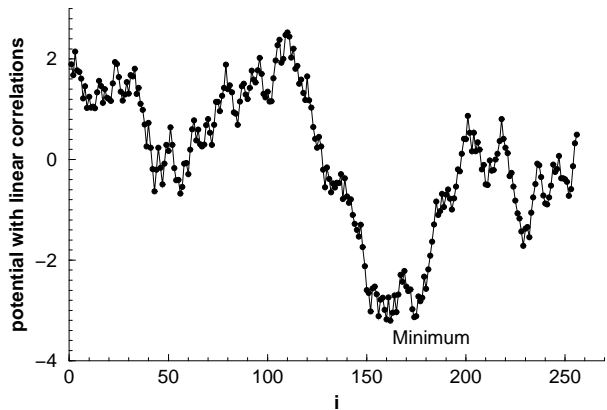}}
\caption{ \narrowtext A typical random potential configuration for
algebraically growing correlations \label{figlr} }
\end{figure}

\medskip

(iii) {\it marginal case, logarithmic correlations}: 
the most interesting case is when correlations grow 
as $\tilde{\Gamma}({\bf r}) \sim 4 \sigma \ln|{\bf r}|$.
A typical logarithmically correlated landscape is illustrated 
in Fig. \ref{figlogr}. One can already see
that, contrarily to Fig. \ref{figlr}, it has states with similar
energies far away. 
\begin{figure}[thb] 
\centerline{\fig{8cm}{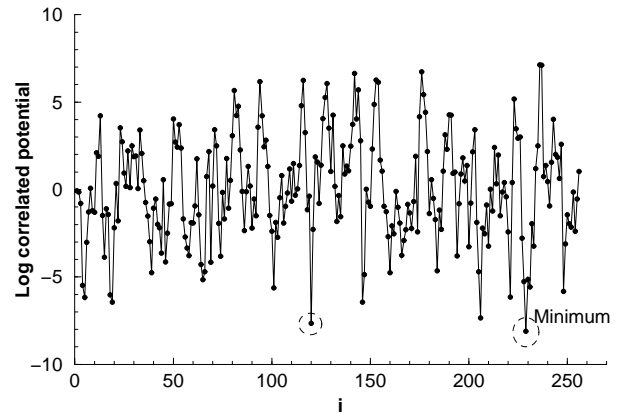}}
\caption{ \narrowtext A typical random potential configuration for
logarithmic correlations \label{figlogr} }
\end{figure}

Given the growth of correlations one sees that the {\it typical}
energy differences over a distance $L$ scale as 
$(V(0) - V(L))_{typ} \sim \pm \sqrt{ 4 \sigma \ln L }$. Computing the
minimum energy is a harder task here, but if one estimates it 
as in \cite{korshunov_nattermann_energy} through
the REM approximation $1/L^d = \int_{-\infty}^{V_{min}} P_1(V) dV$
(which neglects correlations), one finds that it 
behaves as $V_{min} \sim - 2 \sqrt{\sigma d} \ln L$ (for Gaussian disorder).
This estimate appears rather uncontrolled here since correlations {\it grow} with
distance, while the theorems for uncorrelated random variables 
apply a priori only for correlations {\it decaying} slower than $1/\ln r$. In fact
the situation is a bit more complex, and as we will find below from the RG 
and our numerics, the leading behaviour of $V_{min}$ with $\ln L$ 
is still correctly given by the REM approximation, although the
next subleading -universal- correction is not.
Thus the energy of the minimum $- 2 \sqrt{\sigma d} \ln L$
can now balance the entropy of typical sites $T d \ln L$ which
yields the possibility of a transition. The REM approximation
of the model indeed yields a transition at $T_c = \sqrt{\sigma/d}$
between a high temperature phase for $\beta < \beta_c = \sqrt{\sigma/d}$
and a frozen phase $\beta > \beta_c$. This scenario is confirmed by various
approaches in the following sections.

An interesting feature of this model is that the
low temperature phase exhibits a non trivial structure.
Unlike long range disorder discussed above, 
for logarithmic correlations we find that the low temperature phase 
is dominated not by one, but by {\it a few} states in the thermodynamic
limit. This is in stark contrast with the standard droplet picture and
is reminiscent of the replica symmetry breaking phenomenology, even though
we are dealing here with a very simple finite dimensional system.

One can visualize the transition, and the peculiar nature of the low
temperature 
phase in Fig. \ref{fight},\ref{figlt}, 
where a typical Gibbs measure $p({\bf r})$ is shown in both phases : 
is fairly delocalized at $T >T_c$ (Fig. \ref{fight}) 
but peaks around a few states when $T<T_c$ (Fig. \ref{figlt}) separated
by a distance of the order of the system size.  

\begin{figure}[thb] 
\centerline{\fig{8cm}{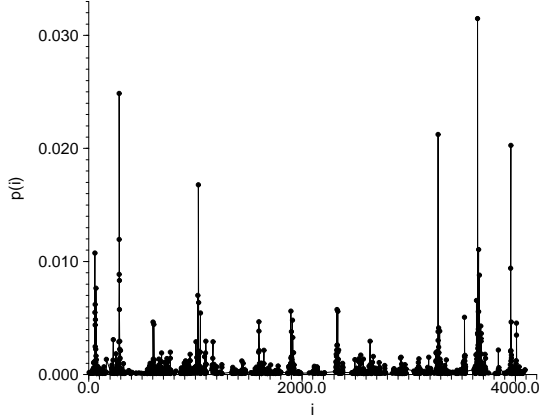}}
\caption{\narrowtext Gibbs measure in a typical sample in the
high temperature phase ($\beta=0.5< \beta_c=1$), $L=4096$ \label{fight}}
\end{figure}

\begin{figure}[thb] 
\centerline{\fig{8cm}{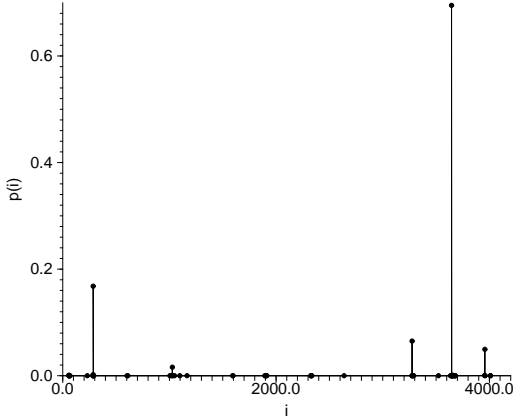}}
\caption{\narrowtext Gibbs measure in a typical sample in the
low temperature phase ($\beta=3. > \beta_c=1$), $L=4096$. Only points 
such that $p_i > 10^{-7}$ are indicated. \label{figlt}}
\end{figure}

This peculiar nature of the frozen phase can be tested by showing that
distant secondary local minima with a finite $\Delta E$ exist with
finite probability in the thermodynamic limit. Thus we have investigated
numerically the distribution $P_{R,L}(\Delta E)$ of the lowest excitation.
As illustrated in (Fig. \ref{fig1}), if the phase is non trivial, we
expect that this distribution has a well defined limit for 
e.g. $R=L/3$ when $L \to \infty$ with a
finite typical $\Delta E$. Contrarily to the LR disorder, we expect the
probability that e.g. $\Delta E(L/3,L)$ be smaller than a fixed number
to saturate (not to decrease) as $L \to \infty$, i.e that there
is a fixed probability that a second state within $\Delta E$
exists far away (as was already apparent in Fig. \ref{figlogr}).
We show in Fig. \ref{fig02}, Fig. \ref{fig023} 
and Fig. \ref{fig022} numerical evidence that this distribution 
has a well defined limit (the details of
the simulation are discussed in Section \ref{numerics}).
Finite size effects are clearly important in this system,
but their magnitude appears compatible with the predictions
of our RG approach, as discussed below. Thus we conclude 
that the numerics are consistent with
the existence of such a limit distribution and hence with 
a frozen phase with a non trivial structure.

\begin{figure}[thb] 
\centerline{\fig{8cm}{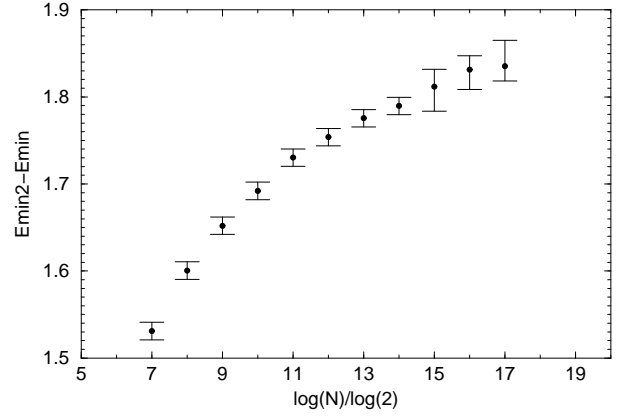}}
\caption{ \narrowtext Averaged energy difference $\overline{\Delta E}$
between the absolute minimum at $r_{\min}$ and the constrained
secondary minimum (i.e the minimum over the set $|r - r_{min}| > L/3$), 
as a function of the system size \label{fig02}}
\end{figure}

\begin{figure}
\centerline{\fig{9cm}{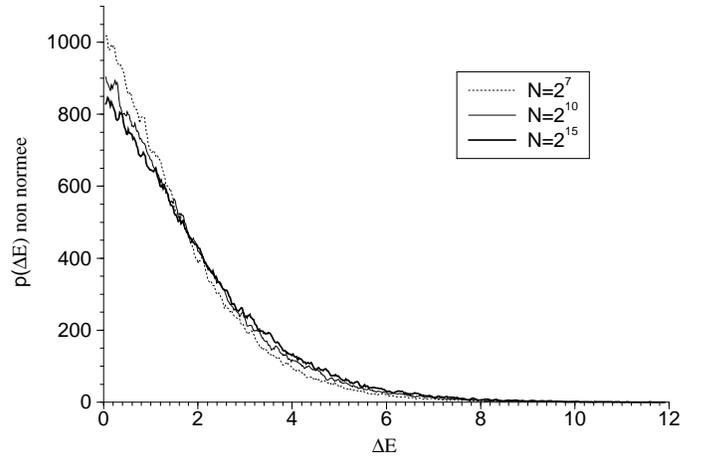}}
\caption{
\narrowtext
Distribution of $\Delta E$ , the energy difference between
the absolute minimum at $r_{\min}$ and the constrained
secondary minimum (i.e the minimum over the set $|r - r_{min}| > L/3$)
for different system sizes. \label{fig023} }
\end{figure}

\begin{figure}[thb]
\centerline{\fig{8cm}{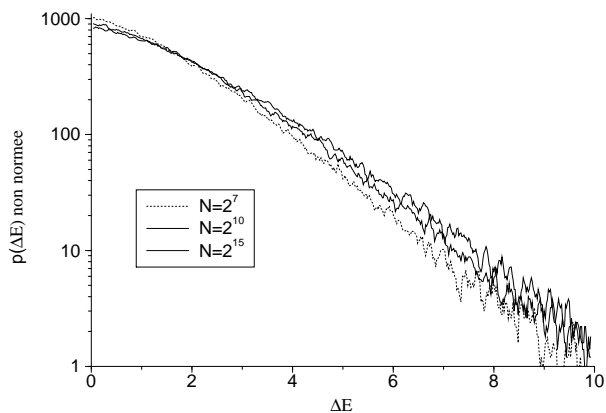}}
\caption{\narrowtext
Distribution of $\Delta E$ in log scale. \label{fig022}}
\end{figure}

\section{renormalization group approach}
\label{renormgroup}

\subsection{Idea of the method}

We now study the model (\ref{def-1part}, \ref{correlog}) 
using a renormalization approach introduced by us to study 
$d=2$ disordered XY models \cite{carpentier_xy_prl,carpentier_pld_long}.
There, one
is led to study a neutral collection of interacting $\pm1 $ charges 
(XY vortices) in a random potential $\pm V({\bf r})$ 
with (\ref{correlog}).
The single particle problem studied here amounts to restrict the 
Coulomb gas RG of \cite{carpentier_xy_prl,carpentier_pld_long} to the sector 
of a single $+ 1$ charge. Here however there is no charge neutrality
and one must be careful to study a system of finite size $L$,
as some quantities (such as $\overline{V({\bf r})^2}$) 
explicitly depend on $L$, while appropriately defined quantities 
have a well defined thermodynamic limit.

The idea is first to formulate the problem in the continuum,
with a short distance cutoff $a$:
\begin{equation}
Z=\int \frac{d^{d}{\bf r}}{a^{d}}
e^{-\beta V({\bf r})}
\end{equation}
and an appropriately defined cutoff-dependent distribution for
$V({\bf r})$, and second, by coarse graining infinitesimally,
to relate the problem defined with a cutoff $a'=a e^{dl}$
to the problem with a cutoff $a$. In general, this
implies to be able to follow under this transformation
the full probability measure of the potential $V({\bf r})$, 
which is quite difficult, as complicated correlations can be 
generated under coarse graining. In some very favorable
cases, for instance in the $d=1$ Sinai
landscape (where $V(r)$ performs a random walk as
a function of $r$ - case $\delta=1$), it is possible to
follow analytically an asymptotically {\it exact} RG transformation 
(in the statics and in the dynamics \cite{fisher_pld_monthus}).
There  a very specific real space decimation procedure is required,
which can in principle be extended here, although it may not be
tractable beyond numerics. The present case of the logarithmically
correlated potential 
is thus a priori less favorable but still, thanks to some known
properties of the Coulomb potential, a RG method a la Kosterlitz
can be constructed which, we argue, should
describe correctly all the universal properties 
of the model. There are two possible derivations, one which
uses replicas and is more precise, and the other one without.
We start with the latter, which is physically more
transparent.

The key observation is that before (and also after)
coarse graining, the logarithmically correlated
disorder studied here can naturally be decomposed into 
two parts as:
\begin{eqnarray}  \label{decomposition}
V({\bf r}) = V^{>}({\bf r})  + v({\bf r})
\end{eqnarray}
where $V^{>}({\bf r})$ is a smooth gaussian disorder
with the same LR correlations as the initial $V({\bf r})$ 
which represents the contribution of the scales larger 
than the cutoff $a e^l$, and $v({\bf r})$ is a local short range 
random potential which represents the contribution of scales
smaller than, or of the order of, the cutoff $a e^l$. In the starting
model $v({\bf r})$ appears naturally as a gaussian variable (see below).
After coarse graining, $v({\bf r})$ {\it does not} remain gaussian, but it 
{\it does} remain uncorrelated in space
(i.e correlations of short range $a$).
The decomposition (\ref{decomposition}) allows to 
 follow the distribution of the $V({\bf r})$ under 
coarse graining in a tractable way.

The precise way of decomposing the disorder in 
(\ref{decomposition}) depends on the details of
the cutoff procedure, but should not matter as far as
universal properties are concerned. For illustration let
us indicate a simple way to do it, a more detailed
discussion is given in \cite{carpentier_pld_long}. It starts with the
well known continuum approximation in $d=2$ of the 
lattice Coulomb potential $\tilde{\Gamma}({\bf r} - {\bf r}') \approx 
4 \sigma \left(  \ln \left(\frac{| {\bf r} - {\bf r}'|}{a}\right)
+ \gamma\right)(1-\delta^{(a)}({\bf r} - {\bf r}'))$
where $\delta^{(a)}({\bf r} - {\bf r}')=1$ 
for $|{\bf r}-{\bf r}'|<a$ and $0$ otherwise
($\gamma=\ln(2 \sqrt{2} e^C)$ and $C=0.5772$ is the Euler constant).
This decomposition can be performed more generally, e.g. with other
short-distance regularization of the potential $\tilde{\Gamma}({\bf r})$
(which preserve the large distance logarithmic behaviour)
and in any $d$, which amounts to modify the
value of $\gamma$. Using this approximation
the bare disorder (\ref{correlog}) can indeed be
rewritten equivalently as a sum (\ref{decomposition}) of two gaussian disorder 
$V^{>}({\bf r})$ and $v({\bf r})$ with no cross
correlations and with respective correlators:
\begin{eqnarray}
\label{long-range}
\overline{(V^{>}({\bf r}) - V^{>}({\bf r}'))^2} 
& = & 4 \sigma  \ln \frac{|{\bf r}-{\bf r}'|}{a} 
 (1-\delta^{(a)}({\bf r} - {\bf r}')) \\
\label{short-range}
\overline{v({\bf r}) v({\bf r}')} & = &
2 \sigma  \gamma \delta^{(a)}({\bf r} - {\bf r}')
\end{eqnarray}

With this definition, the problem to be studied is
rewritten as:
\begin{equation}  \label{defv}
Z = \int \frac{d^{d}{\bf r}}{a^{d}}~~ z({\bf r})~
e^{-\beta V^> ({\bf r})}  \qquad z({\bf r}) = e^{- \beta v({\bf r})}
\end{equation}

We can now study the behaviour of the model under a change of
cut-off. 
There are two main contributions from eliminated short length scales
variables. The first one can be 
seen most simply by rewriting the correlator 
in (\ref{long-range}):
\begin{eqnarray}  \nonumber
\overline{(V^{>}({\bf r}) - V^{>}({\bf r}'))^2} 
& = & 4 \sigma  \ln \frac{|{\bf r}-{\bf r}'|}{a'}
 (1-\delta^{(a')}({\bf r} - {\bf r}')) \\ \label{change}
& + & 4 \sigma dl ~ (1-\delta^{(a')}({\bf r} - {\bf r}'))
\end{eqnarray}
explicitly as the sum of a new LR disorder 
correlator with cutoff $a'=a e^{dl}$ and 
a SR disorder correlator (we have discarded terms of
order $O(dl^2)$). Thus the original problem with
cutoff $a$ can be rewritten as one with cutoff $a'$
with (i) a new gaussian LR disorder with identical
form of the correlator (\ref{long-range}) with $a$ replaced
by $a'$ (ii) a new short range
disorder $v({\bf r}) \to v({\bf r}) + dv({\bf r})$
with $\overline{dv({\bf r}) dv({\bf r}')} = 2 \sigma dl
\delta^{(a)}({\bf r} - {\bf r}')$
since it is clear from (\ref{change}) that
when $a \to a e^{dl}$ the LR disorder 
produces an additive {\it gaussian} 
contribution $dv$ to the SR disorder.

The second contribution resulting from a change of cutoff is
that neighboring regions will merge.
Points ${\bf r}_1$ and ${\bf r}_2$ previously separated as 
$a < |{\bf r}_1 - {\bf r}_2| < a e^{dl}$ should now be considered
as within the same region. The second important observation
is that the resulting transformation can only affect the SR part 
$v({\bf r})$ of the disorder. Indeed, in the region
$a < |{\bf r}_1 - {\bf r}_2| < a e^{dl}$ the LR part $V^> ({\bf r})$
can be considered as constant up to higher order terms of
order $dl$. One must view this coarse graining as
resulting in a ''fusion of local environments'' : 
the two local partition sum variables $z({\bf r}_1)$
and $z({\bf r}_2)$ combine into a single one $z({\bf r})$
according to a rule which we will write as
$z({\bf r}) = z({\bf r}_1) + z({\bf r}_2)$.
The exact choice of the form of this fusion rule
is again dependent of the cutoff procedure and
thus to a large extent arbitrary.

Putting together these two contributions we
obtain the following RG equation for the distribution
$P_l(z)$ of the local disorder $z=e^{- \beta v}$
variable (also called ''fugacity'' in the Coulomb Gas context).

\passage 
\begin{eqnarray} \label{rg-pz}
\partial_{l}P (z)=\beta^{2} 
\sigma \left(1 + z\partial_{z} \right)^2 P - d  P(z)
+d \int_{z'z''} P (z')P (z'')\delta (z- (z'+z''))
\end{eqnarray}
\retour

This equation also describes the evolution of the
universal part of the total free energy distribution 
with the system size. Indeed, the total
partition function can be written at any scale
as:
\begin{equation} \label{Z-coarse}
Z(\beta) = \int \frac{d^{d}{\bf r}}{a^{d}} e^{-\beta V({\bf r})}
\approx \int \frac{d^{d}{\bf r}}{(a e^l)^{d}} z_l({\bf r}) e^{-\beta
V_l^> ({\bf r})}   
\approx z_{(l^{*})}
\end{equation}
where the $z_l({\bf r})$ are independent variables 
distributed with $P_l(z)$ and the $V_l^> ({\bf r})$
are gaussian distributed as (\ref{long-range}).
In the last equality we have coarse grained up to 
the system size : 
$L= a~ e^{l^{*}}$. At this scale, there remains a single site 
of (random) fugacity
$z_{l^{*}}$. Thus the distribution function of the 
partition function $Z(\beta)$ can be
deduced from the distribution of the 
random fugacities at scale $l^{*}$. 
The distribution of the free energy $F=- T \ln Z$ is thus given 
by $\tilde{P}_{l^*}(v=F)$ (where $\tilde{P}(v) dv = P(z) dz$ from
the change of variable from $z$ to $v=-T \ln z$). Note that
the $\approx$ in (\ref{Z-coarse}) means that these
distributions are the same a priori only up to
subdominant non universal terms (multiplicative for
$Z$ and additive for $\ln Z$).

For a fixed system size $L$, the above RG equation
describes the evolution with the scale $l$ smaller that $l^*$
of the distribution of $z({\bf r})$, which is
the local partition sum over scales around ${\bf r}$
smaller or equal to $a e^l$ (i.e of 
a ``local free energy'' $ -T \ln z_l({\bf r}) = v_l({\bf r})$).
The remaining long wavelength disorder at that scale, 
$V^>_l(\bf{r})$ should
still be taken into account when computing the total
partition sum.

It is striking that the equation (\ref{rg-pv}) is identical to
the RG equation for the partition function of a continuum version of a
directed polymer on a Cayley tree (a so called branching process
\cite{derrida88}). We note that
it has been derived here for a problem with complete (statistical)
translational invariance, with no ad-hoc assumption about an
underlying tree structure and simply adapting to the present problem the
Coulomb gas renormalization a la Kosterlitz.
That the correspondence between the two problems
naturally appears within the RG with no
additional assumptions, is even more apparent on the 
derivation using replicas of the next section. Thus we consider
that this establishes on a firm footing the strong connection between the 
two problems. 

Before analyzing the consequences of the above RG equation
let us sketch the more precise derivation using replicas.
Other derivations without replicas are also possible and
we refer the reader to \cite{carpentier_pld_long} 
for more details.

\subsection{derivation of the RG equation using replicas}

Let us consider the whole set of moments $\overline{Z^{m}}$
which encode for the distribution function $P[Z]$. They
can be written as:
\begin{eqnarray}
\overline{Z^{m}}=
\int \frac{d^{d}{\bf r}_{1}}{a^{d}}\dots \frac{d^{d}{\bf r}_{m}}{a^{d}}
e^{\frac{\beta^{2}}{2}\overline{\left( \sum_{i=1,..m} V ({\bf
r}_{i})\right)^{2}} }
\end{eqnarray}
This can be rewritten as:
\begin{multline}
\overline{Z^{m}}=
\int \frac{d^{d}{\bf r}_{1}}{a^{d}}\dots \frac{d^{d}{\bf r}_{m}}{a^{d}}\\
e^{- \frac{\beta^{2}}{4} \sum_{i \neq j=1,..m} 
\tilde{\Gamma}({\bf r}_{i} - {\bf r}_{j})}
e^{m^{2} \sigma \beta^2 \ln \frac{L}{a} }
\label{part}
\end{multline}
We have used that $\Gamma({\bf r},{\bf r})= \Gamma_L(0) 
= 2 \sigma \ln \frac{L}{a}$. One can choose a regularisation,
e.g. $\Gamma({\bf r} - {\bf r}')
= \overline{V ({\bf r})V ({\bf r}')}=-
\sigma \ln \frac{|{\bf r}-{\bf r}'|^{2}+a^{2}}{L^{2}}$.
Notice that only the large distance behaviour of the above correlator is
important for the following renormalization.

We now switch to another representation of the replica
partition sum. (\ref{part}) is a partition sum of
$m$ particles located at ${\bf r}_1,..{\bf r}_m$ corresponding to 
$m$ replicas. Now instead
we will index the configurations using (vector) {\it columnar
replicated charges}. To each point ${\bf r}$, within a hard core
size $a$, we associate a $m$-component vector ${\bf n}$
whose components $n^i({\bf r})$ are either $1$ or $0$ depending 
on whether the particle corresponding to the $i$-th replica is present 
within $a$ of ${\bf r}$
($|{\bf r} - {\bf r}_i| < a$) or not. These charges thus correspond to
 ${\bf n}=(0,1,0,..0,1,1)$ since several replicas 
can be present near a given point. Choosing a columnar hard core 
for the vector charges corresponds to a choice of cutoff,
which is arbitrary, but the universal features of the
renormalization should not depend on it
\cite{footnote8}.

The $m-$th moment of $P[Z]$ then read
\begin{eqnarray}\label{partcg}
&&\overline{Z^{m}}=
\left(\frac{L}{a} \right)^{\beta^{2} \sigma m^{2}}
\sum'_{\{n_{\alpha}^{i} \}} \prod_{\alpha} Y[{\bf n}_{\alpha}]\\
\nonumber 
&&
\int_{|{\bf r}_{\alpha}-{\bf r}_{\alpha'}|\geq a}
\frac{d^{d}{\bf r}_{\alpha}}{a^{d}}
\exp \left(- 2 \beta^{2} \sigma \sum_{\alpha < \alpha'}
n_{\alpha} n_{\alpha'}
\ln \left(\frac{|{\bf r}_{\alpha}-{\bf r}_{\alpha'}|}{a} \right)\right)
\end{eqnarray}
where the primed sum correspond to a sum over all distinct non zero
configurations of
replica charges
${\bf n}_\alpha$ at sites ${\bf r}_\alpha$. We have defined
$n_{\alpha}=\sum_{i} n^{i}_{\alpha}$ as the total number
of replicas present in a given charge ($n_{\alpha }^{i}=1$). The quantities
$Y[{\bf n}]$ are functions of the local vector
charge and are the so-called vector charge fugacities.
In the bare model they appear as soon as the
continuum approximation to the lattice Green function is used
and read $Y[{\bf n}]= e^{ - 2 \sigma \gamma n^2}$. Since we are 
studying a single particle problem, there is also
an important
global constraint on the configuration sum 
that only one particle in any replica $i$ 
is present in the system, i.e:
\begin{eqnarray}
\sum_{\alpha} n^{i}_{\alpha} = 1
\label{constraint}
\end{eqnarray}
which is preserved by the RG.

The RG equations for this model read:
\begin{eqnarray}\label{RGreplicaFug}
\partial_{l}Y[{\bf n}]&=& \left(d+ \beta^{2} \sigma n^2
\right)Y[{\bf n}]\\
\nonumber &&
\quad \quad \quad \quad +\frac{S_{d-1}}{2}
\sum_{{\bf n}' + {\bf n}'' = {\bf n}}
Y[{\bf n}']Y[{\bf n}'']
\end{eqnarray}
where the sum is over ${\bf n}'$ and ${\bf n}''$
non zero vector charges (also ${\bf n}$ is non zero)
and $S_{d-1}$ is the volume of the unit sphere in dimension
$d$. We recall that $n =\sum_{i=1}^m n_i$.
These equations are obtained by a generalization 
of the Kosterlitz procedure \cite{nienhuis87} as follows. 
The first term comes from explicit cutoff dependence in
(\ref{partcg}). Upon increasing the cutoff infinitesimally $a\to a'=a e^{dl}$
the integration measure and the $a$ dependence in
all logarithms combine to
give $Y[{\bf n}_{\alpha}] \to Y[{\bf n}_{\alpha}] 
e^{dl (d + \sigma \beta^2 n_{\alpha}^2)}$.
We have used that $2 \sum_{\alpha < \alpha'}
n_{\alpha} n_{\alpha'} = m^2 - \sum_{\alpha} n_{\alpha}^2$
which holds due to (\ref{constraint}). The last term in the above 
equation (\ref{RGreplicaFug})
 comes from the fusion of replica charges upon increase of
the cutoff. The above RG equations hold for any $m$.

We should now look for solutions of this set of equations analytically
continued to $m \to 0$. One way to do that is to find a convenient
parametrization for the set of $Y[{\bf n}]$. Here we preserve 
replica permutation symmetry within the RG and we can thus
choose $Y[{\bf n}]$ to be a function
of $n= \sum_i n_i$ only. Then we define the parametrization 
$Y[n]=\int dz \Phi_{l}(z) z^{n}=\int du \tilde{\Phi}_{l}(v)
e^{- \beta n v}$. The
different terms in the above equation then translate into
\begin{eqnarray}
 n^{2} Y[{\bf n}] &=& \int dv e^{- \beta n v}
(\beta^{-1} \partial_{v})^{2} \tilde{\Phi}_{l} (v)\\\nonumber 
\sum_{{\bf n}^{(1)} + {\bf n}^{(2)} = {\bf n}} 
Y[{\bf n}_{1}]Y[{\bf n}_{2}] &=&
\int_{z',z''} \Phi_{l} (z')\Phi_{l} (z'')\delta (z-z'-z'')\\
&&-2{\cal
N}\Phi_{l}
(z)+\delta (z){\cal N}^{2}
\end{eqnarray}
where ${\cal N}=\int_{z} \Phi_{l} (z)$. One then easily converts the
equation for $\Phi_{l}(z)$ into an equation for 
a normed function $P_{l}(z)=\Phi_{l}(z)/{\cal N}_{>}$
defined only for $z>0$, with ${\cal N}_{>}=\int_{z>0} \Phi_{l}(z)$
(as in \cite{carpentier_pld_long}) by noting that 
${\cal N}_{>}$ converges quickly to ${\cal N}_{>}=2d/S_{d-1}$.
The resulting equation for $P_{l}(z)$ is exactly the
one (\ref{rg-pz}) given above, and its physical interpretation 
in terms of the probability distribution of the fugacity
(i.e the local partition sum) was given in the previous section.

What is the small parameter which controls the validity of
the above RG equations (with and without replicas)?
In conventional Coulomb gas context, these
RG equations are known to become exact in the dilute limit
of non zero (vector) charges \cite{nienhuis87}.
It is easy to see that this corresponds
to the tail of the distribution $P(z)$ for large $z$ (or equivalently
small $v$). This is further confirmed, a posteriori, by the remarkable
universality properties of the resulting non linear RG
equation (\ref{rg-pz}), analyzed in the following section, which arises precisely
in this region of $z$. So to obtain the universal behaviour 
(e.g. of the distribution of
free energy) we are working with sufficient accuracy. On the other
hand the bulk of the distribution $P_l(z)$ seem to be sensitive to 
details of the cutoff procedure (e.g. details in the fusion rule) and 
as discussed below, is thus likely (unless proven otherwise)
to be non universal.

\subsection{analysis of RG equation and results}
\label{part-RGanalysis}

\subsubsection{KPP front propagation equation and velocity selection}

Let us analyze the solutions to the RG equation (\ref{rg-pz}).
In terms of the (local free) energy variable $v({\bf r})=-T \ln z({\bf r})$
(from (\ref{defv}) and its distribution
$P_l(v) = P_l(z=e^{-\beta v}) \beta e^{-\beta v}$)
it has a well defined zero temperature limit, since then 
the fusion rule simply  becomes the extremal rule
$v' = \min(v_1,v_2)$ leading to :
\begin{eqnarray} \label{rg-pv}
\partial_{l}P (v)= \sigma \partial_{v}^2 P + d P(v) 
\left(-1 +
2 \int_{v}^{+\infty} P(v') dv'\right)
\end{eqnarray}
To be able to work at all temperatures, it is in fact
useful to trade the distributions $P_l(z)$ or $P_l(v)$ for the
generating function \cite{derrida88,footnote9}:
\begin{equation}\label{def-Gbis}
G_{l;\beta}(x)= \langle e^{- z e^{\beta x}} \rangle_{P_l (z)} = 
\langle e^{- e^{\beta (x - v)}} \rangle_{P_l (v)}
\end{equation}
We will sometimes drop the index $\beta$.
At zero temperature, the double exponential becomes
a theta function and $G_l(x)$ simply identifies with the distribution
function:
\begin{equation}\label{def-GbisT0}
G_{l;\beta = +\infty}(x)= \int_{x}^{+\infty} P_l(v) dv = \text{Proba}(v>x)
\end{equation}
and for all $\beta$ it is a decreasing function of
$x$ with $G_l(x \to - \infty) = 1$ and $G_l(x \to + \infty) = 0$.
Note the asymptotic behaviour 
\cite{footnote10} at very large 
negative $x$, $1 - G_l(x) \sim <z>_{P_l} e^{\beta x}$.
The temperature appears only via the initial condition
\cite{derrida88} and the problem at hand is thus to determine the large $l$ behaviour of
$G_l(x)$ for a given initial condition.

The equation (\ref{rg-pv}) is easily transformed, at all
temperatures, into the Kolmogorov (KPP)
non linear equation 
\begin{eqnarray} \label{KPP-bis}
&& \frac{1}{d}\partial_{l} G (x)=\frac{\sigma}{d} 
\partial_{x}^{2} G + F{[}G{]} \\
&& F{[}G{]}= - G (1-G)
\end{eqnarray}
which describes the diffusive invasion of a stable state
$G=0$ into an unstable one $G=1$. This class of
equations admits a family of {\it traveling wave solutions} $G_l(x)
=g (x + m(l))$ which describe a {\it front} moving towards 
negative $x$ and located around $x \sim - m(l)$.
This is readily seen by plugging this form in (\ref{KPP-bis})
and assuming that $\partial_l m_{\beta}(l) \to c$ one obtains
the equation for the front shape:
\begin{eqnarray}
\frac{1}{d} c g'(x) = \frac{\sigma}{d} g''(x) + F[g(x)]
\label{fronteq}
\end{eqnarray}
The family of such traveling wave solutions $g_c(x)$
can thus be parametrized by the velocity $c$. (\ref{fronteq})
simplifies for large negative $x$ when $g \approx 1$. Denoting
$\tilde{g}=1-g$ and using that 
$F[g] \sim - \tilde{g}$ for $g \approx 1$, one finds the
linearized front equation for $\tilde{g}_c$ :
\begin{eqnarray}
\frac{1}{d} c \tilde{g}' = \frac{\sigma}{d} \tilde{g}'' + \tilde{g}
\end{eqnarray}
This equation allows to relate the speed of the front $c$ to
the asymptotic decay of the front, since
if $\tilde{g}(x) \sim e^{\alpha x}$
for large negative $x$ one finds:
\begin{eqnarray}
\frac{c}{d} = \frac{\sigma}{d} \alpha + \frac{1}{\alpha}
\end{eqnarray}

The problem at hand now is to determine toward which of these
front solutions $g_c(x)$ will $G_l(x)$ converge at large $l$,
and thus what will be the asymptotic front velocity. This velocity 
will determine the intensive free energy of the original problem.
Indeed, the convergence at large $l$ of the solutions of non linear equations 
of the type (\ref{KPP-bis}) (with a general $F[G]$) towards one 
of such front solutions, and the corresponding problem of the selection 
of the front velocity $c$, is a famous problem, still under current interest 
in nonlinear physics \cite{saarloos89,brunet97,ebert98,ebert98b,saarloos98}.

The simplest argument is to use the fact that for very large negative $x$,
one must have $\tilde{g}(x) \sim e^{\beta x}$ and thus $\alpha = \beta$.
This seems to imply that the front velocity is:
\begin{eqnarray}
c = c(\beta) = \left(\frac{\sigma}{d} \beta + \frac{1}{\beta}\right) d
\label{cbeta}
\end{eqnarray}
This however is not always true. First note that the curve
$c(\beta)$ has two branches, i.e that in this
naive estimate two different $\beta$ would correspond to the same
velocity. 
The special point $\beta_c=\sqrt{d/\sigma}$ 
corresponds to $c=c^*=2 d \sqrt{\sigma/d}$. 
For more general non linear equations one usually relies 
on the so called marginal stability criterion 
(e.g. which
shows that the large $\beta$ branch is unstable and can be eliminated)
\cite{saarloos89,derrida88}
Here there are rigorous results available : the Bramson theorem \cite{bramson83}
ensures the following results, which are
{\it independent} of the precise form of $F[G]$ (up to some
rather weak conditions on $F[G]$ \cite{bramson83}):

(i) At high temperature, $\beta < \beta_c = \sqrt{d/\sigma}$
the asymptotic front is indeed an exponential for large negative $x$
and $G_l(x)$ uniformly converges towards the traveling wave solution
$g_{c(\beta)}(x + m(l))$ where the velocity is given by 
(\ref{cbeta}), thus continuously dependent on temperature.

(ii) At low temperature $\beta \ge \beta_c$ the velocity {\it freezes} 
to the value $c=c^*$ and the front decays as:
\begin{eqnarray}\label{formula43}
\tilde{g}(x) \sim - x e^{\beta_c x}
\end{eqnarray}
for large negative $x$,
thus independent of the temperature. The solution 
$G_l(x)$ uniformly converges towards the traveling wave solution
$g_{c^*}(x + m(l))$. Thus in that regime, 
one must then distinguish two regions in $G_l(x)$ at large $l$, the front
region and the region very far ahead of the front ($x + m(l) \gg \sqrt{l}$) 
where the decay is again as $G_l(x) \sim \exp(\beta x)$ as it should:
this will be discussed again below.

There are additional rigorous results from \cite{bramson83} and in particular 
the remarkable fact that not only the velocity but also the
{\it corrections to the velocity} are
{\it universal} (independent of $F[G]$) 
i.e one has for the position of the traveling wave $m_{\beta}(l)$ at 
``time'' $l$:
\begin{mathletters}
\begin{eqnarray}\label{KPP-m}
&& m(l) = \left( \frac{\sigma}{d} \beta + \beta^{-1}  \right) d l + Cst
\quad \beta < \beta_c= \sqrt{\frac{d}{\sigma}} \\
&& m(l) = \sqrt{\frac{\sigma}{d}} \left(2 d l - \frac{1}{2} \ln l\right)
\qquad \beta = \beta_c \\
&& m(l) = \sqrt{\frac{\sigma}{d}} \left(2 d l - \frac{3}{2} \ln l\right)
\qquad \beta > \beta_c 
\end{eqnarray}
\end{mathletters}

\subsubsection{Results for the fugacity and free energy
distribution and extremal statistics}

These results on the KPP equation (\ref{KPP-bis}) can now 
be translated (via (\ref{def-Gbis})) into results for 
the fugacity distribution $P_l(z)$ and for the distribution 
of free energy $\tilde{P}_l(v)$. One finds that $P_l(z)$ 
and $\tilde{P}_l(v)$ also take the form of 
a front at large $l$, e.g.:
\begin{eqnarray}
\tilde{P}_l(v) \to p(v + m(l))
\end{eqnarray}
with $p(v')$ related to $g(x)$ by 
$g(x)=\int_{v'} p(v') e^{- e^{\beta (x - v')}}$. Thus we obtain that
the local free energy is:
\begin{eqnarray}
-\beta^{-1} \langle \ln z \rangle \sim - m_{\beta} (l)
\end{eqnarray}
up to a finite constant, where the position of the front 
$m_{\beta}(l)$ is 
given above in (\ref{KPP-m}). Using the
result (\ref{Z-coarse}), $N=d \ln (L/a)= d l^{*}$
we obtain using the RG that the free energy $F[V]$ of the
system of size $L$ reads:
\begin{eqnarray}\label{formula47}
F[V] = f_L(\beta) d \ln L + \delta F
\end{eqnarray}
where $\delta F$ is a fluctuating part of $O(1)$
of probability distribution $p(\delta F)$ and the intensive free energy
reads:
\begin{subequations}
\begin{eqnarray} \label{fq}
&& f_L(\beta) = - \left(\frac{\beta}{\beta_{c}^2} + \frac{1}{\beta}\right) 
+ O\left(\frac{1}{\ln L}\right) \quad \beta < \beta_c= \sqrt{\frac{d}{\sigma}} \\
&& f_L(\beta) = - \frac{1}{\beta_c} \left(2 - 
\frac{1}{2} \frac{\ln (\ln L)}{d \ln L}\right) 
+ O\left(\frac{1}{\ln L}\right) \quad \beta = \beta_c \\
&& f_L(\beta) = - \frac{1}{\beta_c} \left(2 - 
\frac{3}{2} \frac{\ln (\ln L)}{d \ln L}\right) 
+ O\left(\frac{1}{\ln L}\right) \quad \beta > \beta_c 
\label{dpctfree}
\end{eqnarray}
\end{subequations}
where the factors $1/2$ and $3/2$ which arise in the
finite size corrections are {\it universal}.

Thus we have found using our RG method that in any dimension 
$d \ge 1$ the original model (\ref{def-1part},\ref{correlog})
exhibits a phase transition at $\beta=\beta_c(d)$. This transition is
very similar to the freezing transition of the continuous version of the 
random directed polymer on the Cayley tree. Our RG thus confirms that
the  REM approximation (\ref{remapp}) to the model does give the transition
at the same $\beta_c$, and with same asymptotic intensive free
energies (\ref{remfree}) as (\ref{dpctfree}). It allows however 
for a more detailed study and shows that the universal finite size 
corrections differ in the two model. In the REM the above formula
with the factor $1/2$ holds in all the low temperature phase, which
is not the case for the present model. Thus the present model is in a 
different universality class than the REM. The physics that we find 
here is much closer to the one of the directed polymer 
on the Cayley tree: it remains to be seen whether this
can be extended to other observables.

The RG method also yields the distribution of the $O(1)$ fluctuating part 
$\delta F$ of the free energy, and in particular at $T=0$ it gives 
a result for the extremal statistics of the correlated variables.
We must now carefully distinguish between what is clearly universal
(and thus for which we can be confident that the RG approach gives the
exact result) and what may not be (as it depends on the details of the cutoff
procedure, yielding e.g. a different KPP non linearity $F[G]$).

Let us start with $T=0$. We find (cf. (\ref{formula47},\ref{dpctfree}))
  that the minimum $V_{min}$ of $L^d$ 
logarithmically correlated variables behaves as:
\begin{eqnarray}
V_{min} = - 2 \sqrt{\sigma d} \ln L + \frac{3}{2} 
\sqrt{\frac{\sigma}{d}} \ln (\ln L) + \delta V
\end{eqnarray}
and $\delta V$ is a fluctuating part of order $O(1)$. 
Since at $T=0$ one has $p(v) = \tilde{g}'(v)$, from
the result (\ref{formula43}) 
we get that the tail of the distribution 
of $u = \delta V - <\delta V>$ for $u \to - \infty$ 
is universal and behaves as:
\begin{eqnarray} 
p(u)   \sim  - u e^{ \beta_c u }
\label{tail}
\end{eqnarray}
with $\beta_c = \sqrt{d/\sigma}$. Thus we find 
a distribution {\it different from the Gumbell distribution},
and thus correlations do matter. 

The question of what is universal in this distribution is
non trivial. We find from our method that the full distribution
of $P(u)$ depends on the detailed form of the front
(and thus on $F[G]$ and a priori on the cutoff procedure)
and is thus less likely to be universal (although 
this remains to be investigated). Hence we believe that universal 
features include {\it at least} the tail of the distribution
(\ref{tail}).

The above result (\ref{tail}) carries through the tail of the
distribution of the free energy $u = F - <F>$ for $u \to - \infty$ 
for $T< T_c$ and it was shown in 
\cite{derrida88} that for $T>T_c$ one has:
\begin{eqnarray} \label{fqbis}
&& p(u) \sim  e^{ u \beta_c^2/\beta}  \qquad \beta < \beta_c
\end{eqnarray}

\subsection{More on fronts, REM via nonlinear RG and extremal statistics}
\label{fronts}

To illustrate how the previous results fit in a broader context,
let us show how the simpler properties of extremal statistics of
uncorrelated variables and of the random energy model can be 
recovered within the same RG framework. This provides, en passant,
yet another solution of the REM.

\subsubsection{uncorrelated variables with fixed distribution:
Gumbell via RG}

Let us consider $N=e^{ld}=(L/a)^d$ independent random variables $V(r)$ 
$r=1,..N$ with a fixed distribution $P(V)$ ($d$ here does not
play any role as the true variable is $ld$ but we keep
it for the sake of comparison). The generating function
of the distribution of the partition function $Z[V]=\sum_r e^{- \beta V(r)}$ 
of model (\ref{def-1part}) reads:

\begin{eqnarray}\nonumber 
&& G_l(x) = < \exp( - Z[V] e^{\beta x}) >_{P(V)} \\
&& = \left( \int dV P(V) \exp\left( - e^{\beta (x - V)} \right) \right)^{e^{ld}}
\end{eqnarray}

It satisfies the equation:

\begin{eqnarray}
\frac{1}{d} \partial_l \ln \ln \frac{1}{G}  = 1
\label{lnln}
\end{eqnarray}

Or, interestingly enough, it obeys a KPP type equation
with no diffusion term:

\begin{eqnarray}
&& \frac{1}{d} \partial_l G = F{[} G {]} \\
&& F{[} G {]} = G \ln G
\end{eqnarray}

The Gumbell distribution now emerges naturally
from the front solutions of this equation.
Writing $G_l(x) \sim g( \alpha_l (x + m_l))$ 
and assuming $\partial_l (\alpha_l m_l) \to c$ yields 
$c g' = g \ln g$ whose solutions
with the above boundary conditions 
are $g(y) = \exp( - \gamma e^{y/c})$
($\gamma$ being a positive constant).
We have assumed $\partial_l \alpha_l \to 0$.
Since there is some freedom of choice 
for $\alpha_l$ and $m_l$, one can always 
set $c=\gamma=1$. The determination of the
rescaling factors $\alpha_l$ and $m(l)$ 
is performed in Appendix \ref{app-indep}.
At $T=0$ one has $P(V_{min}) = - G'(V_{min})$
and one recovers the known results from probability
theory for the convergence to the Gumbell distribution
detailed in Appendix \ref{galambos-app}, but the generating function
$G_l(x)$ takes a Gumbell form also at finite $T$.

\subsubsection{REM via RG}

We now turn to an alternative derivation of the 
solution to the Gaussian REM model using a
RG approach and a traveling wave analysis.
This allows to make some connections with the
correlated case studied previously. Let $l= \ln L$
and $\ln N=l d$.

We want to write a RG equation for:
\begin{eqnarray}
G_l(x) = \left(\left< e^{- e^{\beta (x-V)}}\right>_{P_l(V)}\right)^{e^{ld}}
\end{eqnarray}
where the single site distribution $P_l(V)$ now is scaled with $l$. 
We introduce
\begin{eqnarray}
\tilde{G}_l(x) = \left< e^{- e^{\beta (x-V)}}\right>_{P_l(V)} =
\exp \left(e^{-ld} \ln G_l(x)\right)
\end{eqnarray}
Let us choose the single site distribution $P_l(V)$ 
which corresponds to the REM approximation (\ref{remapp}) of the model
studied here (\ref{def-1part},\ref{correlog}), i.e the gaussian:
\begin{eqnarray}
P_l(V) = \frac{1}{\sqrt{4 \pi \sigma l}} e^{- \frac{V^2}{4 \sigma l}}
\end{eqnarray}
It satisfies:
\begin{eqnarray}
\partial_l P_l(V) = \sigma \partial_V^2 P_l(V)
\end{eqnarray}
One easily checks that it implies that:
\begin{eqnarray}
\partial_l \tilde{G}_l(x) = \sigma \partial_x^2 \tilde{G}_l(x)
\end{eqnarray}
This leads to the equation for $G_l(x)$:
\begin{eqnarray}
\partial_l G = \sigma \partial_x^2 G + d G \ln G - \sigma (1-e^{-l d})
\frac{1}{G} (\partial_x G)^2
\label{remkpp}
\end{eqnarray}
Thus the RG equation of the REM, for large $l$ reads:
\begin{eqnarray}
\partial_l G = \sigma \partial_x^2 G + d G \ln G - \sigma 
\frac{1}{G} (\partial_x G)^2
\end{eqnarray}
and is almost a KPP equation, except that it has an additional
gradient (KPZ type) term. This term here plays an important
role and yields a different universality class than KPP.
We now search for the front solutions. 

Let us rewrite the exact equation (\ref{remkpp}) 
using the function $h = - \ln G$ (remember that $0<G<1$):
\begin{eqnarray}
\partial_l h = d h + \sigma h'' + \sigma e^{-l d} {h'}^2 
\label{equah}
\end{eqnarray}
For large $l$ we can neglect the decaying nonlinear part,
and we now look for a solution of the linear equation.
The only front solution
of the form $h(x) = \tilde{h}(x + m(l))$ with $\partial_l m(l)
\to c$ which satisfies the boundary conditions $h(-\infty)=0$
and $h(+\infty)=+\infty$ is the exponential:
\begin{eqnarray}
&& h_l(x) = e^{\alpha (x + m(l))} \\
&& \partial_l m(l) = c = \frac{d}{\alpha} + \sigma \alpha
\end{eqnarray}
By using again the $h_l(x) \sim e^{\beta x}$ 
boundary condition at $x \to - \infty$, we find
$\alpha=\beta$ and:
\begin{eqnarray}
c(\beta) = \frac{d}{\beta} + \sigma \beta
\label{cb}
\end{eqnarray}
as in (\ref{cbeta}). This is correct in the high $T$ 
phase and yields the correct REM value for the intensive free
energy $f(\beta)= c(\beta)/d + O(1/\ln L)$ as in (\ref{remfree})
(and also correctly yields 
the absence of non trivial finite size corrections). 
Thus for the REM in the high $T$ phase we find:
\begin{eqnarray}
G_l(x) \approx \exp( - e^{ \beta ( x + m(l) )} )
\end{eqnarray}
thus again a Gumbell form, with $\alpha_l = \beta$ and 
$m(l)=(\frac{d}{\beta} + \sigma \beta) l$.

To see the transition to a low T phase for $\beta \ge \beta_c=\sqrt{d/\sigma}$
and the freezing of the velocity at 
$c=c^* = 2 \sqrt{d \sigma}$, one needs to carry a slightly
more detailed analysis (discarding again the decaying nonlinear part).
The general solution of the linear part of the equation (\ref{equah}) is:
\begin{eqnarray}
h_l(x) = \int dx' \frac{1}{\sqrt{4 \pi \sigma l} }
e^{ld - \frac{(x-x')^2}{4 \sigma l}} h_0(x')
\label{integ}
\end{eqnarray}
where $h_0(x')$ can be interpreted as the 
$h_l(x')$ at earlier time $l_0$ such that the
nonlinear terms can already be neglected
and decays as $h_0(x') \sim e^{\beta x'}$ for
$x' \to - \infty$.

This formula nicely exhibits the REM transition. 
In the high $T$ phase, using the asymptotic form $h_0(x') \sim e^{\beta x'}$
we find that there is a saddle point at $x'=x + 2 \sigma \beta l$.
This gives $h_l(x) \sim e^{\beta(x + c(\beta) l)}$ with $c(\beta)$ 
given in (\ref{cb}). The front $h_l(x)$ is centered at $x^*=
- c(\beta) l$ and consistency requires that the corresponding 
saddle point ${x'}^*$ moves to $- \infty$ 
so that the asymptotic form of $h_0(x')$ can
indeed be used. Hence we have ${x'}^* \sim (\sigma \beta - \frac{d}{\beta}) l$.
Thus the saddle point become inconsistent and the high $T$ solution
ceases to hold, for $\beta \ge \beta_c = \sqrt{d/\sigma}$.

The solution in the low $T$ phase is easy to find.
Setting $x = - m(l) + y$ one finds for large $l$:

\begin{eqnarray}
h_l(y) \sim e^{ld - \frac{1}{4 \sigma l} m(l)^2 - 
\frac{1}{2} \ln (4 \pi \sigma l) }
e^{ \frac{c^*}{2 \sigma} y}
\int dx' e^{- \frac{c^*}{2 \sigma} x'} h_0(x')
\label{limit}
\end{eqnarray}
where we have denoted $c^* = \lim_{l \to +\infty} m(l)/l$
and neglected the additional factor $e^{-{x'}^2/(4 \sigma l)}$ in the
integral. This is correct provided the integral:
\begin{eqnarray}
\int dx' e^{- \frac{c^*}{2 \sigma} x'} h_0(x')
\end{eqnarray}
is convergent, i.e $c^* < 2 \beta \sigma$. The consistent choice for $c^*$
and $m(l)$ must be:
\begin{eqnarray}
c^* = 2 \sqrt{\sigma d} \qquad m(l) = \sqrt{\frac{\sigma}{d}} 
\left( 2 l d - 
\frac{1}{2} \ln(4 \pi \sigma l) \right) + O(1)
\end{eqnarray}
which ensures that (\ref{limit}) has a proper limit
$h_l(y) \sim A e^{\frac{c^*}{2 \sigma} y} = A e^{\beta_c y}$
which is again a Gumbell form for $G_l(x)$ but now is
temperature independent. This holds for $\beta \ge \beta_c = \sqrt{d/\sigma}$.

From this method of solving the REM we have recovered the
result of \cite{derrida81} namely that for $\beta \ge \beta_c$
the free energy behaves as:
\begin{eqnarray}
f_L(\beta) = - \frac{m(l)}{d l} = - \frac{1}{\beta_c} 
\left(2 - 
\frac{1}{2} \frac{\ln (\ln L)}{d \ln L}\right) 
+ O\left(\frac{1}{\ln L}\right)
\end{eqnarray}
In addition we recover, for $T=0$ the result for the minimum $V_{min}$
in the REM approximation:
\begin{eqnarray}
V_{min} = - 2 \sqrt{\sigma d} \ln L + \frac{1}{2} 
\sqrt{\frac{\sigma}{d}} \ln (\ln L) + \delta V
\end{eqnarray}
with $u=\delta V - <\delta V>$
distributed with a Gumbell distribution:
\begin{eqnarray}
\text{Proba}(u > x) = \exp( - A e^{\beta_c x})
\end{eqnarray}
where $A$ is a constant.

\subsubsection{conclusion on RG fronts and extremal statistics}

Thus we have seen on two examples that extremal statistics problems
(and their $T>0$ thermodynamic model counterpart) can be
studied using non linear RG equation with traveling wave solutions.
In one example (uncorrelated rescaled variables, i.e the REM)
the RG equation is exact,
while in the second (logarithmically correlated variables) 
we only know it presumably in the tails. 
The front position 
represents the typical value of the minimum $V_{min}$
as a function of $l = d^{-1}\ln N$
while the shape of the front gives the distribution of the $V_{min}$
(resp. of the free energy $F$).
This suggests that a broader class of such models can be 
approached by these methods, and raises the question of universality.

Studies of such non linear equations \cite{saarloos98} usually distinguish
between pushed fronts where the velocity relaxes exponentially in
$l$ (velocity selection by non linear terms) and pulled fronts 
(velocity selection by the marginal stability criterion).
The extremal statistics
(and the glassy phase) correspond to the pulled fronts.
There one expects
a very broad universality as stressed in \cite{ebert98,ebert98b}:
not only is the asymptotic front universal but also the 
velocity and its corrections. In a nutshell, the argument
for the universal $\frac{3}{2} \ln l$ corrections to the front position
comes from matching of the universal tail of the front 
$g(y) \sim (A y + B) e^{- \beta_c \xi}$ with $y=x + m(l)$
with the far tail region, so far ahead of the front that 
one can linearize the KPP equation and get:
\begin{eqnarray}
1-G_l(x) \approx e^{- \beta_c y} \psi(y) 
\qquad  \partial_l \psi_l(y) = \sigma \partial_{y}^2 \psi_l(y)  
\end{eqnarray}
The only matching solution is $\psi_l(y)=y/l^{3/2} e^{-y^2/(4 \sigma l)}$.
Inserting $y=x + m(l) = x + c^* l + C \ln l$ immediately yields
$C=3/2$ for proper matching. As discussed in \cite{ebert98b}
this universality extend for {\it pulled fronts} 
in a very broad class of non linear (or coupled non linear) equations
and holds for steep enough initial conditions (i.e in the glass
phase in our language).

This argument fails in some cases, such as at the bifurcation between
pushed and pulled fronts (e.g. at the glass transition 
$\beta=\beta_c$ or equivalently when the initial condition has slow decay
$\sim \exp(- \beta_c x)$) (see e.g analysis in \cite{brunet97}).
Interestingly, it clearly fails also for the non linear equation
corresponding to the REM model, which is thus in a different universality
class
(this may be related to the fact that fronts are here unbounded
\cite{spc}). Presumably what happens there is that the coefficient
$A$ vanishes, and the solution is exactly $e^{- \beta_c y}$,
hence the $\frac{1}{2} \ln l$ (since the above matching function is now
$\psi(y) \sim l^{-1/2} e^{- y^2/(4 \sigma l)}$).

Next is the question of universality. We will address it only for
our model of gaussian variables with logarithmic correlations.
We have recast the RG equation (\ref{rg-pz}) into a KPP equation with a
specific non linear term $F[G]$. From our RG we have obtained
$F[G] = - G (1-G)$. The structure of the RG derivation suggests that we have
obtained correctly the two lowest orders of $F[G]$. From the above
discussion this is enough for the universality. Thus, and we call
it the restricted universality scenario, it is likely that
higher order terms $F[G] = - G (1-G) +  O((1-G)^3)$ are non universal
and thus that only the tail of the distribution of the minimum
of log-correlated variables is universal.

Let us mention however that we were not able to rule out another scenario,
the broad universality scenario, such that the true distribution of
the minimum of log-correlated variables is indeed universal. If this was true
the following conjecture would be tempting: since we know that for
{\it uncorrelated} variables the KPP RG equation is exact
with $F[G] = G \ln G$ and $\sigma=0$
(and see Appendix \ref{galambos-app} is asymptotically exact even for weakly correlated 
ones), one could conjecture an interpolating KPP equation (\ref{KPP-bis}) 
with $F[G] = G \ln G$ and $\sigma >0$ which would gives exactly the
distribution of the minimum of log-correlated variables.
Unfortunately we have been unable to confirm (but also to strictly 
rule out) numerically this conjecture,
due to the very large finite size corrections, as discussed in
Section \ref{numerics}.

\subsection{structure of low temperature phase and replica
symmetry breaking}
\label{part-RSB}

Let us now return to the structure of the low temperature phase
for the particle in the $d$-dimensional random potential with
logarithmic correlations. We argue that (i) it has a non trivial
structure, with a few states (ii) this structure is reminiscent of
the so-called ``replica symmetry breaking'' \cite{replica_ref}
This non trivial structure can be caracterised more
precisely here as the various states of the model correspond
to the different positions of the particle, and have thus a
natural meaning in real space. In particular, the minima
of the "energy landscape" (or metastable states) are nothing
but the local  minima in the sample of the random potential
for our problem. A precise caracterisation of these 'local
minima' is given below. Also, approximate replica solutions
of our model are shown in the following to exhibit RSB at low T.

\subsubsection{Spatial distribution of secondary minima}

Let us start with a simple argument: for a given realization
of disorder,  we divide our system into two subsystems of size $L^d/2$,
and call $V_{min1}$ and $V_{min2}$ the two corresponding minima in
each subsystem.  

Within the REM approximation,
we know from (\ref{remgumbell}) that $V_{min1} - V_{min2}
\sim (y_1 - y_2) \sqrt{\sigma/d} \sim O(1)$ where
$y_1$ and $y_2$ have independent Gumbell distributions. Thus clearly in that
case there is a non trivial structure: the secondary minimum (defined as being 
constrained to lie within the other subsystem) is typically 
within $\Delta E = O(1)$ in energy of the absolute minimum (and within
this approximation the distribution is also easily computed).

The RG analysis performed in this paper indicates that adding correlations will
not change this conclusion. Indeed, one first coarse grains up to scale 
$l_0=\ln(L)-\frac{1}{d} \ln 2$. At this scale, the system can be
described by two local energies (one for each half) of minima 
 $v_1$ and $v_2$ distributed
according to $P_{l_0}(v)$, to which should be added a term $\delta V$
which correlates the two halfes and is gaussian of variance 
$\sim \frac{2 \sigma}{d} \ln 2$. 
This however does not change the fact that the difference  
$V_{min1} - V_{min2} \sim O(1)$. Thus one still finds
that there exist secondary minima of $O(1)$ in energy from the
minimum, and a typical
distance $L$ away from the absolute minimum.
As discussed in Section \ref{qualitative}
this property was also confirmed by numerical simulations.

It is natural, in view of the analogy with the directed polymer on the Cayley tree,
to introduce the "overlap" between two different states (i.e positions of particles)
${\bf r}_1$ and ${\bf r}_2$ as: 
\begin{eqnarray}
q({\bf r}_1,{\bf r}_2) = 1 - \frac{\ln(a + |{\bf r}_1 - {\bf r}_2|)}{\ln L}
\end{eqnarray}
We expect it to be non self averaging and characterized 
by the ``overlap distribution'':
\begin{eqnarray}
P_2(q) = \overline{ \sum_{{\bf r}_1,{\bf r}_2} p({\bf r}_1) p({\bf r}_2) 
\delta(q - q({\bf r}_1,{\bf r}_2)) }
\end{eqnarray}
Although we have not attempted to compute this function directly
using our RG it is natural to expect that,
as in the REM and the DPCT, it is non trivial for $T<T_c$ and reads:
\begin{eqnarray}
P_2(q) = \frac{T}{T_c} \delta(q) + (1- \frac{T}{T_c}) \delta(1-q)
\end{eqnarray}
Similarly one expects that in a given disorder environment,
the probability of finding an overlap $q$
between two thermal realizations becomes in the large $L$ limit:
\begin{eqnarray}
&& \tilde{Y}(q) dq = (1-Y) \delta(q) + Y \delta(q-1) 
\end{eqnarray}
with $\overline{Y} = 1 - \frac{T}{T_c}$ and $Y$ has the same distribution
as in the REM. Thus the natural expectation, from the DPCT analogy,
is that the overlap in the low $T$ phase will be either
$1$ or $0$ (i.e secondary minima - of energy difference
of order $T$ - will be either near the absolute one $\ln r_{12}/\ln L \to 0$,
or a distance $r_{12} \sim O(L)$ typically a fraction of the system size away)
It would however be of interest to investigate further these
properties in the present model, 
in particular to obtain more detailed information at intermediate scales,
e.g. correlations probing the whole range $\ln r_{12} \sim (\ln L)^a$ 
with $0 \leq a\leq 1$.

\subsubsection{Approximate replica symetry breaking solutions of the
model}

Let us now turn to the replica representation and discuss
how the present model exhibits a form of ``replica symmetry breaking''.
The replicated partition sum reads:
\begin{eqnarray}
\overline{Z^{m}}=
\int \frac{d^{d}{\bf r}_{1}}{a^{d}}\dots \frac{d^{d}{\bf r}_{m}}{a^{d}}
e^{- 2 \sigma \beta^{2}  \sum_{i<j} 
\ln \frac{|{\bf r}_{i} - {\bf r}_{j}|}{a} }
e^{m^{2} \sigma \beta^2 \ln \frac{L}{a} }
\label{partrep}
\end{eqnarray}
It turns out that various approximations of this partition function 
(specifically the REM and the DPCT approximations) are dominated,
in the limit $m \to 0$ by replica symmetry breaking configurations.

In the context of 2d Dirac fermions with random vector potential
(see Section \ref{part:dirac}) an estimate of (\ref{partrep}) was given 
in \cite{grinstein}. For small $\beta$ it is clear that
the exponential containing
the logarithmic attraction between replicas does not decay fast
enough and thus the integral is dominated by the configurations
where the replicas are all far away $O(L)$ apart thus:

\begin{eqnarray}
\overline{Z^{m}} \sim  \left(\frac{L}{a}\right)^{\beta^{2} \sigma m^2 
+ d m - \beta^2 \sigma m (m-1) } =
\left(\frac{L}{a}\right)^{m d (1 + \frac{\sigma}{d} \beta^2 ) } \label{zrep}
\end{eqnarray}

This estimate of Ref. \cite{grinstein} is in fact
incorrect as it misses the glass transition.
Indeed, one can redo this argument using configurations
where $m/p$ packets of $p$ replicas are $O(L)$ far apart 
(while in each packet the replica (independent particles) 
are close to each other). This estimate was
performed in Ref. \cite{scheidl97} and gives instead:

\begin{eqnarray}
\overline{Z^{m}} \sim  \left(\frac{L}{a}\right)^{\beta^{2} \sigma m^2 
+ d \frac{m}{p} - \beta^2 \sigma m (m-p) } 
\end{eqnarray}
The interaction term being proportional to the number of
pairs of replicas in different packets, which is
$m(m-p)/2$. In the limit $m \to 0$
one can then optimise over $0 < u = p <1$, i.e:

\begin{eqnarray}
\overline{Z^{m}} \sim 
\exp \left( d \ln \frac{L}{a} 
\max_{0<u<1} \left( \frac{1}{u} + \frac{\sigma}{d} \beta^2 u \right) \right)
\end{eqnarray}
For $\beta < \beta_c = \sqrt{\frac{d}{\sigma}}$ the saddle is 
for $p=1$ and one recovers the above expression. For 
$\beta > \beta_c = \frac{d}{\sigma}$ one finds that
the saddle is for $u= \beta_c/\beta = T/T_c$
which gives:

\begin{eqnarray}
\overline{Z^{m}} \sim e^{ d \ln \frac{L}{a} 2 \frac{\beta}{\beta_c} }
\end{eqnarray}
Thus this calculation yields a transition. In Ref. \cite{scheidl97}
it was claimed that it does not correspond to replica symmetry breaking.
We believe that this is incorrect and that this {\it is} a (one step) RSB
estimate of the above partition sum. This is clear since 
this calculation exactly amounts to the corresponding one 
for the REM approximation of the model, i.e
replacing in (\ref{zrep})  $\sum_{i<j} \ln 
\frac{|{\bf r}_{i} - {\bf r}_{j}|}{a}$ by
$\sum_{i<j} (1 - \delta_{r_i, r_j}) \ln(L/a)$. In the REM we know from 
Ref. \cite{derrida81}
that the correct solution for $T<T_c$ can be obtained by
performing the analytical continuation to $m\to 0$ on a RSB saddle point
(note that the REM finite size correction $\frac{1}{2} \ln \ln L$ is
also obtained from the saddle integration).

One can go one step further and use an argument based on universality,
which puts the present problem in the DPCT universality class (for some
observables such as the free energy distribution).
For the DPCT, it was shown in Ref. \cite{derrida_dpct_replica}
that one can also recover 
the correct result for the averaged free energy by considering 
directed polymer configurations which break replica symmetry as $m \to 0$. 
It remains to be demonstrated how to obtain other universal
quantities, e.g. the $\frac{3}{2} \ln \ln L$ finite size
corrections, via a RSB saddle point calculation.

It is interesting to see how the features associated to RSB arise 
from the RG developed here, despite the fact that it is {\it explicitly 
replica symmetric}. Quite generally, if one can find independent local free energy 
variables with an exponential distribution $P(f) \sim e^{\beta_c f}$
one naturally obtains a RSB picture. This is the case here, up to
some more detailed universal preexponential structure in $P(f)$.
The important feature of our RG is thus that it follows
the full distribution $P_l(z)$ of local disorder 
(i.e. of local Boltzman weights $z$) 
which becomes algebraically broad as $l \to +\infty$.
Here this property is sufficient to show that the low $T$ phase
has a structure reminiscent of RSB. Indeed,
let us again coarse grain the system up to an already large scale
$L_0=a e^{l_0}$ but still much smaller than $L$, the ratio
$L/L_0 = e^{l_1} = M$ being {\it large} but {\it fixed} as $L \to +\infty$.
Assuming that $L_0$ is so large that $P_{l_0}(z)$ has reached its fixed point
already (except in a remote tail region corresponding to very rare events).
Since one has the decomposition (\ref{decomposition}) the
RG tells us that the sample is divided in $M$ subsystems with free
energies $F_i = v_i + V^>_i$, $i=1,..M$ where the 
variables $z_{i=1,..M}=e^{- \beta v_i}$ are
independently drawn from the common distribution $P_{l_0}(z)$
and the $V^>_i$ are still correlated but gaussian.
Neglecting first the $V^>_i$ we are left with a system of $M$ subsystems of
Gibbs measure:
\begin{eqnarray}
\frac{z_i}{\sum_j {z_j}} \label{fewpart}
\end{eqnarray}
Since the $z_i$ are 
drawn from a distribution 
with algebraic tails $P(z) \sim 1/z^{1+\mu}$ with $\mu=T/T_c$
one has $<z> = +\infty$ for $T<T_c$ and, as is well known, 
the partition sum (\ref{fewpart}) is dominated by {\it a few} 
of the $z_i$ variables \cite{derrida_rw,footnote18}
(which in essence is the physics associated to RSB).
Since the correlated $V^>$ variables are in 
finite number and with gaussian tails they cannot change the
exponential tails of the $F_i$ and thus adding them back 
should not change the above conclusions. 

Thus here, although the RG is replica symmetric, since it allows for generation
of broad tails it can capture features usually associated with RSB. 

\section{numerical study}
\label{numerics}

Since we found via the RG and other arguments that
there should be a transition in any dimension $d \ge 1$
it is particularly convenient to perform numerical
simulations in the ''extreme case'' of $d=1$ (i.e the
further away from mean field). However, even in $d=1$
numerical simulations are delicate because the 
finite size corrections are very large (and interesting
to study, in order to distinguish various universality class). 
Indeed we have found that the main numerical 
uncertainties come from the finite size effects and
not from the number of averages. In most of the numerical
work averaging over $\sim 10^{4}$ realizations of disorder was sufficient,
while a simulation of a system of size $2^{21} \sim 2~10^{6}$
leads to important corrections to the thermodynamic behaviour 
of the model. In view of this, we believe that the previous 
numerical investigation \cite{chamon96} was at best approximate.

We have considered a lattice model in $d=1$ with $L=2^n$ sites. The
potential $V(r)$ on each site ($r=1,..L$) was computed from its
Fourier components $V(r) = w_{L/2} (-1)^r +
\sum_{k=1}^{L/2-1} w_k \cos(2 \pi k r/L - \phi_k)$, eliminating the
$k=0$ mode, with $w_k$ independent gaussian variables
$\overline{w_k w_{k'}} = \Delta(k) \delta_{k,k'}$ ($k,k'=1,..L/2$)
and each $\phi_k$ independently distributed uniformly in
$[0,2 \pi]$. We choose $\Delta(k)$ such that:
\begin{eqnarray}\nonumber 
 \Gamma(r-r') &=& \overline{V(r) V(r')} \\
&=&  \sigma \frac{2 \pi}{L}
\sum_{k=1}^{L-1}
\frac{\cos( \frac{2 \pi k}{L} (r-r'))}{
|\sin(\frac{\pi k}{L}) | \sqrt{6 - 2 \cos(\frac{2 \pi k}{L})}}
\end{eqnarray}
so that $\overline{(V(r) - V(r'))^2} = 4 \sigma \ln(r-r')$
for $1 \ll r-r' \ll L/2$. This is the choice which also
corresponds to correlations along the axis $y=0$ 
on a 2d square lattice.

The behaviour of the model has been studied, without loss of
generality, at zero and at finite
temperature for a disorder strength $\sigma=1$ (others value of $\sigma$ can
be incorporated in the definition of the temperature scale).
We have first computed the average minimum $e_{min}=\overline{V_{min}}/\ln N$
(with $N=L$) for system sizes ranging from $L=2^{7}=128$ to
$L=2^{21} \sim 10^{6}$ and for each size we have taken the average
over $10^{4}$ realizations of disorder. 
An estimate of the uncertainty on the disorder average 
was made by measuring the variance of a series of average over $10^{4}$
realizations. This variance was found to be of the order of $10^{-3}$ for all
the value of $\overline{V_{min}}$. The results are plotted
in Fig. \ref{fig32}. We recall that the RG prediction reads for $\sigma=1$:
\begin{eqnarray}
\frac{1}{\ln N}\overline{V_{min}} = 2 \ln N - \frac{3}{2} \ln(\ln N)
+ O(1) 
\label{rgprediction}
\end{eqnarray}

We should first note that if one does not assume {\it anything} about the finite
size corrections, the resulting uncertainty on the ratio $e_{min}
=\overline{V_{min}}/\ln N$ is very large even for sizes $L=2^{21}$
since the ratio $\frac{3}{2} \ln(\ln N)/\ln N \approx 0.3$. Hence
with no assumption it is hard to estimate $e_{min}$ to better than
$10$ per cent accuracy.

However, if one assumes that $e_{min}=2$, the plot in Fig. \ref{fig32}
shows the existence of the $\ln(\ln N)$ corrections with a slope
definitely larger than $1$ and consistent with $3/2$ (although the
accuracy is not excellent). It is however sufficient to rule
out a REM type behaviour and is consistent with the RG prediction
(\ref{rgprediction}).

\begin{figure}
\centerline{\fig{9cm}{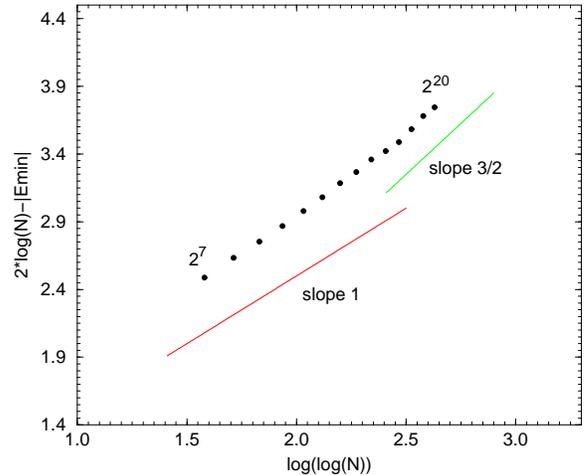}}
\caption{
\narrowtext
Zero temperature limit: finite size corrections to
the minimal energy. Plotted is $2 l - |V_{min}|$
versus $\ln l$ ($l= \ln N$) \label{fig32}}
\end{figure}

Next, we have plotted the distribution of $V_{min}$
in Fig. \ref{fit}
and compared with the prediction of the RG for the tails.
Here also the agreement is satisfactory.

\begin{figure}
\centerline{\fig{9cm}{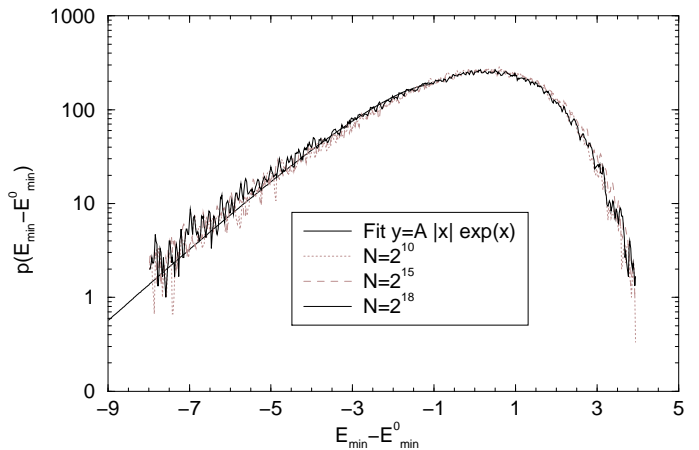}}
\caption{
\narrowtext
Distribution of $E_{min}$ \label{fit}}
\end{figure}

Finally, we have plotted the ``glass order parameter'' 
$Y_2 = \overline{\sum_r p_r^2}$ which is non zero when the
system is dominated by a few states. It is consistent with a very slow
convergence towards $Y_2=(1-T/T_c) \theta(T_c-T)$ but clearly
other forms cannot be ruled out.

\begin{figure}
\centerline{\fig{9cm}{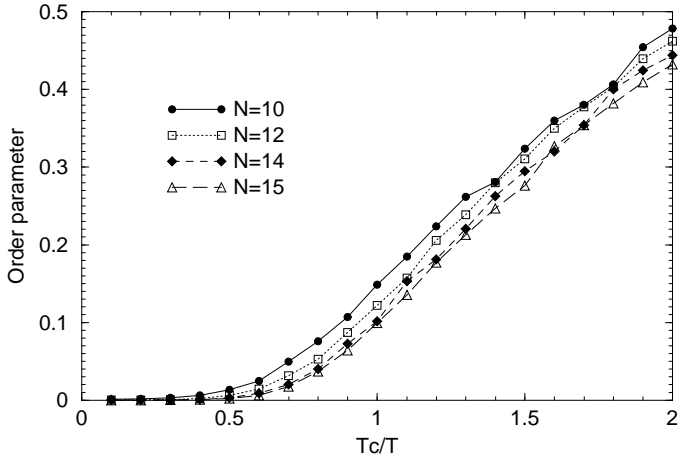}}
\caption{
\narrowtext
Plot of $Y_2 = \overline{\sum_r p_r^2}$ as a function of
temperature for different system sizes $L=2^N$ }
\end{figure}

\section{Relations with Liouville and Sinhgordon models}
\label{part-Liouville}

In this Section we describe the relation between
the problem of the particle in the log-correlated random
potential and the Liouville and sinh-Gordon models.
Exact results on the sinh-gordon model are compatible with (and 
also point out towards) the existence 
of the transition at $\beta=\beta_c$.

\subsection{Relations with the sinh-Gordon model in $d=2$ and $d=1$}

Let us start with the correspondence with the sinh-Gordon model.
Although less direct, it is also simpler to analyze,
as the model does not contain subtle boundary conditions
problems. The interest of the connexion is that the 
sinh-Gordon model is integrable in $d=2$ and $d=1$ (Boundary
sinh-Gordon) \cite{lukyanov,fateev,korepin}. 

The connexion requires introducing a slightly different
version of the initial problem, defined by the partition function:
\begin{eqnarray}
Z_{sh}[V] = Z[V] + Z[-V] = \sum_{\bf r} ( e^{- \beta V({\bf r})} +
e^{ \beta V({\bf r})} ) 
\label{sinh}
\end{eqnarray}
which corresponds to a particle in a random potential which can explore both
$V({\bf r})$ and $-V({\bf r})$. A physical realization would be
a particle with an Ising spin in a random field. As it turns out the
physics of this disordered model is very similar to the original
problem. At low temperature, it is now related to the distribution 
of the minimum of $- |V({\bf r})|$.

We define the generating function of this model  
$G_{sh}(x) = \left<\exp(- \mu Z_{sh}[V])\right>$, with
$\mu = e^{\beta x}$, which is related to the distribution of the
free energy of the particle. In the continuum limit and in $d=2$, it
can be rewritten as:
\begin{eqnarray}
&& G_{sh}(x) = H_{sh}[\mu] = \int DV e^{- S_{sh}[V] }  \\\nonumber 
&& S_{sh}[V] = \int_{-\infty}^{+\infty}  \int_{-\infty}^{+\infty} dxdy 
\left( \frac{1}{8 \pi \sigma} (\nabla V)^2  + 2
\mu \cosh(\beta V({\bf r}))\right)
\label{sinh2}
\end{eqnarray}
i.e the partition function of the sinh-Gordon model in $d=2$. 
Similarly, the $d=1$ version of our model is related to the
well studied boundary sinh-Gordon model \cite{fateev}
defined as:
\begin{eqnarray}
&& G_{shB}(x) = H_{shB}[\mu] = \int DV e^{- S_{shB}[V] }  \\
\nonumber 
&& S_{shB}[V] = \int_{0}^{+\infty} dy 
\int_{0}^{L} dx \biggl( \frac{1}{4 \pi \sigma} (\nabla V)^2  \\
&& \quad \quad \quad \quad \quad \quad \quad \quad  \quad \quad \quad \quad + 
2 \mu \cosh(\beta V(x,0))  \biggr)
\end{eqnarray}
Indeed one has, as required, that $(V(x,0) - V(x',0))^2 \sim
4 \sigma \ln \frac{|x-x'|}{a}$ at large $|x-x'|$,
and one only studies (boundary) observables defined at $y=0$. 

In the limit of $\beta=+\infty$ one has in both case:
\begin{eqnarray}
G_{sh}(x) &=& \text{Proba}( x < \text{min}(V_r , - V_r) ) \\ 
&=&
\text{Proba}( x < - \text{max}_{r} |V_r| )
\end{eqnarray}
and thus the (properly discretized) partition function of the 
(boundary) sinh-Gordon model
becomes related, in that limit, to the distribution function of
the maximum of the set of positive random variables $|V({\bf r})|$.
The results described in the previous Sections about the statistics 
of extrema of such variables imply that some transition must occur
as a function of $\beta$ corresponding to a related ``change of behaviour'' 
 in the sinh-Gordon and boundary sinh-Gordon models as well. 
This is a prediction, as we are not aware of such a change of behaviour
at $\beta = \beta_c$ being mentionned in the literature. 
As we now discuss, examination of known results is perfectly
compatible with the transition at $\beta = \beta_c$.

Let us first describe the known exact results both in $d=2$
and $d=1$. The extensive free energy of the bulk sinh-Gordon model
is defined as: 
\begin{eqnarray}
f_{sh} = \lim_{L \to + \infty} - L^{-2} \ln G_{sh}
\end{eqnarray}
where the model defined in (\ref{sinh2}) is considered in
finite size $L$. The model is studied usually using the field
$\phi = V \sqrt{2/\sigma}$, the nonlinear term
being $2 \mu \cosh(\beta V) = 2 \mu \cosh(b \phi)$
and its free energy depends on the single variable 
$b = \beta \sqrt{\sigma/2} = \beta/\beta_c$,
where $\beta_c=\sqrt{d/\sigma}$ is dimension dependent.
Using the variable $b$, its exact expression, proposed in Ref. 
\cite{lukyanov}, 
reads when explicited \cite{footnote17}:
\begin{subequations}
\begin{eqnarray}
&& f_{sh}(\mu) = C_2(b) \mu^{\frac{1}{1+ b^2}} \\
&& C_2(b) = \frac{2 \pi}{
\left(\Gamma[\frac{1}{2 + 2 b^2}] \right)^2 
\left(\Gamma[1 + \frac{b^2}{2 + 2 b^2}] \right)^2 
\sin(\frac{b^2 \pi}{1 + b^2})}\nonumber \\
&& \quad \quad \quad \quad  \quad \quad \quad \quad \times
\left(\frac{\pi \Gamma[1 + b^2]}{- \Gamma[-b^2]}
\right)^{\frac{1}{1+ b^2}} \nonumber
\end{eqnarray}
\end{subequations}
These results
are a priori only valid for $b<1$ ($|b| < 1$), as they 
were obtained in \cite{lukyanov} from an analytical continuation of
the sine-Gordon model (performing $\mu \to -\mu$ and $b^2 \to - b^2$,
$M$ being the soliton mass). The constant $\mu$ was defined 
in the continuum model by fixing the 
normalization of the field $<\cos(b \phi({\bf r}) )\cos(b \phi{\bf r}')>
= \frac{1}{2} |x-y|^{- 4 b^2}$ of the sine-Gordon model.

The $d=1$ version corresponds to the boundary sinh-Gordon model
usually studied using the $\phi =V/\sqrt{\sigma}$ and
$2 \mu \cosh(\beta V) = 2 \mu \cosh(b \phi)$, with 
again $b=\beta/\beta_c$ ($\beta_c = 1/\sqrt{\sigma}$). The analogous
expression for the free energy reads, from \cite{fateev}:
\begin{eqnarray}
&& f_{shB}(\mu) = \lim_{L \to + \infty} - L^{-1} \ln G_{shB} =
C_1(b) \mu^{\frac{1}{1 + b^2}} \\
&& C_1(b) = \frac{1}{8 \pi^{3/2}}
\Gamma\left[\frac{1+2 b^2}{2 + 2 b^2}\right]
\Gamma\left[\frac{- b^2}{2 + 2 b^2}\right]
\left( - \frac{2 \pi}{\Gamma[- b^2]} \right)^{\frac{1}{1 + b^2}}
\nonumber
\end{eqnarray}

Let us now comment on these results. The power law dependence
in $\mu$ of the free energy is just the naive dimensional result
$\sim \mu^{\frac{1}{1 + b^2}}$ in both cases. This result should hold
for $\beta < \beta_c$. However, there is clearly, in both $d=2$ and
$d=1$ cases, a singularity as $\beta \to \beta_c^{-}$ as the amplitude 
$C(b)$ diverges as $b=\beta/\beta_c \to 1^{-}$. This is thus in perfect
agreement with the existence of a phase transition in the particle
model. In the sinh-Gordon model itself, we do not expect strictly
speaking a phase transition, as the model is massive both below 
and above $b=1$, however we do expect some ``change of behaviour'',
which may be related to a change of nature of the excitations 
around the ground state. This is not ruled out by exact results \cite{vega}
as it clearly comes here from the physical mass acquiring a nontrivial
dependence in the bare mass parameter $\mu$ (contrarily to sine-Gordon
model, for sinh-Gordon model there is no presently known exact solution
of a lattice version).

Let us now interpret these results for our model. They mean that the
generating function $G_{sh}(x)$ of the free energy distribution,
with $\mu = e^{\beta x}$, takes
indeed the form of a traveling wave:
\begin{eqnarray}
G_{sh} \sim \exp\left( - L^2 C_d\left(\frac{\beta}{\beta_c}\right)
\mu^{\frac{2}{(1 + \frac{\beta}{\beta_c})^2}} \right)
= g(x + c l + \gamma)
\label{velocsh}
\end{eqnarray}
with $l=\ln L$ and a velocity:
\begin{eqnarray}
c = \frac{d}{\beta} + \sigma \beta
\end{eqnarray}
This is exactly the velocity given by the KPP equation
for the particle model, in the high temperature phase.
It also yields a front $g(y) = \exp(- e^{a y})$ with 
$a = \frac{\beta}{1 + (\beta/\beta_c)^2}$ and
$\gamma = \frac{\beta}{1 + (\beta/\beta_c)^2} \ln C_d(\beta/\beta_c)$.
This form however should be taken with caution as strictly speaking
formula (\ref{velocsh}) is valid only in the limit where 
$L$ goes to infinity first (at fixed $\mu = e^{\beta x}$). 
It should be compared with the asymptotic behaviour of the
front in the region of large positive $y$. 
We expect universality in the other region of the front (of 
very negative $y$ i.e. $x << c l$) and exact knowledge about this
region would be equivalent to exact knowledge of the sinh-Gordon
model at finite size, which is not yet available.

The physics of the problem of the particle in the random potential
leads us to conjecture that the 2d sinh-Gordon model 
(as well as the boundary sinh-Gordon model) will exhibit
a change of behaviour, the algebraic $\mu$-dependance
of its free energy
will freeze for $\beta \ge \beta_c$, which corresponds
to the low temperature glassy phase of the particle model.
We thus expect:
\begin{subequations}
\begin{align}
& f_{sh}(\mu) \sim \mu^{\alpha} \\
& \alpha = \frac{1}{1 + (\beta/\beta_c)^2} \qquad
&&\beta<\beta_c=\sqrt{d/\sigma} \\ 
& \alpha = \frac{1}{2} \qquad &&\beta > \beta_c=\sqrt{d/\sigma}
\label{decadix}
\end{align}
\end{subequations}
and presumably log corrections (at least at $\beta=\beta_c$, and maybe
for all $\beta > \beta_c$).

This is confirmed by a renormalization group analysis directly on
the Sinh-Gordon and Liouville models discussed below.

\subsection{Relation with the Liouville model in $d=2$}

The relation between our original model (\ref{def-1part}) of the particle 
in the random potential and the Liouville model proceeds 
via the generating function:
\begin{eqnarray}\nonumber 
G(x) &=& \left< \exp( - e^{\beta x} Z[V])\right>_V \\
&=&
\left< \exp( - \sum_{\bf r} e^{\beta (x-V({\bf r})) } )\right>_V
\label{derrida}
\end{eqnarray}
which encodes the full probability distribution of the
free energy of the particle. In the case of the $d=2$ potential
with logarithmic correlations it is identical to the partition function of a Liouville model,
which one can write either on the original lattice or
in the continuum (with UV and IR
cutoffs $a$ and $L$) as ($\mu= e^{\beta x}$) :
\begin{eqnarray} \nonumber 
&& G(x) = H[\mu] = \int DV e^{- S[V] }  \\ 
&& S[V] = \int d^2 {\bf r} \left( \frac{1}{8 \pi \sigma} (\nabla
V({\bf r}))^2   
+ \mu ~e^{- \beta V({\bf r})} \right) 
\label{lm}
\end{eqnarray}
where the functional integral is normalized such that $H[\mu=0]=1$
(equivalently 
one redefines $H[\mu] \to H[\mu]/H[\mu=0]$). We call it the Liouville
Model (LM) 
to distinguish it from the continuum Liouville field theory LFT whose
definition 
is recalled below. A relation also exists between the correlation functions 
of the Gibbs measure and some correlation functions
in the Liouville model:
\begin{eqnarray}
&&< p({\bf r}_1) .. p({\bf r}_n) >
= \int_{\mu >0} \mu^{n-1}
e^{- \beta ( V({\bf r}_1) + ... V({\bf r}_n))} e^{-S[V]}
\end{eqnarray}
Strictly speaking the model (\ref{lm}) above is not well defined because
of the zero mode $V({\bf r}) \to V({\bf r}) + w$ and must be complemented
with boundary conditions. In the particle problem studied here we
have chosen periodic boundary conditions with the additional constraint
$\sum_{\bf r} V({\bf r}) = 0$ to pin the zero mode.

On the other hand, many results are known for the (related) continuum 
Liouville field theory (LFT), of great interest in quantum gravity
\cite{polyakov,seiberg,gervais,david,kazakov}.
It is usually defined on a arbitrary genus 
$h$ manifold with background metric $g$ and associated curvature 
$R$ by the action \cite{z2}:
\begin{eqnarray}\label{SLFT}
S_{LFT} = \int d^2 x \left( \frac{1}{4 \pi} (\partial_a \phi)^2 + 
\mu e^{2 b \phi} + \frac{Q}{4 \pi} R \sqrt{g} \phi\right)
\end{eqnarray}
in conventional notations, with $\phi = - V/\sqrt{2 \sigma}$ and
$b=\beta/\beta_c$. The (standard) choice is $Q=b+ 1/b$ for which
the theory is critical and has local conformal invariance
(with a central charge $c_L = 1 + 6 Q^2 = 25 + c$) \cite{z2}.
It can also be formulated as the theory of (liquid) random surfaces
\cite{david,cates} e.g. as random triangulations.
There one defines the total area
$A = \int_{\bf r} e^{- 2 b \phi(r)}$ which is nothing but the partition
function $A=Z[V]$ of the particle problem, and studies the distribution
${\cal Z}(A) \sim e^{- \mu_c A} A^{\gamma_{\text{string}}-3}$
which is nothing but $P(Z)$.

The particle model allows us to make precise statements on the
Liouville model defined above. The LFT allows for exact calculations
(e.g. of correlation functions) and in principle one could hope to
translate those in the particule model (Gibbs measure correlations).
The relation between the two however is rather subtle. For instance,
the boundary conditions chosen in the particle problem would correspond to
Liouville on a torus $h=0$, except that the additional 
pinning condition spoils it. We will thus not explore here all these intricacies 
but give a few general remarks, mostly about the behaviour of
the Liouville model under coarse graining.

First we know that $G_l(x)$ satisfies a RG equation of the KPP type.
Thus upon coarse graining (i.e as a function of $l=\ln (L/a)$ 
see Section \ref{renormgroup})
the Liouville model partition function satisfies a KPP non linear RG equation.
The corresponding front velocity gives the scaling of the partition function with $\mu$.
The glass transition, with freezing of the front velocity, corresponds
exactly in the Liouville model to the transition between two regimes
(the so-called $c=1$ barrier in the LFT):

(i) {\it weak coupling Liouville}: $b = \beta/\beta_c <1 $. In that regime
there is no problem to define a continuum limit.
The KPP non linear RG front solution of velocity $c(\beta)=2(\frac{1}{\beta}+
\frac{\beta}{\beta_c^2})\equiv c(b)$ yields
$G_l(\mu) \sim \mu^{1/(1 + b^2)}$ and in the Liouville model 
the scaling dimensions are the one obtained by naive dimensional
counting. e.g. :
\begin{eqnarray}
&& \mu \int d^2 {\bf r}  e^{- \beta V({\bf r})} = \mu e^{\ln Z[V]}
 \sim \mu L^{\beta c(\beta)} = \mu L^{2(1 + b^2)}
\end{eqnarray}
In the LM only one of the two branches (deduced by $b \to 1/b$) is selected
as $l \to + \infty$. 

The regime $b \leq 1$ is the one where the continuum LFT is well defined:
there the role of the $Q$ term in (\ref{SLFT}) 
is to shift the conformal dimension of the fields
$e^{2 \alpha \phi}$ to $\Delta(\alpha)=\alpha (Q-\alpha)$ 
(while the naive power counting dimensions in the 
LM are $\Delta(\alpha)=-\alpha^2$, see above) which renders the
Liouville term $\int e^{2 b \phi}$ {\it exactly
marginal} (and thus the theory critical). It was argued in 
\cite{tsvelik_prl}
that the LFT gives the correlations of the Gibbs measure, the operator
$p({\bf r})$ (eq. (\ref{def-gibbs}))  
corresponding to the Liouville field $e^{2 b \phi}$
(and is thus of conformal dimension $\Delta(b)=1$ (i.e $p({\bf r}) \sim L^{-2}$).
A hint in favor of this conjecture was that the corresponding LFT 
conformal dimensions of the composite fields $p({\bf r})^{q}
\sim e^{2 q b \phi}$ is
simply $\Delta(q)=q (1+ b^2 - b^2 q)$ which correctly reproduces
the multifractal spectrum:
\begin{eqnarray}
\int d^2 {\bf r}  ~ p({\bf r})^q \sim L^{2(1- \Delta(q))} 
\end{eqnarray}
given in \cite{chamon96} and Section \ref{part:dirac}
(in the weak disorder regime
$q<q_c$). This is not a very strong test since the same 
multifractal spectrum can also be obtained within the LM model 
by considering the dimension of the normalized Gibbs measure 
(rather than the unnormalized one $e^{- \beta V}$). Indeed the
effect of the $Q$ term is to shift:
\begin{eqnarray}
e^{2 b q \phi} \to L^{2 b q Q} e^{2 b q \phi} 
\sim Z[V]^{-q} e^{2 b q \phi}
\end{eqnarray}
To convincingly establish the conjecture of \cite{tsvelik_prl}
the effect of the additional $Q$ term should be checked on
the many point correlations \cite{z2}, where it is rather more subtle,
and further investigation is needed. In particular, the RG described
here suggests by extension that the Gibbs measure correlations 
(or at least some limits of them)
should also be computable within the DPCT model. This suggests a direct
relation between LFT and DPCT, a check of which 
would be of great interest. Note also that a critical model,
which mimics the effect of the $Q$ term (as adding an average value
to $\phi$ \cite{z2}) is studied in Appendix \ref{extended}.

(ii) {\it strong coupling Liouville}: $b = \beta/\beta_c >1$.
This corresponds to the glass phase for the particle which,
interestingly, has a non trivial structure. As is well known there are
serious difficulties in defining the continuum LFT in that
regime. Using what we know from the particle problem we
can gain some idea of what happens in the Liouville theory. First let
us note that since $G(x)$ is such that
for $\beta=+\infty$ and fixed $L$ it is equal to the distribution function
of the minimum of the set of $V({\bf r})$:
\begin{eqnarray}
G(x) = \text{Proba}( x < \min_{{\bf r}} V({\bf r}) )
\end{eqnarray}
The (infinitely strong coupling) Liouville model can be
recast as {\it an extremal statistics} problem in that limit. The partition sum
$Z[V]$ of the particle model being dominated, for $b>1$ by
a few regions of space where $V({\bf r}) \sim V_{min}$
(with little dependence in $\beta$, i.e the Liouville
wall become a hard wall for all $\beta < \beta_c$ with thickness
of order $O(1)$) we expect this spatial heterogeneity to show up in
LM as well. From what we have learned in previous Sections
we know that upon coarse graining the following {\it effective Liouville
model} action $S_{eff}$ is generated:
\begin{eqnarray}
&& G_l(x) = H_l[\mu] = \int D\tilde{V} <e^{- S[\tilde{V},z] }>_{P(z)} \\
&& S_{eff,l}[V] = \int d^2 {\bf r} \left( \frac{1}{8 \pi \sigma} 
(\nabla \tilde{V}({\bf r}))^2  
+ z({\bf r}) e^{- \beta \tilde{V}({\bf r})} \right)
\label{lmcg}
\end{eqnarray}
i.e {\it a new field $z({\bf r})$ is dynamically generated}, and has
short range correlations {\it but} has a broad power law distribution:
\begin{eqnarray}
P(z ) dz \sim z^{-1 + \frac{1}{b}}
\end{eqnarray}
while $\tilde{V} \equiv V^>$ is the smooth field introduced in
(\ref{decomposition}). For $b<1$ this dynamically generated local field
can be averaged out without changing significantly the 
action (note that even for
$b<1$ it changes properties of operators $e^{- q \beta V}$
for $q>q_c$) while for $b>1$ it changes crucially the physics.
One can define the effective Liouville potential $U[V]$ for
the smooth field $\tilde{V}$ after averaging over the
$z$ field as:
\begin{eqnarray}
U_l[\tilde{V}] = - \ln <\exp(- z e^{- \beta \tilde{V}})>_{P_l(z)}
= - \ln G_l(x= - \tilde{V})
\end{eqnarray}
the bare Liouville potential being $U[V]=\mu ~ \exp(- \beta V)$.
We can now use the front solution of the KPP equation (i.e the
scaling region in the large $L/a$ limit) described in previous
Sections. For $b<1$,
since $<z>_{P(z)} < + \infty$ we have that for large $V$
\begin{eqnarray}
U_l[V] \approx c \mu e^{2(1+b^2)l} \exp(- \beta V)
\end{eqnarray}
and thus the coarse grained potential is similar to the bare one.
However, for $b>1$ one has for large $V$
\begin{eqnarray}
U_l[V] \approx c \mu e^{4 l} V \exp(- \beta V)
\label{decayfrozen}
\end{eqnarray}
because of the broad distribution of the $z$ field.

Since the $z({\bf r})$ are highly heterogeneous on short scales
it is not surprising that a continuum limit is hard to obtain
for $b>1$. These heterogeneities are linked to the
structure of the glass phase reminiscent of replica symmetry
breaking.
It is tempting to conjecture that it
may also be related to the branched polymer structure 
which appear in LFT for $b>1$, i.e beyond the $c=1$ barrier
\cite{david}, or to the spike instability \cite{cates}
of fluid membranes. 

Furthermore, let us notice that the LFT theory at $b=1$ is known to
have two marginal operators  
whose dimensions are degenerate $e^{2 b \phi}$ and $\phi e^{2 b \phi}$.
This is in exact parallel with the behaviour of the KPP front
solution, which develops at $b=1$ two degenerate linear eigenmodes
$\exp(- \beta V)$ and $V \exp(- \beta V)$.

Thus we have seen that the Coulomb gas RG can be used to understand the
behaviour of the Liouville model. A scenario is obtained where for 
$b \ge 1$ new short scale degree of freedom are generated
(short scale instability). Averaging over these changes the
effective Liouville potential. The parallel with the particle model 
suggests that the short scale instability
in Liouville may be related to the generation of strong inhomogeneities in the
Gibbs measure $p({\bf r})$, analogous to structures discussed in the context of
replica symetry breaking. Thus, if the mapping onto the LFT is confirmed
it suggests to also investigate RSB type effects in strong coupling LFT.

\subsection{Direct renormalisation group analysis
of sinh-Gordon and Liouville models and traveling waves}
\label{part-RGLiouville}

Let us now illustrate how one can see explicitly the
freezing of the free energy exponent in the strong coupling phase 
from renormalisation group approaches {\it directly} on the sinh-Gordon
and Liouville models. Such functional RG methods have been applied
to study the analogous problem \cite{fisher_wetting} 
of the wetting of an interface of height $V$.
Related exact RG methods, with various truncation schemes, have also been
applied to the Liouville model, and in the context of quantum gravity 
to the LFT \cite{reuter}. In all cases we will illustrate how the main
physics lies in the selection mechanism for the
traveling wave solutions of the non linear RG equation.

\begin{figure}
\centerline{\fig{9cm}{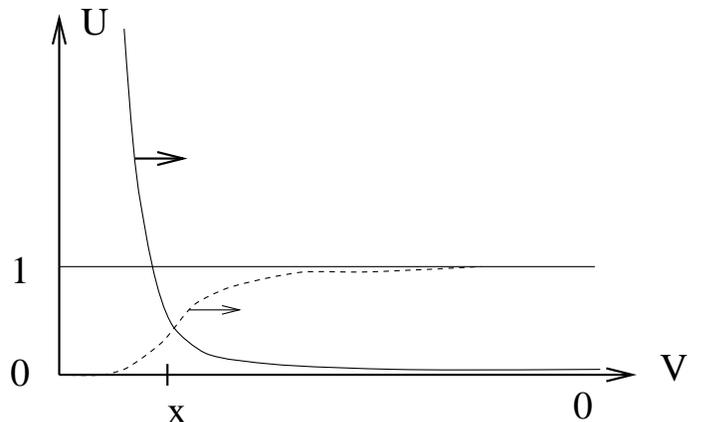}}
\caption{
\narrowtext
Liouville wall moving under RG as a travelling wave.
Represented is $U[V]$ on the negative $V$ side, 
of original form $U[V]= e^{- \beta(V-x)}$ and 
also (dashed line) $G(x) = \exp(- U[V])$. Both move
under RG forming a travelling front, whose velocity determines
the ``free energy'' exponent. For sinh-Gordon a 
second, mirror image wall is also moving symmetrically 
towards $0$. Freezing in the front velocity occurs
at and below the transition at $\beta=\beta_c$.
\label{wall}}
\end{figure}

The study proceeds as follows. We consider
\begin{eqnarray}
&& G(x) = H[\mu= e^{\beta x}] = \int DV e^{- S[V] }  \\ \nonumber 
&\text{with }& S[V] = \int d^2 {\bf r} \left( \frac{1}{8 \pi \sigma} (\nabla V({\bf r}))^2  
+ U[V] \right)
\end{eqnarray}
One can perform a Wilson RG analysis (or if one prefers a
suitably truncated exact RG analysis), and one finds
in $d=2$ the flow for the local part $U_l[V]$ as:
\begin{eqnarray}
\partial_l U = 2 U + \sigma U'' + O(U^2) 
\label{rgeq}
\end{eqnarray}
There may also be corrections to $\sigma$ to $O(U^2)$
(in the Sinh-Gordon model), but we focus for now on the
RG to linear order. Let us recall that the initial condition 
is $U_{l=0}[V] = \mu e^{-\beta V}$ for Liouville and 
$U_{l=0}[V] = \mu e^{-\beta V} + \mu e^{\beta V}$ for sinh-Gordon,
and that we are interested in the small $\mu$ limit.
In this limit the initial condition corresponds to a 
a very wide well $U[V]$ (e.g. in the sinh-Gordon model) with a very small
curvature $U''[0]$. To obtain the free energy exponent as
$\mu \to 0$, one simply iterates the RG until $U_l''[0] \sim O(1)$
at a scale $l^*= \ln (L^*/a_0)$ (more precisely 
$U_l''[0] \sim 1/(\sigma a_0^2)$ where $a_0$ is the bare UV cutoff
of the model). At this scale, the free energy is $O(1)$, as
can be estimated from gaussian fluctuations (straightforwardly at
least in the SG model) and thus the
initial free energy is:
\begin{eqnarray}
F \sim A_d(\beta) \left(\frac{L}{L^*}\right)^2
\label{freeen}
\end{eqnarray}

Remarkably, it is now possible to use what we learned 
in the previous Sections and demonstrate the ``freezing'' transition 
at $\beta=\beta_c$ (corresponding to the glass transition for the particle)
simply from the RG to this order. Indeed the
solution of the truncated equation is:
\begin{eqnarray}
U_l[V] = e^{2 l} \frac{1}{\sqrt{4 \pi \sigma l}} \int dV' 
\exp\left(- \frac{(V-V')^2}{4 \sigma l}\right) U_{l=0}[V]
\label{integral}
\end{eqnarray}
A straigthforward conclusion would then be that 
the exact solution corresponding to Liouville is:
\begin{eqnarray}\label{UlV}
U_l[V] = \mu e^{(2 + \sigma \beta^2) l } e^{- \beta V}
\end{eqnarray}
and similarly for the sinh-Gordon:
\begin{eqnarray}
U_l[V] = 2 \mu e^{(2 + \sigma \beta^2) l} \left(e^{- \beta V} +
e^{\beta V}\right) 
\end{eqnarray}
since $\exp(\pm \beta V)$ are exact eigenvectors of the
linear RG equation for any $\beta$. From (\ref{freeen}) this
immediately yields the ``naive dimensional'' result for the free
energy
\begin{eqnarray}
F \sim L^2 \mu^{\frac{1}{1 + (\beta/\beta_c)^2}}
\end{eqnarray}
with $\beta_c=\sqrt{2/\sigma}$. As we know from the above exact
result this is correct for $\beta < \beta_c$.
Note how the potential $U_l[V]$ evolves. Using the
notation $\mu = e^{\beta x}$ (natural from our extremal
statistics interpretation) it forms a ``Liouville wall'',
which can be seen as a ``front solution''
moving as $\exp( - \beta(V - x - c l))$ towards $U=0$ 
(and in the sinh-Gordon model there are two symmetric walls
moving towards $U=0$ and reaching it at $l=l^*$).
The Liouville front velocity is:
\begin{eqnarray}
c = \frac{2}{\beta} + \sigma \beta
\end{eqnarray}
which plotted as a function of $\beta$ is the famous parabola,
such that two values of $\beta$ corresponds to the same $c$,
which is also a well known property of Liouville theory.

As we now show, (\ref{UlV}) is {\it incorrect} for $\beta \ge \beta_c$.
This is so for a subtle reason, as apparently the statement that
$\exp(- \beta V)$ is an exact eigenvector of the linear RG (and
of (\ref{integral})) cannot fail ! However, by now we are well used
to fronts: in fact we have encountered exactly 
the same equation in our previous solution of the REM
model via RG ($h_l(x)$ in (\ref{integ}) is identical to
$U_l(V)$ in (\ref{integral})). To describe correctly the
bare Liouville (or equivalently the Sinh-Gordon) model
one should generalize the initial condition $U_{l=0}[V]$,
still assuming that $U_{l=0}[V] \sim \exp(- \beta (V-x))$ for $V \gg x$
($x$ here is very negative corresponding to a small
$\mu$). Then one can use the saddle point method to estimate
(\ref{integral}) as was done in (\ref{integ}) to evaluate 
$h_l(x)$
and one discovers that for $\beta > \beta_c$ the velocity
freezes into:
\begin{eqnarray}
c = 2 \sqrt{2 \sigma}
\end{eqnarray}
which yields a free energy
\begin{eqnarray}
F \sim L^2 \mu^{\frac{1}{2}}
\end{eqnarray}
instead of the naive dimensional estimate, thus in agreement with
our expectation for the SG model (\ref{decadix}).
In addition we find that the decay of the renormalized potential
$U_{l}[V] \sim e^{-\alpha V}$ is frozen at $\alpha=\beta_c$
for all $\beta>\beta_c$ consistent with (\ref{decayfrozen}).

What has happened is that although $U_{l=0}[V]= \exp(- \beta (V-x))$
is indeed formally an exact eigenvector,
it is {\it dynamically unstable}, i.e if one chooses another function with
the same large positive $V-x$ behaviour one gets a different velocity
(which is not the case for $\beta < \beta_c$).
It is easy to see that the choice $U_{l=0}[V]= \exp(- \beta (V-x))$ 
exactly for all $V$ does not make sense for $V \to - \infty$.
Indeed it is immediately spoiled by the slightest amount of coarse graining
(as would appear also by considering the non linearities in the RG equation).
The simplest way to see it is to notice that the coarse grained potential:
\begin{eqnarray}
\tilde{U}[V] = - \ln \left(\int dv \exp(\left[- \mu e^{- \beta
(V+v)}\right] - \frac{v^2}{2 s} )\right)
\end{eqnarray}
does not grow as $\sim \exp(- \beta V)$ for large negative $V$ but much slower
as $\sim V^2$. To illustrate further the point let us consider the
initial condition:
\begin{eqnarray}
U_{l=0}[V] = \frac{e^{- \beta (V-x)}}{1 + e^{- \beta (V-x)}}
\end{eqnarray}
It behaves as $e^{- \beta (V-x)}$ for large positive $V-x$ (and thus corresponds
to the Liouville model) but goes to $1$ on the other side. For $\beta=+\infty$
it is easy to compute $U''_l[V=0]$ from (\ref{integral}) 
since $U_{l=0}[V] = \theta(x-V)$. One finds:
\begin{eqnarray}
U''_l[V=0] \sim e^{2 l - \frac{x^2}{4 \sigma l}} 
\end{eqnarray}
and thus one has that $l^*$ defined above is such that:
\begin{eqnarray}
c l^* = x \qquad c = 2 \sqrt{2 \sigma}
\end{eqnarray}
This is in fact valid for all $\beta > \beta_c$ as was shown in
detail in previous sections.

Thus the freezing transition can be obtained from the linearized 
(i.e lowest order) RG equations, using only elementary insight from
coarse graining or the existence of higher order non linear terms.
It provides an interesting example where the naive dimensions hold
in some regime but are modified in another.
Of course, as we have seen in Section \ref{fronts}
from the study of fronts,
to really establish the transition and determine the universality class
one needs to consider higher order non linearities in (\ref{rgeq})
which goes beyond this paper. For the LFT in quantum gravity the
reader can find some exact functional RG studies in Ref. \cite{reuter}. 
Although not discussed in this reference, 
the non linear RG there seems to also exhibit
traveling front solutions, whose physics may be
important in understanding the problem of the $c=1$ barrier.

\section{Critical Dirac fermions in a random gauge field} \label{part:dirac}

In this section we relate our RG study of the previous section to 
the study of the critical wave
functions of 2D Dirac fermions in a random magnetic field. We first confirm
the results of \cite{castillo97} for the multifractal spectrum, and obtain
their finite size corrections. Then we study the transition from the weak
disorder to the strong disorder phase, related to the glass transition
for the particle, and find that the 
strong disorder phase has a new and non trivial structure,
leading to what we call {\it quasi-localized} eigenstates.

\subsection{Critical wave function of 2D random Dirac}
\label{part:Dirac1}

Let us first recall the problem of a massless two dimensional Dirac fermion in a 
static random magnetic field
\cite{chamon96,castillo97,comtet98}. This model,
and its non abelian generalizations, has received a lot of
attention in connection with the integer quantum Hall effect
transitions with disorder. As discussed in \cite{grinstein} two dimensional
Dirac fermions can experience three generic types of disorder: random gauge,
random mass and random potential. Random gauge disorder is believed to
be a line of fixed points in this general model and is still not yet fully
understood. Here we address only the random gauge disorder model of
hamiltonian:
\begin{equation}\label{dirac-eq}
H =  \sigma_{\mu}\left(i v_F \partial_{\mu}-A_{\mu}({\bf r}) \right)
\end{equation}
where the $\sigma_{1,2}$ are the $2$x$2$ Pauli matrices and $\mu=1,2$
(we set the Fermi velocity $v_F=1$ from now on). 
The random magnetic field ${\bf B}$ corresponding to the gauge potential 
${\bf A}$ is chosen to be gaussian with mean value 
$\overline{{\bf B} ({\bf r})}=0 $. The type of correlations studied here
correspond to the most interesting case where the gauge potential has
short range correlations. In the
Coulomb gauge, we can introduce the scalar potential $\phi$  such that 
$A_{\mu}=\epsilon_{\mu \nu} \partial_{\nu} \phi , 
B ({\bf r})=-\partial_{\mu}^{2}\phi ({\bf r}) $. The gaussian distribution of
$\phi ({\bf r})$ is thus given by 
\begin{equation}\label{phi-distrib}
P[\phi]= cte\times e^{-\frac{1}{4 \pi g}\int_{{\bf r}} 
(\partial_{\mu}\phi ({\bf r}))^{2}}
\end{equation}
where $g$ parametrises the strength of the random magnetic field $B$.
The correlator of the function $\phi({\bf r})$ is thus:
\begin{eqnarray}
\overline{(\phi({\bf r}) - \phi({\bf r}'))^2} \sim 
2 g \ln \frac{|{\bf r} - {\bf r}'|}{a} 
\end{eqnarray}

In this model, the wave functions at energy $E$ are 
localized for all energies other than the
critical energy $E=0$. We restrict our study to 
the $E=0$ critical eigenstate, which satisfy:
\begin{equation} \label{groundst}
H \Psi_0({\bf r}) = 0
\end{equation}
For a system of finite size $L$ with appropriate boundary conditions
there are two independent {\it normalized} solutions of (\ref{groundst}):
the first one can be written $\Psi_{0,1}({\bf r}) = (\Psi_{0}({\bf r}),0)$
with:
\begin{equation}  \label{wave0}
\Psi_{0}({\bf r})^2 = \frac{e^{- 2 \phi ({\bf r})}}{
\sum_{\bf r} e^{- 2 \phi ({\bf r}')}}
\end{equation}
the second one being $\Psi_{0,2}({\bf r}) = (0,\tilde{\Psi}_{0}({\bf r}))$
where $\tilde{\Psi}_{0}({\bf r}))$ is given by (\ref{wave0}) changing
$\phi ({\bf r}) \to - \phi ({\bf r})$. We denote $\sum_{\bf r}$ having in mind
either a discrete problem, or a continuous problem with some short scale
cutoff $a$. 

\subsection{participation ratios and multifractal spectrum}

Thus in a given configuration of disorder $\phi ({\bf r})$ the 
quantum probability $|\Psi_{0}({\bf r})|^2$ is {\it identical}
to the Gibbs probability $p({\bf r})$ defined in (\ref{def-gibbs}) for 
the particle in the logarithmically correlated
random potential $V({\bf r})$ with the correspondence:
\begin{eqnarray}
&& |\Psi_{0}({\bf r})|^2 = p({\bf r})  \\
&& 2 \phi({\bf r}) = \beta V({\bf r}) 
\label{corr}
\end{eqnarray}
and thus the model depends on a
single parameter $g = \frac{1}{2} \beta^2 \sigma$. As we have discussed 
in previous Sections the particle in the logarithmically correlated
random potential undergoes a transition at $\beta_c = \sqrt{2/\sigma}$
at which its Gibbs measure changes from being dominated by 
many sites (high $T$ phase) to being dominated to a 
few sites (low $T$ phase). Thus in the quantum problem we expect
a transition at:
\begin{eqnarray}
g=g_c = 1
\end{eqnarray}
with a weak disorder phase for $g<1$ and a strong disorder
phase for $g>1$. In the weak disorder phase the quantum probability 
(and thus observables such as the mean squared position fluctuations 
$<r^2> - <r>^2$) is delocalized ($<..>$ means averages over 
$\Psi_0$). In the strong disorder phase the quantum probability
is more concentrated, but it cannot be called localized in the usual 
sense (of an exponential decay around a single center) and in 
fact both phases have rather peculiar properties.

Properties of wave functions can be described by
the inverse participation ratios defined from the normalized wave function
$\Psi_{0} ({\bf r})$ in a system of size $L$ by 
\begin{equation}\label{def-Rq}
R_{q} (L) = \int d^{2}{\bf r} |\Psi_{0} ({\bf r})|^{2q}
= \int d^{2}{\bf r} ~ \left( p({\bf r})\right)^q
\end{equation}
At a very qualitative level, the nature of the eigenfunction can be inferred
from the scaling behaviour of
the inverse participation ratio with the system size $L$ : for an
exponentially localized state $R_{q}(L)$ scales \cite{footnote11}
as $R_{q} (L) \sim \text{const}$ for all $q>0$,
while for a plane wave delocalized state we get  
$R_{q}(L)\sim L^{- 2 (q-1)}$.
In addition to the localized and delocalized states, there exist states 
such that $\tau(q)= - \ln R_{q}(L)/\ln L$ is a non linear function of $q$:
they correspond to multifractal wave functions whose moments cannot be described
by a single length as usual but rather by a spectrum of exponents.
Here, as in \cite{castillo97}, we also find intermediate
multifractal behaviour.

To compute the finite size inverse participation ratios
we can use the information of previous Sections since:

\begin{equation}
s_q(L) = - \ln R_{q} (L) = - \ln Z_{q^2 g} + q \ln Z_{g}
\end{equation}
where we have defined $Z_{g}= Z(\beta = \sqrt{2 g/\sigma})$ where
$Z(\beta)$ is the partition function for the particle at
inverse temperature $\beta$. In particular we will be interested 
in the multifractal asymptotic scaling exponent 
$\tau(q)$ defined by  
\begin{equation}\label{def-tau}
\tau (q)=\lim_{L\to \infty} \frac{s_q(L)}{\ln L}
\end{equation}
These exponents were computed previously in \cite{castillo97} using the
REM approximation. Here we use our RG results and also 
obtain finite size corrections. Note that these correspond to properties of
$\Psi_0$ defined above and could be changed if other boundary conditions
were used.

From the previous Sections we obtain:
\begin{subequations}
\begin{align}
& \ln Z_g = 2 (1 + g) \ln L + \Delta_g  &&,g < 1  \\
& \ln Z_g = (\sqrt{g} (4 \ln L - \frac{1}{2} \ln (\ln L) ) + \Delta_g  
&& ,g = 1  \\
& \ln Z_g = (\sqrt{g} (4 \ln L - \frac{3}{2} \ln (\ln L) ) + \Delta_g  
&& ,g > 1   
\end{align}
\end{subequations}
where $\Delta_g$ is a sample dependent variable of order $O(1)$ with
a $g$ dependent distribution (whose tails we have characterized previously).
From there we obtain $s_q(L)$, which have different behaviours in
the two phases.

{\it (i) weak disorder phase}

For $g<1$ we find, denoting $q_c = \frac{1}{\sqrt{g}}$ : 
\begin{subequations}
\begin{align*}
& s_q(L) = 2 (q-1) (1 - g q) \ln L + A_{q,g} \\
&\qquad \qquad \qquad \qquad \qquad \qquad \qquad \qquad\text{for }
|q|<q_c   \\ 
& s_q(L) = \frac{2}{\sqrt{g}} (1- \text{sgn}(q) \sqrt{g})^2 \ln L +
\frac{1}{2} 
\ln \ln L + A_{q,g} \\
&\qquad \qquad \qquad \qquad \qquad \qquad \qquad \qquad\text{for }|q|=q_c   \\
& s_q(L) = 2 q (1- \text{sgn}(q) \sqrt{g})^2 \ln L + \frac{3}{2}  
|q| \sqrt{g} \ln \ln L  + A_{q,g}\\
&\qquad \qquad \qquad \qquad \qquad \qquad \qquad \qquad\text{for } |q|>q_c 
\end{align*}
\end{subequations}

where $A_{q,g}$ is a fluctuating part of order $O(1)$.

{\it (ii) strong disorder phase}

For $g>1$ we find:
\begin{subequations}
\begin{align}
& s_q(L) = - 2 (q \sqrt{g} - 1)^2 \ln L - \frac{3}{2} 
q \sqrt{g} \ln \ln L  + A_{q,g} \\\nonumber 
&\qquad \qquad \qquad \qquad \qquad \qquad \qquad \qquad\text{for }
  |q|<q_c \\
& s_q(L) = - \frac{1}{2} \ln \ln L  + A_{q,g}  \qquad\text{for }
  q=q_c \\
& s_q(L) = A_{q,g}  \qquad \text{for } q>q_c \\
& s_q(L) = - 2 |q| \sqrt{g} \left( 4 \ln L - \frac{3}{2} \ln \ln L\right) +
A_{q,g} \\\nonumber  
&\qquad \qquad \qquad \qquad \qquad \qquad \qquad \qquad\text{for }q < - q_c \\
& s_q(L) = - 2 \left(4 \ln L - \frac{1}{2} \ln \ln L\right)   + A_{q,g} 
\qquad\text{for } q = - q_c 
\end{align}
\end{subequations}
where $A_{q,g}$ is a fluctuating part of order $O(1)$. 

The corresponding scaling exponents $\tau(q)$ are thus
identical to the one found in \cite{castillo97} and in
addition we have obtained their finite size corrections.
In the weak disorder phase one for $q>0$:
\begin{equation}\label{tq}
\tau (q) =
\left\{
\begin{array}{lll}
2 (q-1) \left(1-\frac{q}{q_{c}^{2}}\right)
 &\text{ for }& q \leq q_{c}=\sqrt{\frac{1}{g}}\\
2q \left(1-\frac{1}{q_{c}} \right)^{2} &\text{ for }
& q \geq q_{c}
\end{array} \right.
\end{equation}
which means a parabolic form with a termination point at
$q=q_c$ as represented in the Fig. \ref{multi1}.

\begin{figure}
\centerline{\fig{5cm}{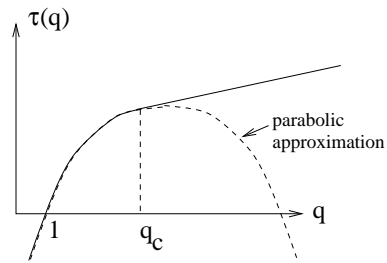}}
\caption{
\narrowtext Multifractal spectrum in the weak disorder phase
\label{multi1}}
\end{figure}

In the strong disorder phase $g >1$, {\it i.e} when $q_{c}\leq 1$,
the above expression becomes (for $q>0$):

\begin{equation}\label{tq2}
\tau (q) =
\left\{
\begin{array}{lll}
-2   \left(1-\frac{q}{q_{c}}\right)^{2}
 &\text{ for }& q \leq q_{c}=\sqrt{\frac{1}{g}}\\
0  &\text{ for }
& q \geq q_{c}
\end{array} \right.
\end{equation}
Since the inverse participation ratio does not scale
with the system size $L$ for each integer $q$, one could naively conclude 
that it is the sign of a localized state (see however below).

As was discussed in \cite{castillo97} these results can be translated into
spectrum for exponent $\alpha$. If one assumes that $p({\bf r})$
is of order $L^{-\alpha}$ in a number $L^{f(\alpha)}$ sites then
the above spectrum is recovered if:
\begin{eqnarray}
f(\alpha) = 8 \frac{(d_+ - \alpha)(\alpha - d_-)}{(d_+ - d_-)^2}
\end{eqnarray}
with $d_{\pm} = 2 (1 \pm \sqrt{g})^2$ for $g>1$ and $d_{+} = 8 \sqrt{g}$,
$d_-=0$ for $g>1$. It is easy to see that:
\begin{eqnarray}
< (r - <r>)^{2 k} > \ge L^{\max_\alpha ((k + 1) f(\alpha) - \alpha)}
\end{eqnarray}
showing that the eigenstate is never localized in the usual sense
(exponential decay around a single center) since the exponent is
always positive for large enough $k$. Since 
$\lim_{q \to \pm \infty} s_q(L)/q = \ln p_{max,min} $
one obtains that the maximum of the Gibbs measure $p_{max} =
\max_{{\bf r}} p({\bf r})$ and the minimum behave for large
$L$ as:
\begin{eqnarray}
&& p_{max} \sim L^{-2(1-\sqrt{g})} (\ln L)^{- \frac{3}{2} \sqrt{g}} \\
&& p_{min} \sim L^{-2(1+\sqrt{g})} (\ln L)^{+ \frac{3}{2} \sqrt{g}} 
\end{eqnarray}
in the weak disorder phase.

\subsection{nature of the strong disorder phase: quasi-localization}

Let us now concentrate on the case $g>1$. There, we know from previous
Sections that the Gibbs measure of the particle is concentrated in
{\it a few} sites. Thus from (\ref{corr})
the quantum probability $|\Phi_0({\bf r})|^2$
is also concentrated in {\it a few} sites, analogous to 
the RSB picture. This is a very peculiar type of eigenstate.
Indeed if one computes the quantum
average $<r^2> - <r^2>$ in a given sample, it has a finite probability to
be of order $O(L^2)$. Thus the eigenstate cannot be considered as localized in
the usual sense. Since it is peaked around a few sites we call it
``quasi-localized''. Around these centers the wavefunction decays fast
enough to be normalizable. It would be interesting to investigate further
the typical spatial decay of such eigenstates around their (multiple) centers,
which we expect to be slower than exponential.

\section{conclusion}

In this paper we have studied the equilibrium problem of a particle in
a random potential with logarithmic correlations, through exact bounds,
numerical simulation, qualitative arguments and a renormalization group
method that we have developed specifically for this problem.
We have shown that it exhibits a glass transition at finite
temperature $T_c = 1/\beta_c >0$ in any dimension.
This confirms earlier conjectures and allows for a more detailed study of the problem.
The RG method allowed to obtain the universal features of the free energy
distribution at low temperature. The relation to the problem
of extremal statistic of correlated variables was investigated. It has
been found that it exhibits universal finite size corrections, 
consistent with our numerical calculations.

Most interestingly, we found that this logarithmic model
provides a particularly simple example (maybe the simplest)
of a finite dimensional model - i.e with {\it translationally invariant
disorder correlations}- 
such that the low temperature phase is non trivial. It is non trivial in
the sense that in the thermodynamic limit $L \to +\infty$, there are,
with a finite probability, several low lying states 
({\it i.e} possible positions of the particle)
with energy differences 
of order one, and separated in space by distances of order $L$. 
Thus the Gibbs measure at low 
temperature is dominated by ``a few'' spatially well separated
states. Interestingly, this transition and this type of glass phase
occurs only for logarithmically growing correlations, faster growth 
(e.g. as in Sinai model) yielding only a glass phase with
single ground state dominance,
while slower growth yielding only a high temperature phase.

Although oversimplified in some respect (it has no internal space)
it does provide one example of a model where the usual droplet picture 
(which assumes dominance of a single ground state - or several related by a symmetry)
does not apply. Rather, it provides one example where some features of
the physics usually associated to RSB, namely dominance by a few states 
with exponential free energy distributions, can be explicitly exhibited.
In fact, due to the finite dimensional correlations, there are some departures 
from the behaviour observed in the simplest prototype mean field models (such as
the REM), as can be seen for instance from the free energy distribution
which has more structure than a simple exponential. It would
of course be interesting
to explore further the additional features specific to finite dimensions.

Although the present model is already of obvious physical interest (in 2D it
describes e.g. a single vortex in a random gauge XY model) its non trivial
properties provide a motivation to search for models with more degrees of freedom
and with similar features. One way to proceed
would be to search for interface models via an internal dimensional expansion 
around the present model. The key feature however 
appears to be the {\it marginality} of the model, i.e the logarithmic growth 
of typical energy fluctuations. This corresponds 
to a fluctuation energy exponent $\theta =0$, i.e the situation
where the temperature (i.e the entropy) is {\it marginal} in the RG sense.
The droplet arguments indeed assume that $\theta>0$, consistent with the
single ground state dominance (and activated behaviour typical of 
a zero temperature fized point where $T$ is formally irrelevant). In the
situation $\theta =0$ one does expect more generally
power laws with $T$ dependent exponents, reminiscent of mean field.
It would thus be of great interest to similarly exhibit other
non trivial marginal models (e.g. spin models with $\theta =0$) 
with similar features \cite{footnote13}. Spin models where (domain wall) excitations
(in root mean square and in average) also scale 
logarithmically (as vortices in the random gauge XY model) are presumably
good candidates. 

On one hand we have developed a specific (Coulomb gas) renormalization
group (RG) approach 
to describe the model. From the study of the resulting nonlinear (KPP) RG
equation, we found explicitly that a freezing phenomenon occurs
at the glass transition temperature, and that in the glass phase a
broad (power law) 
distribution of fugacities develop - or equivalently exponential distribution
of local free energy.  It is different from more conventional perturbative
RG (e.g. the one which was used to study the dynamics of this model)
in the sense that the full distribution of probability is followed.
This turns out to be crucial to describe the low temperature phase.

On the other hand, as we have discussed, two {\it approximations} of
the present model, the REM approximation and the 
DPCT hierarchical version can both be solved using replica and 
do require considering the analytical continuation to $m \to 0$
of contributions of replica symmetry breaking saddle points 
\cite{derrida_dpct_replica}.

This shows that a RG approach which is explicitly replica symmetric but allows to
treat broad disorder distributions can be consistent with 
(approximate) approaches based on RSB saddle points
\cite{footnote16}. We have illustrated this on the REM 
which can be recast in terms of non linear RG equations, with a freezing transition.
In fact one of the striking
property of the model is that the RG equations derived here are similar - to the
order we have been working - to the one which hold for a continuous
version of the DPCT problem, the branching process. In particular it
indicates that both problem share the same universal finite size 
corrections.

We have also analyzed some connections in 2D (and via boundaries in 1D)
between the model of the particle and the Liouville and Sinh Gordon models. The
intensive free energy of the particle corresponds to the
scaling dimension in these models with $b=\beta/\beta_c$.
The glass transition corresponds to the weak to strong coupling
transition at $b=1$. Beyond, corresponding to the
glass phase, the scaling dimension
freezes as we have also shown via a direct RG approach on these models.
We have seen that under coarse graining an additional local field
appears in the LM and SGM, with broad distribution, and corresponds to
inhomogeneous configurations being generated (and broad fluctuations
of the local area since the local partition function corresponds to local area).

The present study raises interesting issues to be explored concerning the relations
with the continuum Liouville Field Theory (LFT). An outstanding question is
whether the conjecture of \cite{tsvelik_prl} is correct for the
correlations. Since we have obtained another result linking
the problem to the DPCT, the direct comparison of the LFT and the
DPCT remains to be studied. If it holds it means that the conformally
invariant many point correlations can be related in some limits (large
separations with fixed ratios $\ln r_{ij}/\ln r_{kl}$) to the results from the 
tree problem. It would also raise interesting issues
about the continuation of the LFT beyond $b=1$, and its relation with the
non trivial structure of the glass phase (with RSB features) 
in the equivalent particle model.

We have also extracted from our approach 
some consequences for the problem of
the $E=0$ critical eigenstate of 2D Dirac fermions in a random magnetic field.
We have confirmed, via our RG method, previous results concerning the
multifractal spectrum and extracted their finite size corrections. We have
found that the non trivial low $T$ phase of the particle translates
into peculiar quasi-localized eigentstates for the quantum problem,
peaked around a few distant centers. 
It raises the question of whether this property can be present in other quantum
systems. 

Another interesting question is whether the transition studied here has a signature in
the dynamics as well. Note that a similar non trivial structure at low temperature is
also present in the the Sinai model with a bias, which renormalizes onto
a random walk with algebraic waiting times distribution \cite{footnote18}.
However this is a driven system and it would be interesting to see whether non driven
systems in low dimension can exhibit similar features.

Finally, an outstanding question is how the present model can be studied using
2D conformal field theory (CFT). In particular one wonders what is the signature 
in this context, of 
the physics which was unveiled here, reminiscent of RSB, using RG with broad distributions.
The freezing phenomenon within the non linear RG, which transforms the
naive scaling dimensions into non trivial ones, should correspond to a similar
mechanism in CFT. Recent progress on CFT classification of 
disordered models where supersymmetry can be
used allows to hope that such progress is within sight. We hope that
the present RG method will apply to study other two dimensional
models with similar features and shed light on the more formal
field theoretic methods.

We acknowledge useful discussions with B. Derrida.
This work was supported in part by NSF grant DMR-9528578 (D.C.).

\appendix

\section{Existence of a transition}
\label{proof}

We use the same method as Derrida and Cook \cite{derridacook}
for the directed polymer problem \cite{derrida_private}.
It is easy to compute the first two moments pf $P[Z]$, 
using translational invariance 
and periodic boundary conditions:
\begin{eqnarray}
\overline{Z} = L^d e^{\frac{\beta^2}{2} \Gamma_L(0)} \sim L^{d + \beta^2 \sigma}
\end{eqnarray}
and 
\begin{eqnarray}
\overline{Z^2} = L^d e^{\frac{\beta^2}{2} 4 \Gamma_L(0)} 
\sum_{\bf r} e^{- \frac{\beta^2}{2} \tilde{\Gamma}_L({\bf r})} \\
\sim B L^{2 d + 2 \beta^2 \sigma}   \qquad \beta < \beta_2
\end{eqnarray}
the last estimate being valid as long as the sum over ${\bf r}$ is divergent,
i.e $\beta < \beta_2 = \sqrt{d/(2 \sigma)}$. The constant $B>1$
depends on the details of the model, e.g. for $d=1$ one
can write $B = \lim_{L \to + \infty}
\int_0^1 dy \exp(- \frac{\beta^2}{2} (\tilde{\Gamma}_L(L y) - 4 \sigma
\ln L))$. Thus for $\beta < \beta_2$ the ratio 
\begin{eqnarray}
\frac{\overline{Z^2}}{\overline{Z}^2} \to B
\label{b}
\end{eqnarray}
as $L \to +\infty$. In \cite{derridacook} it is shown that the property (\ref{b})
implies that:
\begin{eqnarray}
\text{prob}( \frac{1}{d \ln L} \ln Z = \frac{1}{d \ln L} \ln
\overline{Z} ) \ge 1/B 
\end{eqnarray}
as $L \to +\infty$. If we take for granted that the
free energy is self averaging, it implies 
that for $T > T_2 = \sqrt{2 \sigma/d}$ the quenched and annealed
(intensive) free 
energies coincide exactly $f(T)=f_A(T)$. Thus for $T > T_2$ the
(intensive) entropy 
is $s(T) = s_A(T) = - \partial_T f_A(T)$ and thus one has:
\begin{eqnarray}
s(T) = 1 - \frac{\sigma}{d T^2}   \qquad T>T_2   \label{entropy}
\end{eqnarray}
Since $s_A(T)$ becomes negative below $T=T_g = \sqrt{d/\sigma}$ it implies 
that there must be a temperature $T_c<T_2$ at which (\ref{entropy}) breaks
down and thus a phase transition. Although this is harder 
to prove, it seems that here (\ref{entropy}) holds down to $T_c=T_g$.

Awaiting a rigorous mathematical proof, we have not attempted to prove
self averaging of $f$. Not only is it highly reasonable in view of
our other results but in fact if it were not the case,
the above argument would imply a rather curious - and unphysical - distribution for $f$
(with a delta peak of non zero weight smaller than one).
In addition, as noted in \cite{derridacook},
by adjusting the small scale details of the model, 
the constant $B$ can be chosen as close to $1$ as wanted.

\section{extremal statistics of correlated variables}

\label{galambos-app} 

In this Section we summarize some results on the extremal
statistics of a set of random variables. We selected the
ones which are useful in putting the problem studied here
in a broader context. We recall some of the classic results 
from probability theory and we have chosen to illustrate them 
by adding a few simple arguments which emphasize the importance 
of some of these results to the physics of disordered systems.
We denote the $N$ random variables 
either $X_r$, $r=1,..N$ when they are normalized in a particular
way, or $V_r$ when they can be readily interpreted as the random potential
variables studied here (the two differing by a trivial uniform
rescaling $V(r) \equiv V_r  \propto X_r$). They apply 
directly to describe $d=1$ ($N=L$)
and can be usually extended to $d>1$ ($V({\bf r})$ and
$N=L^d$).

\subsection{uncorrelated variables}

It is natural to start with the case of $N$ uncorrelated variables
of identical probability distribution $P(V)$. The distribution
$P(V)$ can belong to three classes of extremal statistics, but we will 
recall only the Gumbell class. Schematically for this class, a 
well known theorem \cite{galambos} states that
there exist constants $a_N$ and $b_N$ such that for a fixed $\tilde{y}$

\begin{eqnarray}
Prob(V_{min} > b_N \tilde{y} - a_N ) \to \exp(-e^{\tilde{y}})
\label{theogumbell}
\end{eqnarray}

The constants $a_N$ and $b_N$ are determined as:
\begin{eqnarray}
&& \ln \int_{-\infty}^{- a_N} dV P(V) = \frac{1}{N}  \\
&& b_N = N \int_{-\infty}^{-a_N} dy \int_{-\infty}^{y} dV P(V)
\end{eqnarray}

For variables $X_r$ chosen from a centered Gaussian of unit variance 
$P(X)=\frac{1}{\sqrt{2 \pi}} e^{-X^2/2}$, one can choose $a_N$ and $b_N$ as:
\begin{eqnarray}
&& b_N = \frac{1}{\sqrt{2 \ln N}} \\
&& a_N = \sqrt{2 \ln N} - \frac{1}{\sqrt{2 \ln N}} \frac{1}{2} \ln (4 \pi \ln N)
\label{gumbgauss}
\end{eqnarray}
and thus one can then write schematically that:
\begin{eqnarray}
X_{min,N} \approx - \sqrt{2 \ln N} + \frac{1}{\sqrt{2 \ln N}} 
\left( \frac{1}{2} \ln (4 \pi \ln N)
+ \tilde{y} \right)
\label{approxmin}
\end{eqnarray}

where $\tilde{y}$ is distributed 
with the Gumbell distribution $p(\tilde{y}) = e^{\tilde{y}} \exp(- e^{\tilde{y}})$.

It is useful to note the property of reparametrization
associated to a monotonous function $\psi(V)$.
If one has (\ref{theogumbell}) for the minimum $V_{min}$ of the variables $V_r$ with the
constants $a_N$ and $b_N$, one also has (under some weak conditions)
that (\ref{theogumbell}) for the minimum $\psi(V_{min})$ of the variables
$\psi(V_r)$ with the constants $a'_N = - \psi(- a_N)$ and $b'_N = b_N/\psi'(-a_N)$.
Note also that we have illustrated how to show convergence to 
Gumbell (and generalized it to finite temperature) in the text.

For completeness we recall the necessary conditions for 
the convergence to Gumbell (i.e $P(V)$ belonging to the
Gumbell class). First $P(V)$ must decay fast enough
at $V \to - \infty$ so that there exists $y_0$ such that:
\begin{eqnarray}
\int_{-\infty}^{y_0}  dy \int_{-\infty}^{y} P(V) dV < + \infty
\end{eqnarray}
and second, defining
\begin{eqnarray}
R(t) = \frac{1}{\int_{-\infty}^{t} P(V') dV' } \int_{-\infty}^t dy 
\int_{-\infty}^{y} P(V) dV 
\end{eqnarray}
one must have for all $x < y_0$:
\begin{eqnarray}
\lim_{t \to - \infty} \frac{\int_{-\infty}^{t + x R(t)} P(V) dV}{
\int_{-\infty}^{t} P(V') dV'} = e^x
\end{eqnarray}
These conditions are in fact rather broad. Finally, note also the very 
powerful theorem 2.10.1 
of \cite{galambos} for the rate of convergence to the Gumbell fixed point.

\subsection{correlated variables}

\subsubsection{general lower bound}

We now consider correlated variables with distribution
$P(V_{1},...V_{N})$. Let us start with a
simple but very general bound and extract the consequences.
One has:
\begin{eqnarray}
G(x) = \text{Proba}( V_{min} < x) \le \sum_{r=1,N} \text{Proba}( V_{r} < x)
\label{generalbound}
\end{eqnarray}
since the reunion of all events $V_{r} < x$ implies the event 
$V_{min} < x$ and that $Prob(A ~U ~B) \le Prob(A) + Prob(B)$
(the bound is exactly saturated e.g. when there are strong 
correlations such that $V_{r}-V_{r'} > x$ for all $r \neq r'$).
For variables which have identical one particle distribution
$P(V_{r}) = \int \prod_{r' \neq r} dV_{r'} P(V_{1},...V_{N})$
one has: 
\begin{eqnarray}
G(x) \le N \int_{- \infty}^x P(V) dV
\end{eqnarray}
Let us illustrate the consequences for correlated variables
$X_1,..X_N$ such that the one particle distribution is a 
unit centered Gaussian. Then it implies for $x \to - \infty$:
\begin{eqnarray}
G(x) \le \frac{N}{\sqrt{2 \pi} x} e^{-x^2/2}
\end{eqnarray}
from which one immediately sees that it implies:
\begin{eqnarray}\nonumber 
&& \text{Proba}\left(X_{min} < x_N = - \sqrt{2 \ln N} 
+ \alpha \frac{\ln( 4 \pi \ln N) }{\sqrt{2 \ln N}}\right) \\
&&\qquad \qquad  \qquad \qquad  
\le \frac{1}{(4 \pi \ln N)^{(\frac{1}{2} - \alpha)}}
\xrightarrow[N \to +\infty]{} 0
\end{eqnarray}
by choosing $x=x_N$, for any $\alpha < 1/2$.
Thus one has a general lower bound for the minimum of correlated
variables. In particular for gaussian variables such that 
$\overline{V_{r}^2}  = 2 \sigma \ln N = 2 d \sigma \ln L$ one gets:
\begin{eqnarray}
V_{min} > - 2 d \sqrt{\sigma} \ln L +
\sqrt{\sigma} \alpha \ln( 4 \pi d \ln L)
\end{eqnarray}
with probability $1$ in the large $L$ limit for any $\alpha<1/2$.
Moreover choosing $\alpha=1/2$ and writing:
\begin{eqnarray}
V_{min} = - 2 d \sqrt{\sigma} \ln L + \sqrt{\sigma} ( \frac{1}{2}  \ln( 4 \pi d \ln L)
+ \tilde{y} )
\end{eqnarray}
one finds that:
\begin{eqnarray}
Prob(\tilde{y} < y)  \le e^{y}
\end{eqnarray}
This yields a lower bound which can be compared with 
the REM approximation defined in the text.
Note that the above upper bound is the exact
behaviour of the Gumbell distribution at large negative $y$,
so in a sense the REM approximation saturates the
bound in the tails. Consequently, to allow for a
larger tail (such as $y e^{-y}$ one needs at 
least a coefficient of $\ln \ln N$ strictly larger than $1/2$).

\subsubsection{short range correlations and convergence to Gumbell}

Let us now consider $N$ centered gaussian variables $X_{r}$ with a 
fixed correlation matrix $\Gamma_{r r'} = \overline{X_{r} X_{r'}}$, normalized so
that $\Gamma_{r r}=1$. A powerful bound, which refines
(\ref{generalbound}) above, allows to easily demonstrate convergence to
the Gumbell distribution for a large class of ``short enough range''
correlations. It compares two arbitrary correlators
$\Gamma^{(1)}_{r r'}$ and $\Gamma^{(2)}_{r r'}$ with 
$\Gamma^{(1)}_{r r} =\Gamma^{(2)}_{r r} =1$. Their associated
$G(x)$ functions satisfy \cite{galambos}:
\begin{eqnarray}
|G_1(x) - G_2(x)| \leq \sum_{r \neq r'} 
\frac{|\Gamma^{(1)}_{r r'} - \Gamma^{(2)}_{r r'}|}
{2 \pi (1 - m_{r r'}^2)^{1/2}} e^{- \frac{x^2}{1 + m_{r r'}} }
\label{boundG}
\end{eqnarray}
with $m_{r r'} = max(|\Gamma^{(1)}_{r r'}|,|\Gamma^{(2)}_{r r'})|$.
It is obtained by bounding $\partial G(x)/\partial \Gamma_{r r'}$
and integrating between $\Gamma_1$ and $\Gamma_2$. It will
be used to compare $\Gamma^{(1)}_{r r'} = \Gamma_{r r'}$ with 
the uncorrelated case $\Gamma^{(2)}_{r r'} = \delta_{r r'}$.

To adress the question of the universality of the Gumbell
distribution, let us now consider a ($d=1$) translationally invariant 
correlator $\Gamma_{r r'} = \Gamma(r-r')$ with $\Gamma(0)=1$
where $\Gamma(r)$ is a N-independent function which decays 
to zero as $r-r' \to + \infty$.

Inserting $x=a_N \tilde{y} - b_N$ of (\ref{theogumbell}) and
(\ref{gumbgauss}) into (\ref{boundG}) one easily gets that
if $\Gamma(r)$ decreases fast enough, one has $G(x=a_N \tilde{y} - b_N)
= \exp(-e^{\tilde{y}})$ at large $N$, i.e one has convergence to
the Gumbell distribution with exactly the same coefficients $a_N$ and $b_N$ 
as in the uncorrelated case, so that (\ref{approxmin}) still holds.
As one sees by studying the bound this result holds as long as
$\Gamma(r)$ decreases faster than $1/\ln(r)$ (this is 
theorem 3.8.2. of \cite{galambos}). The limiting 
case (which does not satisfy Gumbell, as discussed below) being 
$\Gamma(r) \sim \tau/\ln(r)$ at large $r$.

Let us give a simple self consistency argument,
more enlightening than the bounds, which explains
why $\Gamma(r) \sim 1/\ln r$ should be the limiting case 
between the short range (Gumbell) universality class 
and other behaviours. Let us split a set of
$2 N$ correlated variables $X_1,..X_{2 N}$ into 
subsystem 1, $X_1,..X_{N}$, and subsystem 2, $X_{N+1},..X_{2 N}$.
If correlations are very short ranged (e.g. exponentially
decaying) it seems reasonable to first neglect
correlations between 1 and 2 and find the
minimum in each subsystem, which read respectively:
\begin{eqnarray}
\tilde{X}_{\min,i}   \approx  \sqrt{2 \ln N} 
- \frac{1}{2} \frac{\ln (4 \pi \ln N)}{\sqrt{2 \ln N}}
+ \frac{x_i}{\sqrt{2 \ln N}}
\label{twomin}
\end{eqnarray}
with $i=1,2$ and where $x_1$, $x_2$ are independently distributed with 
the Gumbell distribution. The symbol $\tilde{V}$ indicates that the minimum (in each 
subsystem) is with respect to a slightly different distribution
than the original one, since all cross correlations between the 
two different subsystems have been set to zero. The second
stage is to add the correlations between the two subsystems.
Typically, the minima $1$ and $2$ will be a distance
$\sim N$ apart and thus their original cross-correlation is $\sim \Gamma(N)$,
and thus, for short range correlations,
very small compared to the fluctuating part $x_i/{\sqrt{2 \ln N}}$.
Thus the distribution of the mininum $X^{(2 N)}_{min}$ of the
original $2 N$ variables should be
given with better and better accuracy at large $N$, as $X^{(2 N)}_{min}
=\min( \tilde{X}_{\min,1}, \tilde{X}_{\min,2})$
(which is automatically satisfied by the approximation
(\ref{twomin})). 
The corrections are irrelevant at large scale provided the typical root mean
cross correlation between the subsystems remain smaller 
than the typical fluctuations of the minimum in each subsystem,
a condition which reads:
\begin{eqnarray}
\sqrt{\Gamma(N)} \ll 1/\sqrt{\ln N}
\end{eqnarray}
which indeed gives correctly the basin of attraction of the Gumbell
distribution. Furthermore, in the limiting case 
$\Gamma(r) \sim \tau/\ln r$ the above argument 
shows that the distribution
of the $x_i$ should be changed, 
which is also the case, as we now examine.

So, to summarize, if correlations are short ranged
with $\Gamma(r)$ decreasing faster than $1/\ln(r)$
this is the ``SR universality class''.
It includes the REM, and one can check that the finite
size corrections in \cite{derrida81} are reproduced (at $T=0$).

\subsubsection{long range correlations and absence of convergence to
Gumbell}

\label{limrange}

There is a simple but instructive model of correlated
variables which can be easily solved and illustrate 
cases where Gumbell does not hold. If one takes:
\begin{eqnarray}
V'_r = V_r + U
\end{eqnarray}
with $V_r$ a set of uncorrelated gaussian variables, 
$U$ a gaussian variable
uncorrelated with the $V_r$. Then clearly, if one chooses 
the variance of $U$ big enough 
(\ref{approxmin}) cannot hold.
To keep using normalized variables ($\Gamma_{rr}=1$) 
one defines:
\begin{eqnarray}
X'_{r} = \frac{1}{\sqrt{1 + w_N}} X_{r} +  u \sqrt{\frac{w_N}{1 + w_N}}
\end{eqnarray}
where $u$ is a centered gaussian random variable with unit variance.
The correlation matrix is then 
$\Gamma'_{r r'} = \frac{1}{1 + w_N} (\delta_{r r'} + w_N)$.
Clearly one has:
\begin{eqnarray}
X'_{min} = \frac{1}{\sqrt{1 + w_N}} X_{min} +  u \sqrt{\frac{w_N}{1 + w_N}} 
\end{eqnarray}
Using the expression (\ref{approxmin}) for $X_{min}$ one sees
that for deviations from Gumbell to arise one needs that
$w_N \sim \tau/\ln N$. In that case one gets from 
(\ref{approxmin}) that
\begin{eqnarray}
X'_{min} \approx \sqrt{2 \ln N} - \frac{1}{2} \frac{\ln (4 \pi \ln N)}{\sqrt{2 \ln N}}
+ \frac{x_i + \sqrt{2 \tau} u + \tau}{\sqrt{2 \ln N}}
\end{eqnarray}

These simple considerations thus allow to understand 
simply the limiting case, that if $\Gamma(r)$ decreases as $\tau /\ln(r)$ 
one has that (\ref{approxmin}) still hold (with the same constants)
but the distribution of $\tilde{y} - \tau$ now converges instead 
to the convolution of the Gumbell distribution and the gaussian 
of variance $2 \tau$ (see e.g. theorem 3.8.2. of \cite{galambos}).

Increasing the range of correlations even further, one gets
into a regime where the fluctuating part (in the $X$ variables)
is larger than $1/\sqrt{\ln N}$ (and thus in the $V \sim X \sqrt{\ln N}$ 
variables the dominant finite size corrections
are non selfaveraging). The fluctuations become then entirely gaussian,
being controlled by the $U$ part, i.e the $q=0$ mode. For instance, 
if $\Gamma(r)$ decreases as $1/(\ln(r))^{\alpha}$ with $1/3 < \alpha < 1$
then (theorem 3.8.4. of \cite{galambos}) one has:

\begin{eqnarray}
&& P(V_{\min} > - \Gamma(N)^{1/2} x - (1 - \Gamma(N))^{1/2} \sqrt{2 \ln N} \\
&&( 2 \ln N - \frac{1}{2} \ln (4 \pi \ln N) )) \to \int_{-\infty}^x 
{2 \pi}^{-1/2} e^{- x^2/2}
\end{eqnarray}

As illustrated below, this behaviour (entirely controlled by the $q=0$ mode)
is in a sense more long range, and further away from Gumbell
than the problem of log-correlated variables that we are
interested in and that we now discuss.

\subsubsection{log-correlated variables}

The case of log-correlated variables is difficult and
little is known. We just make a few comments.

\begin{figure}[thb] 
\centerline{\fig{8cm}{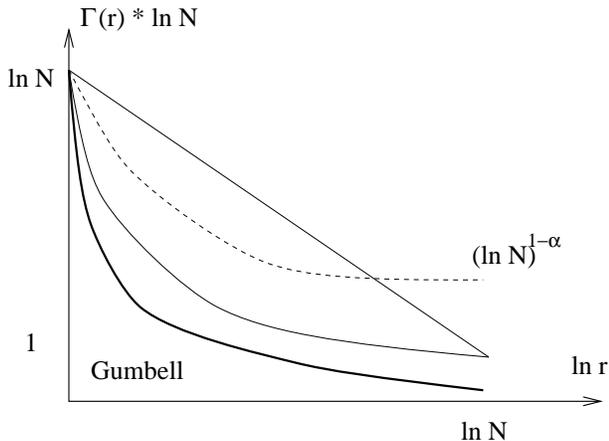}}
\caption{ \narrowtext Correlations as a function of
$\ln r$. The straight line corresponds to the log-correlated
variables studied here. The thick line corresponds to
the limit where the short range Gumbell (and REM) behaviour
holds, with $\Gamma(r) \sim 1/(\ln r)^{\alpha}$
and $\alpha <1$, the curved solid line to the case where a convolution
of Gumbell and gaussian holds (marginal case $\alpha=1$) 
and the dotted line $\alpha>1$ when the mode $q=0$ dominates
the behaviour. \label{fig13}}
\end{figure}

Let us first discuss the form of the correlator.
The correlator (for the normalized variables
$X_r = V_r/\sqrt{2 \sigma \ln N}$ in $d=1$)
is of the form:
\begin{eqnarray}
\Gamma(r) = 1 - \frac{\ln r}{\ln N}
\label{scaling}
\end{eqnarray}
for $1 \gg r \gg N$. One must then distinguish the
two other regions. For small $r$, 
the precise form could vary by adding a short range correlated noise.
This is what we term short scale details, and an important question
is the extent of universality of the results (scaling of minima, distribution)
with respect to the small $r$ form. For $r \sim L$
the behaviour depends on boundary conditions, which may also be
important (see below). For the periodic system in the simulation
$\Gamma(r) = \Gamma(N-r)$ and $\Gamma(r)$ actually becomes negative
at $r=N/2$ and of order $c/\ln N$ (see Section \ref{numerics}). Adding a small
uniform $U$ noise, as described above in \ref{limrange}, could make 
$\Gamma(N/2)=0$, so generally speaking one can discuss
forms such that $\Gamma(N/2)=0$. Seen as a scaling function of
$z=\frac{\ln r}{\ln N}$, $\Gamma(r)$ then converges for large $N$
towards (\ref{scaling}), but it does have boundary layers at
$z=0$ and $z=1$.

It is useful to plot on the same graph the various 
cases studied in this Section. 
This is represented in Figure \ref{fig13}.
We have represented schematically $\Gamma(r) \ln N$ 
versus $\ln r$, for the log-correlated form
(\ref{scaling}) above, and for the various
cases $\Gamma(r) \sim \frac{\ln N}{(\ln r)^{\alpha}}$
with $\alpha>1$ (Gumbell behaviour), $\alpha=1$ and
$\alpha<1$.

As discussed above in the log-correlated case 
the behaviour of $\Gamma(r)$ near $\ln r=\ln N$ can 
be considered as uncertain to order $1/\ln N$.
This can be seen either from the $q=0$ mode, which
depending on boundary conditions one may adjust by
this amount, as discussed above, or even looking at the
first non trivial mode, $q=2 \pi/L$ which has a contribution
of the same order. We know from the previous paragraph that
these contributions can shift the $x$ variable by a 
gaussian, so it makes it unlikely that the Gumbell
distribution would hold in that case.

To conclude, we have given the various behaviours as a function
of the range of correlations. The presence of the 
$\ln \ln N$ corrections seems to be more robust than the
distribution of $x$. For the marginal case with $q=0$ 
LR disorder, the same $\frac{1}{2}  \ln \ln N$ corrections
hold as for the REM, while the distribution is changed.
On the other hand, for log-correlated variables,
we expect a different coefficient $\frac{3}{2} \ln \ln N$
as discussed in the text (and we do not expect the
Gumbell distribution to hold).

\section{Gumbell via RG}
\label{app-indep}

In a more detailed analysis Eq. (\ref{lnln}) yields 
$\ln \ln (1/G_l(x)) = l + \ln \ln (1/G_0(x))$ which can
be rewritten in a front-like form:
\begin{eqnarray}
G_l(x) = \exp( - e^{l + \phi(x)} )
\end{eqnarray}
where $\phi(x) = \ln \ln (1/G_0(x))$. In this Appendix
we set $d l \to l$. The center
of the front is at $x= - m(l)$ solution
of $\phi(- m(l)) = - l$. One can Taylor
expand $\phi(x) = - l + y + \frac{1}{2} \delta_l y^2 + ..$
with $y =  \alpha_l (x + m(l))$, 
$\alpha_l = \phi'(- m(l))$ and $\delta_l = \phi''(- m(l))/\phi'(- m(l))^2$.
Thus in the variable $y$, $G_l$ converges to a 
Gumbell limit distribution ${\cal G} (y) = \exp( - e^{y})$.
It holds provided higher terms in the Taylor expansion
are irrelevant (a necessary, and in simplest
cases sufficient, condition being that the second one $\delta_l \to 0$).

If no rescaling of disorder is performed,
in the relevant large negative $x$ region one 
has $G_0(x) \approx 1$ and thus $\phi(x) \approx \ln (1-G_0(x))$.
Two cases must be distinguished because the
limit $T \to 0$ and $N \to +\infty$ do not commute:

(i) {\it finite fixed temperature $T>0$}:
then for $x \to - \infty$ one has 
$1-G_0(x) \sim C_1(\beta) e^{\beta x} (1 + O(e^{\beta x}) )$ 
with $C_k = \int_V P(V) e^{- k \beta V}$
and we assume that $C_1,C_2 < + \infty$ exists
(distributions falling faster than exponentials).
Then the situation is simple as $\phi(x) = \beta x + \ln C_1(\beta)
+ O(e^{\beta x})$, $m(l) \sim l/\beta + 1/\beta \ln C_1(\beta)$,
$\alpha_l = \beta$ and $\phi''(x)/\phi'(x)^2 \to 0$ exponentially
fast. For a Gaussian distribution:
\begin{eqnarray}
m(l) \sim \frac{1}{\beta} l + \frac{1}{2} \sigma \beta
\end{eqnarray}
There is {\it no transition} to a glass phase.

(ii) {\it zero temperature}: it is an 
extremal statistics problem. Then clearly $1-G_0(x)$ does not
decay as an exponential. Let us consider a class
of distributions such that $1-G_0(x) \sim (A |x|)^{-\gamma}
\exp( - (B |x|)^{\alpha} )$ with $\alpha >1$ (plus exponentially small
corrections). This contains the Gaussian (of variance $\sigma$),
of most interest here, for $\alpha=2$, $B = 1/\sqrt{2 \sigma}$, $\gamma =1$
and $A=\sqrt{2 \pi/\sigma}$.
Then one easily finds from above that

\begin{eqnarray}
&& m(l) \approx \frac{1}{B} (l - \frac{\gamma}{\alpha} \ln l 
- \gamma \ln(A/B) )^{\frac{1}{\alpha}} \\
&& \alpha_l \approx B \alpha 
(l - \frac{\gamma}{\alpha} \ln l - \gamma \ln(A/B))^{1-\frac{1}{\alpha}}
\label{norm}
\end{eqnarray}
and that $\phi''/{\phi'}^2 \sim 1/|x|^\alpha$ thus the
convergence to the Gumbell front holds. Note that the quantity
$\alpha_l m(l) \sim \alpha l - \gamma \ln l + O(1)$ exhibits some
universality.

One thus recovers the standard theorems for extremal
value statistics reviewed in Appendix \ref{galambos-app}, and the
relation to the normalizing constants defined there 
as:
\begin{eqnarray}
m(l) = a_N \quad \alpha_l = \frac{1}{b_N} \quad l = \ln N
\end{eqnarray}
In the Gaussian case, using the values given above one
finds that (\ref{norm}) indeed yields Eq. \ref{gumbgauss} in Appendix \ref{galambos-app}
(up to subdominant terms). Although the distribution is universal, the 
normalizing constants obviously depend on the details of the tail of the
distribution. Note in all cases the presence of finite 
size corrections involving a logarithm.

There is a very small temperature ($\beta_L \sim \sqrt{\ln L}$ for Gaussian)
where the behaviour of $G_0(x)$ changes from (i) to (ii). It can be seen
seen by rescaling temperature or equivalently disorder,
with system size as in the REM.

Let us examine the case where the constants 
$A_l$ and $B_l$ are rescaled and now $l$-dependent
(see also e.g. \cite{bouchaud_mezard}). One
can still use formulae (\ref{norm}). Let us choose 
$B_l = b l^{-1 + 1/\alpha}$ and $A_l/B_l=cst$ 
(which includes the Gaussian REM). One finds 
at $T=0$ that $m(l) \sim  \frac{1}{b} (l - \frac{\gamma}{\alpha^2} \ln l - 
\frac{\gamma}{\alpha} \ln(A/B))$ and $\alpha_l \sim b \alpha$.
In the gaussian case $\sigma_l = 2 \sigma l$ 
one recovers the REM result:
\begin{eqnarray}
&& m(l) \approx \sqrt{\sigma} (2 l - \frac{1}{2} \ln(4 \pi l)) \\
&& \alpha_l \to \frac{1}{\sqrt{\sigma}}
\end{eqnarray}
at $T=0$ (i.e (\ref{KPP-m}) setting $l \to d l$ and $\sigma \to \sigma/d$).
The analysis can be performed at any $T$ and
now yields a transition temperature when the behaviour of
$G_{0,l}(x)$ at large $x$ changes.

\section{An extended model}
\label{extended}

A richer phase diagram can be obtained by adding a 
logarithmic background potential \cite{footnote14}
$V_0({\bf r}) = J \ln \frac{|{\bf r}|}{a}$ to the previous 
random potential $\overline{(V_d ({\bf r})-V_d ({\bf r}'))^{2}} \sim 4 \sigma
\ln \frac{|{\bf r}-{\bf r}'|}{a}$ for $a \ll |{\bf r}-{\bf r}'| \ll L$
and $\overline{V_d(r)} =0$ (i.e writing $V({\bf r}) = V_0({\bf r}) +
V_d({\bf r})$) 
in (\ref{def-1part}). The choice of 
the origin breaks translational invariance. The competition
between the disorder and the binding background potential 
(which if strong enough, tends to favor
localizing the particle in wells far from ${\bf r}=0$) yields
the phase diagram of Fig. \ref{pdiag}.
Another closely related model (model II) which preserves 
statistical translational invariance
and has the same phase diagram is:
\begin{eqnarray}
Z_v[V]= 1 + \left(\frac{L}{a}\right)^{- \beta J} 
\sum_{\bf r} e^{- \beta V_d({\bf r})}
\end{eqnarray}
describes a problem with either zero or one particle (vortex) present,
the energy cost of the vortex being $J \ln(L/a)$. It is thus a one vortex
toy model of the recently studied XY model with random phase shifts
\cite{carpentier_xy_prl,carpentier_pld_long}.

In the absence of disorder the model with a background potential
(model I) trivially exhibits a transition at $\beta= \beta^* = d/J$. At low
temperature $\beta > \beta^*$ the Gibbs measure is 
$p({\bf r}) \sim C (\frac{a}{r})^{\beta J}$ with $C=Z_{L=\infty}$
a finite constant and the particle is bound to ${\bf r}=0$
(it has a finite probability to be within a fixed 
distance of ${\bf r}=0$). At high temperature
$\beta < \beta^*$ the Gibbs measure becomes
$p({\bf r}) \sim (\frac{a}{L})^{d- \beta J} (\frac{a}{r})^{\beta J}$
and the particle is delocalized. This
transition can be seen in the free energy
density $f = F/\ln L= - T \ln Z/\ln L$ since:
\begin{eqnarray}
&& f = 0   \qquad ~~~~~~ \beta > \beta^* \\
&& f = - ( J \beta^* - \beta ) \qquad \beta < \beta^*
\end{eqnarray}
for $\beta < \beta^*$. This first order transition
occurs as $f$ reaches its bound (since $Z > 1$
due to the lattice cutoff, one has that $f \le 0$). The model
II has the same $f$ and a similar transition with either one vortex
present $\beta > \beta^*$ or zero $\beta < \beta^*$.

In the presence of disorder the RG developed 
in this paper can be extended straightforwardly
and leads to:
\begin{eqnarray} \label{KPP-J}
&& \frac{1}{d}\partial_{l} G (x)= \frac{J}{d} \partial_{x} G +
\frac{\sigma}{d} \partial_{x}^{2} G + F{[}G{]} \\
&& F{[}G{]}= - G (1-G)
\end{eqnarray}
The additional term thus results in a simple shift in the front velocity.
The position of the front $m(l)$ thus leads to the free energy
$f = m(l)/(d l)$ which can have three distinct analytical
forms:
\begin{eqnarray} \nonumber 
&& \beta m(l)/l \sim d \beta f(\beta) = \\
&&\qquad \qquad 
 - ( d + \sigma \beta^2 - J \beta )  \qquad \text{high T phase I} \\
&&\qquad \qquad   - ( 2 d \frac{\beta}{\beta_c} - J \beta ) \qquad \text{localized phase II} \\
&&\qquad \qquad  0  \qquad ~~~~~~ \text{bound phase III}
\end{eqnarray}
The phase diagram is represented in Fig \ref{pdiag} 
using the reduced temperature $T/J$ and the
dimensionless disorder parameter $\hat{\sigma}=\sigma/J^2$.
For $4 d \sigma < J^2$ one defines $\beta^*(\sigma)=
\frac{1}{2 \sigma} (J - \sqrt{J^2 - 4 d \sigma})$.
The RG analysis yields three phases. In the model with
the background potential (model I) they are, respectively:

{\it the high temperature phase} (for $\beta < \beta_c$
when $4 d \sigma > J^2$ and for $\beta < \beta^*(\sigma)$
for $4 d \sigma < J^2$): Entropy wins and the particle is 
delocalized over the system.

{\it the localized phase} (for
$\hat{\sigma} > \hat{\sigma}_c = 1/(4 d)$ and $\beta > \beta_c = \sqrt{d/\sigma}$):
The KPP velocity is frozen. The disorder wins and the particle
freezes in wells away from the origin.

{\it the bound phase} (for $\hat{\sigma} < \hat{\sigma}_c = 1/(4 d)$
and $\beta < \beta^*(\sigma)$):
The particule is bound to the origin. Within this phase
near the phase boundaries (where the bound state length is large) a  
a crossover can be distinguished as a remnant of the freezing 
transition. The bound phase arises because of the bound $f \le 0$
(or equivalently the velocity of the KPP equation
must remain positive).

In model II the bound phase correspond to no vortex present. When
one vortex is present it can be either localized in a few wells
or in a high-T phase (as studied in the text of this paper). 

Both transitions away from the bound phase are first order
while the transition between high temperature phase and
localized phase is continuous. An interesting feature is the
multicritical point where the transition
becomes continuous.

\begin{figure}
\centerline{\fig{7cm}{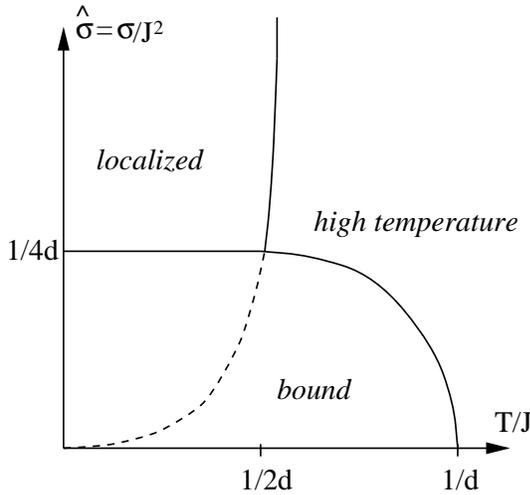}}
\caption{
\narrowtext
Phase diagram in presence of both disorder and external
potential. The freezing of the KPP velocity still occurs
at $\beta=\beta_c$ and is represented by the dashed line 
and its solid prolongation: it remains a transition
line for $\sigma > 4 d J^2$ and becomes a crossover line 
for $4 d \sigma < J^2$. 
\label{pdiag}}
\end{figure}

%
%

\begin{thebibliography}{10}

\bibitem[*]{lptens}  Laboratoire Propre du CNRS, associ{\'e} {\'a} l'Ecole
Normale Sup{\'e}rieure et {\`a} l'Universit{\'e} Paris-Sud. 
      
\bibitem{parisi}
G. Parisi, Phys. Lett A {\bf 73} 203 (1979) J. Phys. A {\bf 13} L 115 (1980),
  Phys. Rev. Lett. {\bf 50} 1946 (1983).

\bibitem{cuku}
L. F. Cugliandolo and J. Kurchan Phys. Rev. Lett. {\bf 71} 173 (1993), J. Phys.
  A {\bf 27} 5749 (1994), Phil. Mag. B {\bf 71} 501 (1995).

\bibitem{fisher_huse}
D. S. Fisher and D. A. Huse Phys. Rev. B {\bf 38} 373 and 386 (1988).

\bibitem{newman_stein}
C. M. Newman and D.L. Stein 
Phys. Rev. Lett. {\bf 76} 515 (1996), 
Phys. Rev E {\bf 57} 1356 (1998), 
J. Stat. Phys. {\bf 94}  709 (1999). 

\bibitem{replica_ref}
For a recent review see also E. Marinari, G. Parisi, F. Ricci-Tersenghi, J, Luiz-Lorenzo
and F. Zuliani, cond-mat/9906076

\bibitem{derrida81}
B. Derrida, Phys. Rev. B {\bf 24},  2613  (1981).

\bibitem{derrida85}
B. Derrida, J. Phys. Lett. {\bf 46},  401  (1985).

\bibitem{derrida88}
B. Derrida and H. Spohn, J. Stat. Phys. {\bf 51},  817  (1988).

\bibitem{derrida_rw}
B. Derrida, Physica D {\bf 107} 186 (1997).

\bibitem{derrida_dpct_replica}
B. Derrida, Physica A163, 71-84 (1990).

\bibitem{chamon96}
C. de~C.~Chamon, C. Mudry, and X. Wen, Phys. Rev. Lett. {\bf 77},  4194
  (1996).

\bibitem{tsvelik_prl}
I. Kogan, C. Mudry and A.M. Tsvelik Phys. Rev. Lett. {\bf 77} 707 (1996).

\bibitem{castillo97}
H. Castillo {\it et~al.}, Phys. Rev. B {\bf 56},  10668  (1997).

\bibitem{nattermann_xy_lowtemp}
T. Nattermann, S. Scheidl, S.~E. Korshunov, and M.~S. Li, J. de Phys. I {\bf
  5},  565  (1995).

\bibitem{scheidl_xy_lowtemp}
S. Scheidl, Phys. Rev. B {\bf 55},  457  (1997).

\bibitem{tang_xy_lowtemp}
L.~H. Tang, Phys. Rev. B {\bf 54},  3350  (1996).

\bibitem{korshunov_nattermann_energy}
S.~E. Korshunov and T. Nattermann, Physica B {\bf 222},  280  (1996).

\bibitem{carpentier_xy_prl}
D. Carpentier and P. {Le Doussal}, Phys. Rev. Lett. {\bf 81},  2558  (1998).

\bibitem{carpentier_pld_long}
D. Carpentier, P. Le Doussal, 
\texttt{cond-mat/9908335}, submitted to Nuclear Physics (1999).

\bibitem{arrays}
E. Granato and J.M. Kosterlitz, Phys. Rev. B {\bf 33} 6533 (1986).

\bibitem{giamarchi_book_young}
T. Giamarchi and P. {Le Doussal},  in {\em Statics and dynamics of disordered
  elastic systems}, edited by A.~P. Young (World Scientific, Singapore, 1998),
  p.\ 321, \texttt{cond-mat/9705096}.

\bibitem{electrons}
C G Grimes and G. Adams Phys. Rev. Lett {\bf 42} 795 (1979) F.I.B. Williams
  Hel. Physica Acta {\bf 65} 297 (1992).

\bibitem{footnote_theta}
The exponent of the growth of correlations $V \sim r^\alpha$ would correspond
  in, e.g. the problem of a manifold in a random potential to
  $\alpha=\theta/\zeta$ where $r \sim x^\zeta$ and $\theta$ is the energy
  fluctuation exponent such that $\delta E \sim x^\theta$. Thus $\alpha=0$
  corresponds to $\theta=0$.

\bibitem{grinstein}
A. Ludwig, M.P.A. Fisher, R. Shankar and G. Grinstein, Phys. Rev. B {\bf 50}
  7526 (1994).

\bibitem{galambos}
J. Galambos, {\it The Asymptotic Theory of Extreme Order Statistics},
R.E. Krieger Publishing. Co., Malabar, Florida (1987). 

\bibitem{rammal}
R. Rammal, J. Physique (Paris) {\bf 46} 1837 (1985).

\bibitem{vinokur}
V.M. Vinokur, M.C. Marchetti, L.W. Chen cond-mat/9604184.
Phys. Rev. Lett. {\bf 77} 1845 (1996) 

\bibitem{bouchaud_mezard}
J. P. Bouchaud and M. Mezard, 
J. Phys. A {\bf 30} 7997  (1997). 

\bibitem{fisher_wetting}
D. S. Fisher D. A. Huse Phys. Rev B {\bf 32} 247 (1985).

\bibitem{majumdar_sire}
Sire C., Majumdar S.N., Rudinger A.,
Phys. Rev. E {\bf 61} 1258  (2000) 

\bibitem{tabar}
E. Abdalla and M. R. Tabar, 
Phys.Lett. B {\bf 440} (1998) 339.

\bibitem{comtet98}
For a review see A. Comtet and C. Texier,  Lect. Notes in
  Physics {\bf 502} (1998), 
\texttt{cond-mat/9707313}, 
and references therein.

\bibitem{diff}
D.S. Fisher, D. Friedan, Z. Qiu and S.J. Shenker,
  Phys. Rev. A {\bf 31} 3841 (1985),
V. E. Kratsov, I. V. Lerner and V. I. Yudson, 
  J. Phys. A {\bf 18} (1985) L 703, 
J.P. Bouchaud, A. Comtet, A. Georges, P. Le Doussal, 
  J. Phys. Paris {\bf 48} (1987) 1445 and 49 (1988) 369.

\bibitem{footnote3}
But not necessarily for stronger correlations.

\bibitem{footnote4}
The need to understand the problem beyond the REM approximation can be
  illustrated as follow. Suppose that one simply shift all variables $V({\bf
  r})$ by a uniform but random gaussian variable of variance $\sim \ln L$. This
  clearly does not change the true transition temperature while it does change
  the REM approximation transition temperature (it adds a constant to
  $\Gamma_L(0)$ without changing $\tilde{\Gamma}_L({\bf r})$).

\bibitem{gnedenko}
B. V. Gnedenko, Ann. Math. {\bf 44} 423 (1943).

\bibitem{footnote5}
The problem is subtle as can be seen by the failure of the following simple
  argument. Since there are, as will be discussed, only few low lying states
  adding a small gaussian noise with only short range correlations (or
  uncorrelated) $V({\bf r}) \to V({\bf r}) + w({\bf r})$ one could conclude
  that the distribution of $\delta V_{\min}$ is changed (since typically
  $V_{min} \to V_{min} + w({\bf r}_{\min})$). This argument is insufficient
  however, as it would yield the same conclusion for the REM approximation
  while there we know (see below) that the distribution does not change: there
  the probability that a secondary minima becomes, after the shift, the
  absolute one compensates exactly the additional gaussian fluctuation.

\bibitem{footnote6}
these results are probably more general than strictly for Gaussian
  distributions, provided the single site distribution $P(V_r)$ decays faster
  than exponentially. Clearly though, one should be careful in making general
  statements about correlated variables: one can always taylor artificial
  correlations to produce more or less pathological exceptions, e.g. one can
  consider a generalized LR Sinai potential $V(r)$ which performs a one sided
  random walk, and which clearly has infinitely many exactly degenerate minima,
  separated by arbitrarily large barriers.

\bibitem{footnote15}
we assume here that the tail of $P_1(V)$ is faster than an exponential.

\bibitem{footnote7}
e.g. scaling $R$ algebraically with $L$).

\bibitem{golosov}
A. O. Golosov Soviet. Math. Dokl. {\bf 28} 19 (1983).

\bibitem{laloux_pld_sinai}
L. Laloux and P. Le Doussal, 
Phys. Rev. E {\bf 57} 6296 (1998)
  and references therein.

\bibitem{fisher_pld_monthus}
D. S. Fisher, P. Le Doussal and C. Monthus Phys. Rev. Lett. {\bf 80} 3539
  (1998) and P. Le Doussal, C. Monthus, D. S. Fisher Phys. Rev. E, {\bf 59}
  4795 (1999).

\bibitem{footnote_x2}
Note that, although the Gibbs probability is typically concentrated around a
  single minimum, some quantities are very sensitive to the rare configurations
  with several almost degenerate low lying minima. For instance the thermal
  width of a packet $\overline{\langle x^2 \rangle - \langle x \rangle^2}$ is
  easily computed (via methods as in \cite{fisher_pld_monthus}) and shown to
  grow as $T L^{3/2}$ (since with probability $1/L^{1/2}$ there are two
  degenerate minima separated typically by $L$).

\bibitem{footnote8}
As detailed in \cite{carpentier_pld_long} it arises naturally if one consider
  first $\overline{Z^m}$ on a lattice, which can be written {\it exactly} as a
  vector CG on the lattice, and then take the continuum approximation of this
  vector CG model).

\bibitem{nienhuis87}
B. Nienhuis,  in {\em Phase Transitions and Critical Phenomena}, edited by C.
  Domb and J. Leibovitz (Academic Press, London, 1987), Vol.~11.

\bibitem{footnote9}
The convention is different from \cite{carpentier_pld_long} with $x \to -x$ and
  $G \to -G$.

\bibitem{footnote10}
It is correct if $<z>_{P_l}$ remains finite which is the case if one starts
  from a well behaved initial condition. However, below the glass transition,
  $<z>_{P_l}$ grows very fast with $l$ (being dominated by very rare events)
  and the intermediate asymptotic behaviour becomes then different from $1 -
  G_l(x) \sim e^{\beta x}$ (only at much larger $x$ is this behaviour
  recovered).

\bibitem{saarloos89}
W. {Van~Saarloos}, Phys. Rev. A {\bf 39},  6367  (1989).

\bibitem{brunet97}
E. Brunet and B. Derrida, Phys. Rev. E {\bf 56},  2597  (1997).

\bibitem{ebert98}
U. Ebert and W. van Saarloos, Phys. Rev. Lett. {\bf 80},  1650  (1998).

\bibitem{ebert98b}
U. Ebert and W.~V. Saarloos, 
\texttt{cond-mat/0003181} (unpublished).

\bibitem{saarloos98}
W. {Van~Saarloos},  in {\em Altenberg summer school on fundamental problems in
  statistical physics (1997)}, 
Phys. Rep. {\bf 301}, 9 (1998). 

\bibitem{bramson83}
M. Bramson, Mem. of the Am. Math. Soc. {\bf 44},    (1983).

\bibitem{spc}
W. Van~Saarloos, private communication, and \cite{ebert98b}. 

\bibitem{footnote18} in the driven Sinai diffusion a very similar
property holds. The process renormalizes onto a directed
walk between traps with random waiting times $\tau_i$
- in direct correspondance with the $z_i$ 
variables here - also with a power law distribution,
see P. Le Doussal PhD thesis 1987 or \cite{laloux_pld_sinai}
and references therein.

\bibitem{scheidl97}
S. Scheidl, Phys. Rev. B {\bf 55},  457  (1997).

\bibitem{lukyanov}
S. Lukyanov and A.B. Zamolodchikov, 
Nucl. Phys. B{\bf 493}, 571-587 (1997). 

\bibitem{fateev}
V. Fateev, S. Lukyanov, A.B. Zamolodchikov, and A.B. Zamolodchikov,
Phys. Lett. B{\bf 406}, 83-88 (1997).

\bibitem{korepin}
V. E. Korepin and N. A. Slavnov,
J. Phys. A{\bf 31}, 9283-9295 (1998) .

\bibitem{footnote17}
obtained in \cite{lukyanov} from $f_{sh}(\mu) = - \frac{1}{4} M^2 \tan( \pi \xi/2)$
with $\xi=\frac{- b^2}{ 1 + b^2 }$, the physical mass of particles
in the sinh-Gordon model being $m=2 M \sin(\pi \xi/2)$ with
$\mu = - \frac{\Gamma[-b^2]}{\pi \Gamma[1+b^2]}
( M \frac{\sqrt{\pi} \Gamma[\frac{1 + \xi}{2}]}{
2 \Gamma[\frac{\xi}{2}]})^{2 + 2 b^2}$. 

\bibitem{vega}
H. De Vega, private communication.

\bibitem{polyakov}
A. Polyakov, Phys. Lett. B {\bf 103} 207 (1981).

\bibitem{seiberg}
N. Seiberg Prog. Theor. Phys. {\bf 102} 319.

\bibitem{gervais}
J.L. Gervais and A. Neveu, 
Nucl. Phys. {\bf B 238} 125 and 396 (1984), 
{\bf B 257} 59 (1985).

\bibitem{david}
F. David, Mod. Phys. Lett. A {\bf 3} (1988) 1651, 
Nucl.  Phys. B {\bf 487} [FS] 633 (1997) and references therein.

\bibitem{z2}
A.B. Zamolodchikov and Al.B. Zamolodchikov,
Nucl. Phys. B{\bf 477}  577-605(1996). 

\bibitem{cates}
see e.g. M.E. Cates, Phys. Lett. B {\bf 251} 553 (1990), J. Correia
  \texttt{hep-th/9805002} and references therein.

\bibitem{reuter}
M. Reuter and C. Wetterich, 
Nucl.Phys. B{\bf 506},  483-520 (1997)

\bibitem{footnote11}
For a simple exponential one gets $R_{q} (L) = \frac{1}{q^2} (2/\pi
  \xi^2)^{q-1}$.

\bibitem{footnote13}
Note that a growth of the onsite variance as $\sigma(L) \sim (\ln L)^b$ with $b
  > 1$ yields also to ground state dominance (although a very weak one, see
  Appendix for extremal statistics). Thus at least for the present toy model,
  $\theta=0$ is not a sufficient condition.

\bibitem{footnote16}
For an attempt to construct a RG which incorporates RSB see
P. Le Doussal and T. Giamarchi, 
 Phys. Rev. Lett. {\bf 74} 606 (1995).

\bibitem{derridacook}
J. Cook and B. Derrida, J. Stat. Phys. {\bf 57} 89 (1989).

\bibitem{derrida_private}
We thank B. Derrida for pointing out this method to us.

\bibitem{footnote14}
(e.g. on a lattice of spacing $a$, such that 
$V_0(|{\bf r}=0|)= V_0(|{\bf r}=a|)=0$).

\end{thebibliography}
%

\unecol
\end{document}